\documentclass[fleqn,usenatbib]{mnras}
\usepackage{newtxtext,newtxmath}

\usepackage{amssymb}
\usepackage{graphicx}
\usepackage{amsmath}
\usepackage[T1]{fontenc}
\usepackage{ae,aecompl}

\def\Msun{\hbox{$\rm\thinspace M_{\odot}$}}

\title[The evolving galaxy stellar mass function]
{The evolution of the galaxy stellar mass function over the last twelve billion years from a combination of ground-based and {\it HST} surveys}
\author[D. J. ~McLeod et al.]
{D. J. McLeod$^{1}$\thanks{Email: mcleod@roe.ac.uk}, R. J. McLure$^{1}$,
  J. S. Dunlop$^{1}$, F. Cullen$^{1}$, A. C. Carnall$^{1}$, K. Duncan$^{1,2}$
\footnotesize\\
$^{1}$Institute for Astronomy, University of Edinburgh, Royal Observatory, Edinburgh EH9 3HJ\\
$^{2}$Leiden Observatory, Leiden University, NL-2300 RA Leiden, Netherlands
}

\date{Accepted XXX. Received YYY; in original form ZZZ}

\pubyear{2021}

\begin{document}
\label{firstpage}
\pagerange{\pageref{firstpage}--\pageref{lastpage}}
\maketitle

\begin{abstract}
We present a new determination of the galaxy stellar mass function (GSMF) 
over the redshift interval $0.25 \leq z \leq 3.75$, derived from a combination of
ground-based and {\it Hubble Space Telescope} ({\it HST}) imaging surveys.
Based on a near-IR selected galaxy sample selected over a raw survey area of 
3 deg$^{2}$ and spanning $\geq 4$ dex in stellar mass, we fit the GSMF with 
both single and double Schechter functions, carefully accounting for Eddington bias 
to derive both observed and intrinsic parameter values.
We find that a double Schechter function is a better fit to the GSMF at all redshifts, although the single and double
Schechter function fits are statistically indistinguishable by $z=3.25$.
We find no evidence for significant evolution in $M^{\star}$, with the intrinsic value
consistent with $\log_{10}(M^{\star}/\Msun)=10.55\pm{0.1}$ over the full redshift range.
Overall, our determination of the GSMF is in good agreement with recent simulation results, although differences persist at the highest stellar 
masses. Splitting our sample according to location on the UVJ plane, we find that the star-forming GSMF can be adequately described by a single Schechter function over the full redshift range, and has not evolved significantly since $z\simeq 2.5$.
In contrast, both the normalization and functional form of the passive GSMF evolves dramatically with redshift, switching from a single
to a double Schechter function at $z\leq1.5$. As a result, we find that while passive galaxies dominate the integrated
stellar-mass density at $z\leq 0.75$, they only contribute $\lesssim 10$\% by $z\simeq 3$. Finally, we provide a simple parameterization
that provides an accurate estimate of the GSMF, both observed and intrinsic, at any
redshift within the range $0 \leq z \leq 4$.
\end{abstract}

\section{INTRODUCTION}
An accurate determination of the evolving galaxy stellar-mass function (GSMF) is crucial for improving our understanding of galaxy evolution.
In addition to tracing the history of stellar-mass assembly, the evolving shape of the GSMF encodes vital information about the impact of different
feedback mechanisms and the physical processes through which star formation is quenched. As a consequence, together with the cosmic star-formation rate density, the evolving
GSMF is arguably one of the most fundamental observational constraints that all theoretical models of galaxy evolution must be able to reproduce.

Over the last two decades, an enormous amount of effort has been invested exploring the evolution of the GSMF.
At low redshifts, numerous studies have exploited the large areas and spectroscopic redshifts provided by the 2dF-GRS \citep{Colless2001}, SDSS \citep{York2000} and
GAMA surveys \citep{Driver2011} to study the form of the local GSMF (e.g. \citealt{Cole2001, Bell2003, Blanton2003, Li2009, Baldry2008, Baldry2012, Weigel2016}). At intermediate redshifts, studies have exploited a combination of photometric and spectroscopic data to study the evolution of
the GSMF out to $z\simeq 1$ (e.g. \citealt{Drory2009, Pozzetti2010, Moustakas2013}), while others have used a combination of increasingly deep ground-based and
{\it HST} near-IR imaging to push the study of the GSMF to $z\simeq 4-5$ and beyond (e.g. \citealt{Fontana2006, Ilbert2009, Ilbert2013, Muzzin2013, Tomczak2014, Mortlock2015, Davidzon2017, Wright2018, Leja2020}). At higher redshifts still, attempts have been made to constrain the GSMF using the deepest
available {\it HST} imaging over the redshift range $5<z<8$ (e.g. \citealt{Duncan2014, Grazian2015, Song2016, Bhatawdekar2019, Kikuchihara2020}).

Based on the wealth of literature studies, several characteristics of the GSMF have been firmly established.
Firstly, it is clear that the local GSMF is well described by a double Schechter function \citep{Schechter1976}, with a characteristic mass of $\log_{10}({M^{\star}/\Msun)}\simeq 10.6$, a low-mass slope of
$\alpha_{2}\simeq -1.4$ and a high-mass slope of $\alpha_{1}\simeq \alpha_{2}+1.0$. Secondly, when the local GSMF is split into star-forming and passive galaxy sub-samples,
it is clear that the passive GSMF requires a double Schechter function, whereas the star-forming GSMF is usually found to be adequately described by a single Schechter function (e.g. \citealt{Baldry2012}). Moreover,
the majority of previous studies have concluded that the evolution of the star-forming GSMF is remarkably modest, at least out to $z\simeq 2$ (e.g. \citealt{Tomczak2014, Davidzon2017}).

A useful insight into the physical information that can be extracted from the GSMF is provided by the analytic model proposed by \cite{Peng2010}, which was motivated by the
observed stability of the star-forming GSMF and evidence that the effects of mass and environmental quenching appear to be
fully separable in the local Universe (e.g. \citealt{Baldry2006}). In the \cite{Peng2010} model, the exponential cut-off and $M^{\star}$
of the star-forming GSMF is established and maintained by a mass quenching rate proportional to star-formation rate (SFR). If the slope of the main sequence of
star-formation is close to unity (i.e. SFR $\propto M_{\star}$) then a natural consequence is the build-up of the high-mass component of the passive GSMF,
with the same value of $M^{\star}$ and a low-mass slope of $\alpha_{1}\simeq\alpha+1.0$, where $\alpha$ is the low-mass slope of the star-forming GSMF. In this model, the
quenching of high-mass ($\log_{10}(M_{\star}/\Msun)\geq 10.5$) galaxies is dominated by mass quenching, usually attributed to some form of active galactic nuclei (AGN) feedback, at all epochs and in all environments.

However, at lower stellar masses, environmental quenching, a combination of galaxy mergers and satellite quenching, becomes increasingly important
and dominates at late times (i.e. $z<1$). Crucially, because environmental quenching is independent of stellar mass,
it naturally produces a second passive-galaxy Schechter function component whose shape, but not normalization, mirrors that of the star-forming GSMF. This apparently simple
model can accurately reproduce the key characteristics of the low-redshift GSMF, and illustrates how accurately determining the evolution of the GSMF offers
the prospect of constraining the relative timing and importance of different quenching mechanisms.

How well the \cite{Peng2010} model performs at higher redshifts is not entirely clear and the observational constraints are inevitably somewhat less stringent. At $z \leq 1$ there is a general consensus that the total GSMF maintains a double Schechter functional form and that the star-forming GSMF remains approximately constant (e.g. \citealt{Pozzetti2010, Ilbert2013, Mortlock2015}).
However, at higher redshifts, studies arrive at different conclusions regarding the shape and evolution of the total GSMF and, in particular,
the detectability, or otherwise, of an environmentally induced upturn in the number densities of low-mass passive galaxies at $z\geq 1$
(e.g. \citealt{Tomczak2014,Davidzon2017,Wright2018}).

Within this context, the primary motivation for this study is to use a combination of the best available ground and space-based photometry, covering a sufficiently
large cosmological volume and dynamic range in stellar mass, to accurately determine both the high and low-mass shape of the GSMF out to $z\simeq 4$. To achieve this,
we exploit the best available near-IR ground-based imaging over a raw survey area of 3 deg$^{2}$ and combine it with the publicly available data over the
five separate {\it HST} CANDELS survey fields \citep{Grogin2011, Koekemoer2011}.
Crucially, in addition to the deepest available optical and near-IR data, the survey fields used in this study also feature the deep mid-IR data from the {\it Spitzer Space Telescope}
that is necessary to derive robust stellar masses at $z\geq 1$.

The ground-based data alone allows us to accurately determine the high-mass end of the GSMF, by accessing a
consistent co-moving cosmological volume of $\simeq 10^{7}$ Mpc$^{3}$ in six redshift bins, spanning the range $0.25 \leq z \leq 3.75$. However, the
addition of the {\it HST} imaging ensures that we have access to sufficient dynamic range in stellar mass, $2.5-3.0$ dex below $M^{\star}$ at all redshifts,
to also accurately determine the evolution of the low-mass end of the GSMF. 

The structure of the paper is as follows. In Section 2 we discuss the
suite of ground-based imaging data utilised in this study, along with
the publicly available {\it HST} CANDELS catalogues. In Section 3 we describe the production of the photometric catalogues, the determination of the photometric
redshifts and the construction of the final galaxy sample. In Section 4 we present our determination of the evolving GSMF over the redshift
interval $0.25\leq z \leq 3.75$ and compare our new results to those of previous studies in the literature and the predictions of the latest theoretical models.
Based on our results, we provide a simple evolving parameterization that can accurately reproduce the total GSMF at any redshift within the range $0\leq z \leq 4$.
In Section 5 we explore the evolution of the star-forming and passive GSMFs and compare to the predictions of the \cite{Peng2010} model. In Section 6 we investigate
the evolution of the integrated stellar-mass density and compare with previous literature results, theoretical models and the integral of the cosmic star-formation rate.
Finally, we present a summary of our results and conclusions in Section 7. All magnitudes are expressed in the AB system \citep{Oke1974, Oke1983} and we assume the following cosmology: $\Omega_{0}=0.3, \Omega_{\Lambda}=0.7$ and $H_{0}=70$ kms$^{-1}$Mpc$^{-1}$.

\section{Data}

\subsection{Imaging data}
The imaging data utilized in this study primarily consist of ground-based UV+optical+near-IR imaging of the
UKIDSS Ultra Deep Survey (UDS), COSMOS and CFHTLS-D1 survey fields. In addition to the ground-based imaging data, we have also made
extensive use of the deep {\it Spitzer Space Telescope} mid-IR imaging available in all three fields. In
Table \ref{tab:depths} we list the data used in each field, along with our determinations of the median global
$5\sigma$-depths in each filter. The depths have been calculated within a circular aperture with a 2--arcsec
diameter and have been corrected to total assuming a point-source aperture correction.

\begin{table}
\begin{center}
\caption{The median global 5$\sigma$-depths for each of the filters used in this study. For all UV, optical and
near-IR filters the $5\sigma$-depths have been calculated using a circular aperture with a 2--arcsec diameter and
corrected to total assuming a point-source aperture correction. The median depths in the two IRAC bands were calculated using the
photometric uncertainties produced by the {\sc tphot} deconfusion software package \citep{Merlin2015}. Note that
the IRAC mosaics used in the COSMOS and UDS fields consist of data from a number of different observing programmes,
leading to significant spatial variations in the depth.}
\begin{tabular}{lcccc}
\hline
&     & UltraVISTA & UltraVISTA & \\
Filter& UDS & deep & ultra-deep & CFHTLS-D1 \\
\hline
CFHT $u^{*}$ & 26.7 & 27.0 & 27.0 & 26.9 \\
CFHT $g$ & - & 27.0 & 27.0 & 27.1 \\
CFHT $r$ & - & 26.4 & 26.4 & 26.5 \\
CFHT $i$ & - & 26.1 & 26.1 & 26.1 \\
CFHT $z$ & - & 25.2 & 25.2 & 25.2 \\
VISTA $Y$ & 24.8 & 24.8 & 25.5 & 24.8  \\
VISTA $J$ & - & 24.6 & 25.3 & 24.4 \\
VISTA $H$ & - & 24.3 & 25.0 & 24.0 \\
VISTA $K_{\rm s}$ & - & 24.7 & 24.9 & 23.7 \\
SSC $B$ & 27.4 & - & - & - \\
SSC $V$ & 27.1 & - & - & - \\
SSC $R$ & 26.8 & - & - & - \\
SSC $i$ & 26.6 & - & - & - \\
SSC $z^{\prime}$ & 25.7 & - & - & - \\
SSC $z^{\prime}_{\rm new}$ & 26.0 & 26.0 & 26.0 & - \\
SSC NB921 & 25.6 & - & - & - \\
WFCam $J$ & 25.4 & - & - & - \\
WFCam $H$ & 24.8 & - & - & - \\
WFCam $K$ & 25.1 & - & - & - \\
IRAC $3.6\mu$m & 24.6 & 25.2 & 25.8 & 23.7 \\
IRAC $4.5\mu$m & 24.8 & 25.2 & 26.0 & 23.9 \\
\hline\end{tabular}
\label{tab:depths}
\end{center}
\end{table}

\subsubsection{The UKIDSS Ultra Deep Survey}
In this study we utilized the $JHK$ near-IR imaging from the latest data release (DR11) of the
UKIDSS Ultra Deep Survey (\citealt{Lawrence2007}; Almaini et al. in prep). Additional $Y-$band near-IR imaging data
was taken from the DR4 release of the VISTA VIDEO survey \citep{Jarvis2013}.

The UV and optical coverage of the UDS field consists of CFHT MegaCam $u^{*}-$band imaging and Subaru Suprime-Cam imaging
in the $BVRiz^{\prime}$ and NB921 filters \citep{Furusawa2008, Koyama2011, Sobral2016}.
Additional $z-$band imaging ($z^{\prime}_{\rm new}$), taken following the refurbishment of Suprime-Cam with CCDs with
improved red sensitivity, was also employed \citep{Furusawa2016}. All of the UV, optical and near-IR imaging data
in the UDS field was PSF-homogenized to a Moffat profile with a FWHM~of~$\simeq0.9^{\prime\prime}$.

At mid-IR wavelengths, {\it Spitzer} IRAC mosaics of the UDS field at 3.6 and 4.5~$\mu$m were constructed by
combining the data from the SPLASH (PI Capak; see e.g. \citealt{Mehta2018}), SEDS \citep{Ashby2013}
and S-CANDELS \citep{Ashby2015} programmes using {\sc mopex} \citep{Makovoz2005}. The overlap region
covered by the full set of UV--to--mid-IR data in UDS is 0.8 deg$^{2}$, which reduces to an effective area of 0.69 deg$^{2}$ when accounting for masking (see Section 3.3).

\subsubsection{UltraVISTA}
The UltraVISTA survey \citep{McCracken2012} provides near-IR $YJHK_{\rm s}$ imaging over an area of $\simeq 1.5$ deg$^{2}$
within the COSMOS \citep{Scoville2007} survey field. The data utilized in the study is comprised
of the 1 deg$^{2}$ overlap region between the latest UltraVISTA data release (DR4) and the optical $u^{*}griz$
imaging of the CFHTLS-D2 field provided by the T0007 data release of the CFHT Legacy Survey \citep{Hudelot2012}. We choose to limit ourselves to this square degree overlap due to the importance to our study of the u-band imaging, which is not available across the rest of the UltraVISTA area.\footnote{We note that during the production of this paper, deep u-band imaging now exists over the remaining 0.5 sq. deg of the UVISTA footprint from the CLAUDS survey \citep{Sawicki2019}. This data is currently scheduled for public release later in 2021.} Our effective area after accounting for masking is 0.86 deg$^{2}$.
In addition to the CFHTLS $z-$ band imaging, we also employed deeper Subaru Suprime-Cam $z^{\prime}_{\rm new}-$band imaging \citep{Furusawa2016}.

While the CFHTLS-D2 and $z^{\prime}_{\rm new}$ imaging is homogeneous, the UltraVISTA imaging is divided into ``deep'' and ``ultra-deep'' stripes,
that account for approximately 45\% and 55\% of the total area, respectively. The $YJH$ imaging in the ultra-deep
stripes is typically $\simeq 0.7$ mag deeper than in the deep stripes, whereas the $K_{\rm s}-$band imaging,
thanks to an on-going homogenization programme, is only $\simeq 0.2$ magnitudes deeper (see Table \ref{tab:depths}). The UV,
optical and near-IR imaging data in the UltraVISTA field was PSF-homogenized to a Moffat profile with a FWHM~of~$\simeq1.0^{\prime\prime}$.

As in the UDS field, {\it Spitzer} IRAC mosaics at 3.6 and 4.5~$\mu$m were constructed by combining data
from the SPLASH, SEDS and S-CANDELS surveys, in addition to data from the SMUVS
\citep{Ashby2018} and S-COSMOS \citep{Sanders2007} programmes.

\subsubsection{CFHTLS-D1}
In the CFHTLS-D1 survey field we utilized the 1 deg$^{2}$ overlap region between the $u^{*}griz$ imaging
from the CFHT Legacy Survey \citep{Hudelot2012} and the $YJHK_{\rm s}$ near-IR imaging from the VISTA VIDEO survey \citep{Jarvis2013}. Our effective area after accounting for masking is 0.8 deg$^{2}$. The UV, optical and near-IR imaging in the CFHTLS-D1 field was
PSF-homogenized to a Moffat profile with a FWHM of~$\simeq1.0^{\prime\prime}$. The mid-IR {\it Spitzer} imaging at 3.6 and 4.5~$\mu$m was provided by the SERVS programme \citep{Mauduit2012}.

\subsection{CANDELS catalogues}
In order to increase the available dynamic range in stellar mass, we have used the publicly available photometric redshifts and stellar masses derived for each of
the five {\it HST} CANDELS fields \citep{Guo2013, Galametz2013, Santini2015, Stefanon2017, Nayyeri2017, Barro2019}.
Although the CANDELS catalogues only cover a total area of $\simeq 0.27$ deg$^{2}$, the depth of the {\it HST} near-IR imaging
plays a crucial role in our ability to properly constrain the low-mass end of the GSMF. A brief description of the relevant properties of each CANDELS catalogue is
provided in Table \ref{tab:CANDELS_cats}.

\begin{table}
\caption{Basic information for the publicly-released CANDELS catalogues employed in this study. Columns one and two list
the survey field and survey area covered by each catalogue. Column three lists the
F160W ($H_{160}$) $5\sigma$ depths quoted in the relevant catalogue papers, as measured in a circular aperture
with a diameter twice the FWHM. Note that the range of $H_{160}$ depths quoted for the GOODS fields reflects the
deep and wide components of the CANDELS imaging \protect\citep{Grogin2011, Koekemoer2011} and, in the case
of GOODS-S, the ultra-deep near-IR data available in the {\it Hubble Ultra Deep Field} (HUDF). The final column
lists the references for the catalogues containing the photometry, photometric redshifts and stellar mass information;
which correspond to (1) \protect\cite{Galametz2013}, (2) \protect\cite{Guo2013}, (3) \protect\cite{Santini2015},
(4) \protect\cite{Nayyeri2017}, (5) \protect\cite{Stefanon2017} and (6) \protect\cite{Barro2019}.}
\begin{center}
\begin{tabular}{lccc}
\hline
Field & Area/arcmin$^{2}$ & $H_{160}$ depth/mag & Reference\\
\hline
UDS         & 202 & 27.5     & 1,3 \\
GOODS-South & 170 & 27.4$-$29.7& 2,3 \\
COSMOS      & 216 & 27.6     & 4 \\
EGS         & 206 & 27.6     & 5\\
GOODS-North & 171 & 27.8$-$28.7& 6\\
\hline\end{tabular}
\label{tab:CANDELS_cats}
\end{center}
\end{table}

\section{Catalogue production and sample selection}
In this section we provide an overview of how the photometry catalogues
for the UKIDSS UDS, UltraVISTA and CFHTLS-D1 survey fields were produced. We also
provide an overview of how the photometric redshifts and stellar masses were calculated and
the processes employed to select the final galaxy sample.

\subsection{Photometry catalogues}

In order to generate photometric catalogues, {\sc sextractor}
\citep{Bertin1996} was run in dual-image mode with the $K-$band
image serving as the detection image across all three fields. For objects detected in the $K-$band,
photometry was extracted from the PSF-homogenized images in all UV--to--near-IR filters using
circular apertures with a 2--arcsec diameter. The photometry was extracted from the PSF-homogenized
images to minimize aperture correction effects for the subsequent spectral energy distribution (SED) fitting.
To further reduce any colour systematics, additional flux corrections were made (at the $1-2$\% level) based on the curves of growth of point sources in each filter.

Accurate flux errors were calculated for each object, in each individual filter, by measuring the aperture-to-aperture r.m.s. of
$\simeq~150-200$ nearby blank sky apertures (see e.g. \citealt{McLeod2015}), where the local value of $\sigma$ is calculated using the
robust median absolute deviation (MAD) estimator.

The photometry in the lower spatial resolution {\it Spitzer} IRAC imaging
at 3.6\ and 4.5~$\mu$m was measured using the {\sc tphot} software package \citep{Merlin2015}. Given that
the {\sc tphot} algorithm uses the isophotal footprint of the objects in a higher spatial-resolution image as
a prior (the original $K-$band detection image in this case), the fluxes generated by {\sc tphot} can be regarded as isophotal. As a consequence, we aperture match the PSF-homogenized photometry at shorter wavelengths to match the {\sc tphot} fluxes, multiplying by $f = K_{\rm iso}/K_{2}$, where $K_{\rm iso}$ is the isophotal flux extracted by {\sc sextractor} from the high-resolution image, and $K_{2}$ is the 2--arcsec diameter flux extracted from the PSF-homogenized $K-$band image.

\subsection{Photometric redshifts}
In order to derive robust photometric redshifts, we undertook six different photometric redshift runs, using three different codes.
Three photometric redshift runs were performed using the {\sc LePhare} \citep{Arnouts2011} SED fitting code, using the
\cite{BC03}, Pegase2 \citep{Fioc1999} and COSMOS \citep{Ilbert2009} template libraries. In each
of these runs a \cite{Calzetti2000} dust attenuation curve was adopted, with colour excess in the range $E(B-V)=0-0.6$. Emission
lines were included in the fits and IGM absorption was accounted for using the \cite{Madau1995} prescription.

Two further photometric redshift runs were performed using the {\sc eazy} SED fitting code \citep{Brammer2008}, with the default PCA and Pegase2 libraries. One final run was
performed with the BPZ code \citep{Benitez2000}, using the default set-up and CWW \citep{Coleman1980} templates.

Before running on the full photometry catalogues, the three different SED fitting codes were trained by fitting to the photometry of
objects with robust spectroscopic redshifts. This process allowed us to apply the necessary zero-point off-sets (e.g. \citealt{Dahlen2013}) and to quantify the performance
of each code/template combination using $\sigma_{\rm z}$ and the catastrophic outlier rate. The value of $\sigma_{z}$ is our preferred measurement of the photometric redshift accuracy, and is defined as $1.483\times \rm{MAD}(dz)$, where MAD is the median absolute deviation and $dz=(z_{\rm phot}-z_{\rm spec})/(1+z_{\rm spec})$. Any object with $|dz|>0.15$ is classified as a catastrophic outlier.

Our best-estimate $z_{\rm phot}$ for each object was taken as the median of our six different photometric redshift estimates (hereafter $z_{\rm med}$). These $z_{\rm med}$ measurements were tested against spectroscopic redshift samples for each of the three fields to ensure their accuracy.

For the UDS field, we used $\simeq 2650$ spectroscopic redshifts obtained from the VIPERS
\citep{Guzzo2014}, UDSz \citep{Bradshaw2013,McLure2013a} and VANDELS \citep{McLure2018,Pentericci2018} spectroscopic surveys. For this sub-sample,
the accuracy of our $z_{\rm med}$ measurements was $\sigma_{z}~=~0.022$, with a catastrophic outlier rate of 2.1\%.

For the UltraVISTA/COSMOS field we compiled a catalogue of $\simeq$11000 high-quality spectroscopic redshifts, the vast majority of which were drawn from the zCOSMOS \citep{Lilly2007, Lilly2009}, 3DHST \citep{Momcheva2016}, PRIMUS \citep{Coil2011}, MOSDEF \citep{Kriek2015} and VUDS \citep{LeFevre2015} spectroscopic surveys. For this sub-sample, the photometric redshift accuracy was $\sigma_{z}=0.019$, with a catastrophic outlier rate of 2.5\%.

Finally, to test the photometric redshifts in the CFHTLS-D1 field, we used a sample of $\simeq 4200$ robust spectroscopic redshifts from the {\sc VIMOS} VLT Deep Survey (VVDS, \citealt{LeFevre2013}). The photometric redshift accuracy for this sub-sample was $\sigma_{z}~=~0.019$, with a catastrophic outlier rate of 2.3\%.

\begin{figure}
\includegraphics[width=\columnwidth]{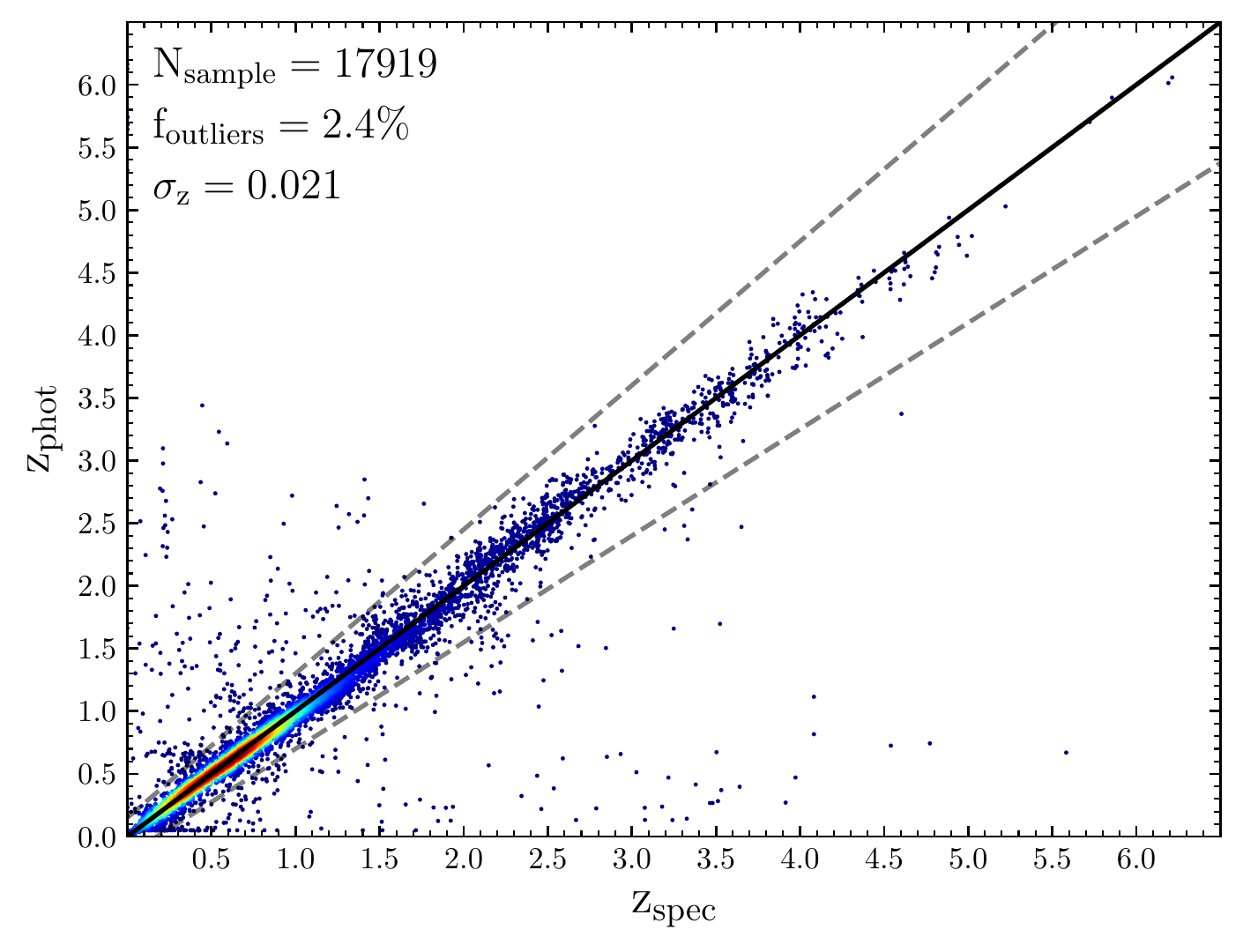}
\caption{A comparison between photometric and spectroscopic redshifts for $\simeq$18000 galaxies with robust spectroscopic redshifts across the
three ground-based data sets. The corresponding values of $\sigma_{\rm MAD}$ and the fraction of catastrophic outliers are provided in the legend.
A density map has been used to aid legibility.}
\label{fig:specz_vs_photoz}
\end{figure}

In Fig. \ref{fig:specz_vs_photoz} we show a comparison between
the spectroscopic and photometric redshifts for our training set of $\simeq 18000$ objects across all three survey fields.
In summary, it can be seen that our $z_{\rm med}$ photometric redshifts are both robust and consistent across all three fields, with a typical
accuracy of $\sigma_{z}=0.021$ and a catastrophic outlier rate of $2.4\%$. We note that the publicly available photometric redshifts available for the five
CANDELS fields (see Table \ref{tab:CANDELS_cats}) are of very comparable quality to those derived here.

Once the photometric redshift training process had been completed, the six different photometric redshift code/template combinations were run on the
full photometric catalogues for the UDS, UltraVISTA and CFHTLS-D1 fields. As before, the final adopted photometric redshift for each object was
taken to be $z_{\rm med}$, the median of our six different estimates.

\subsection{Sample construction}
In order to construct the final sample to be used in the GSMF determination, the ground-based catalogues were initially cut at their global $5\sigma$ limit in the $K-$band and then restricted to those objects in the photometric redshift range $0.25 \leq z_{\rm phot} \leq 3.75$. The lower redshift cut is imposed to ensure that our survey encloses sufficient cosmological volume to constrain the GSMF, whereas the upper redshift cut is imposed to ensure that our $K-$band selection always corresponds to rest-frame wavelengths long-ward of the 4000\AA\, break. Following the initial cuts based on the near-IR signal-to-noise ratio and photometric redshift, the ground-based catalogues were then cleaned to remove stars, AGN, artefacts and objects with contaminated photometry.

In order to reduce the contamination by stars, objects were excluded from the final catalogue using the stellar locus in two colour-colour plots. For the UDS sub-sample, stars were rejected using the stellar locus on the $B-z$ vs $z-K$ colour-colour plot, following \cite{Baldry2010}. For the UltraVISTA and CFHTLS~-~D1 sub-samples, stars were rejected using the stellar locus on the $g-i$ vs $J-K_{\rm s}$ colour plot, following \cite{Jarvis2013}.

Potential AGN were removed from the final sample based on a combination of X-ray and {\it Spitzer} 24 $\mu$m information. The initial rejection of potential AGN was performed using the publicly available X-ray catalogues which cover all three ground-based fields. In the COSMOS field we utilized the Chandra X-ray catalogue published by \cite{Elvis2009}, while in the UDS field we used the XMM-Newton X-ray catalogue produced by the Subaru/XMM-Newton Deep Survey \citep{Ueda2008}. Finally, in the CFHTLS-D1 field we utilized the XMM-XXL north survey catalogue from \cite{Liu2016}.
Based on the information available in these catalogues, potential AGN were excluded using the soft X-ray to optical ratio (X/O), as described
in \cite{Salvato2011}.

Following the exclusion of potential AGN based on their X-ray characteristics, we performed a second round of AGN rejection based on the 24 $\mu$m imaging available
in the UDS (spUDS, PID 40021, PI Dunlop), COSMOS (S-COSMOS, \citealt{Sanders2007}) and CFHTLS-D1 (SWIRE, \citealt{Lonsdale2003}) fields. Objects were removed based on
the specific star-formation rate criteria: sSFR$\geq 10.0$ Gyr$^{-1}$, where the specific star-formation
rate was calculated using the 24 $\mu$m prescription of \cite{Rieke2009}, which was designed to remove those objects whose 24 $\mu$m flux
is dominated by AGN heated dust. The fraction of objects removed from our sample as potential AGN is of order 1\%.

Artefacts and objects with contaminated photometry were removed using a two-step process. Firstly, all objects within the haloes of
bright/saturated stars were removed from the catalogue, with the effective survey area recalculated to compensate. Secondly, within each of
the GSMF redshift bins, those objects whose SED fits produced the worst 5\% of $\chi^{2}$ values were also excluded. The SED
fits for this population of objects were statistically unacceptable, and visual inspection confirmed that they were dominated
by artefacts and objects with badly compromised photometry.

Finally, those objects from the five CANDELS catalogues detected at $\geq 5\sigma$ significance in the $H_{160}$ filter and within the
redshift range $0.25 \leq z \leq 3.75$ were included in our GSMF sample. Objects identified as stars or AGN were once again excluded, based on the flags
provided.

\begin{figure*}
\begin{tabular}{cc}
\includegraphics[width=0.8\columnwidth]{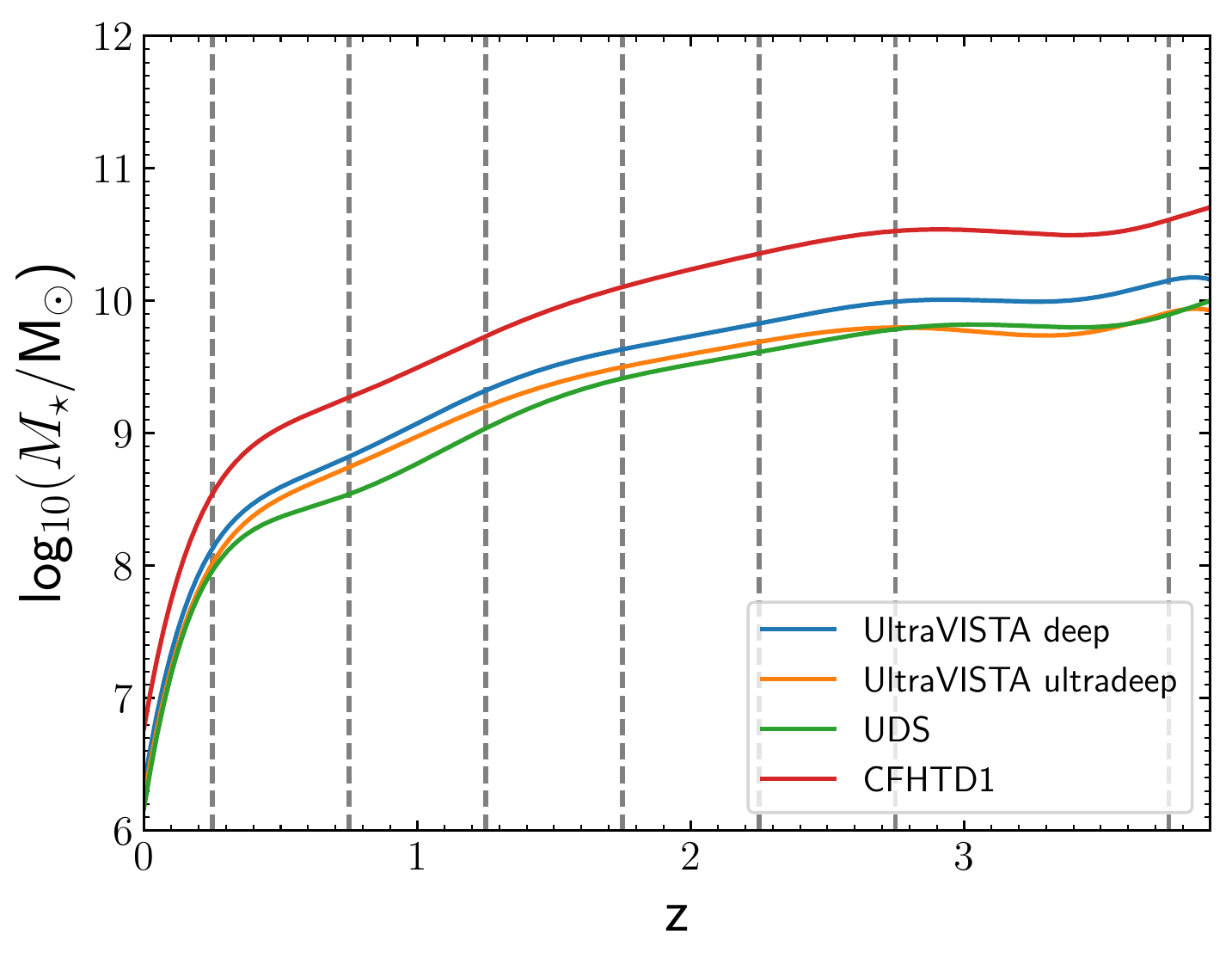} &
\includegraphics[width=0.8\columnwidth]{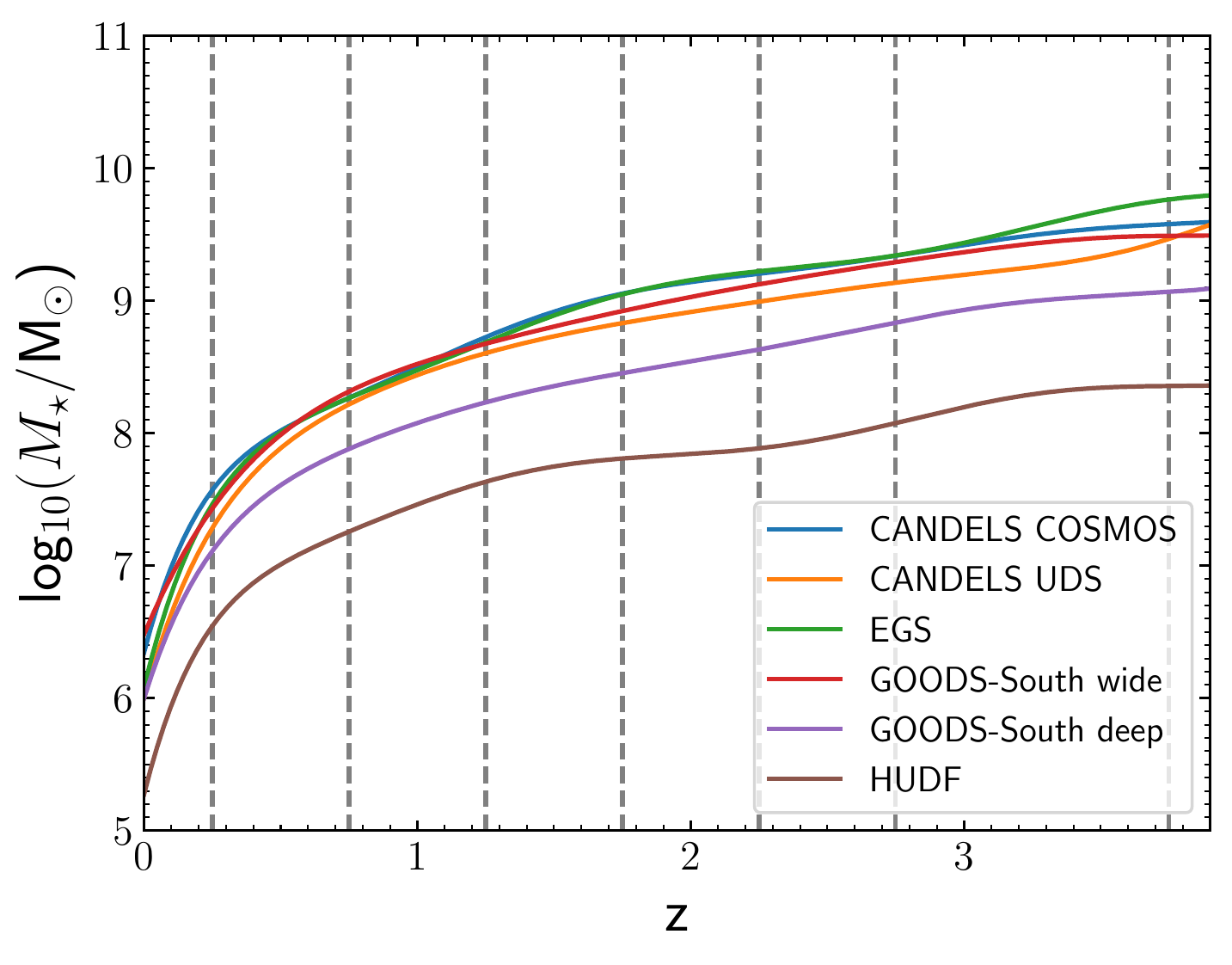}
\end{tabular}
\caption{The left-hand panel shows the 90\% mass-completeness limits as a function of redshift for each of the three ground-based surveys fields, where
the UltraVISTA/COSMOS field has been separated into the deep and ultra-deep components. The grey vertical dashed lines show the limits of the six redshift bins adopted
for the determination of the GSMF. The right-hand panel shows the same information for the CANDELS fields, where the GOODS-South field has been separated into the wide, deep and HUDF components. Note that, for clarity, we do not show the mass-completeness limits for the wide and deep components of the GOODS-North field, which are assumed to be the same as for the equivalent components of GOODS-South.}
\label{fig:mass_completeness}
\end{figure*}

\subsection{Stellar masses}
In measuring the stellar mass for each galaxy, we fix the redshift at its $z_{med}$ as determined in Section 3.2. We then refit the photometry of the galaxy at this fixed redshift, using the photometric redshift code \textsc{LePhare}. The template set is that of \cite{BC03}, with a \cite{Chabrier2003} initial mass function (IMF). A \cite{Calzetti2000} dust attenuation law is used, and IGM absorption is as prescribed in \cite{Madau1995}. Metallicities are m42 and m62, and the star-formation histories are $\tau$ models $SFH=exp(-t/\tau)$, with $\tau$ (Gyr) values 0.1, 0.3, 1, 2, 3, 5, 10 and 15. Finally, we allow $A_{V}$ (mag) values between 0 and 2.8 in steps of 0.05.
At this stage, the stellar masses returned by the SED fitting process were based on isophotal photometry (see Section 3.1). To convert the stellar masses to total, they were multiplied by $K_{\rm auto}/0.9K_{\rm iso}$, where $K_{\rm iso}$ and $K_{\rm auto}$ are the {\sc flux\_iso} and {\sc flux\_auto} fluxes measured
by {\sc sextractor} in the $K-$band, respectively. The factor of 0.9 is necessary to account for the fact that {\sc flux\_auto} typically only captures 90\% of the total flux \footnote{Previous tests based on stacking objects as a function of redshift and apparent $K-$band magnitude in UltraVISTA DR2 confirm that this is a robust assumption \citep{Mortlock2017}.}.

The stellar masses for the objects within the five CANDELS fields were taken from the public catalogues (see Table \ref{tab:CANDELS_cats}). The stellar masses across all
five CANDELS fields were also calculated using BC03 stellar population templates, based on a Chabrier IMF. Cross checks performed using objects in common with
our ground-based photometry catalogues in the UDS and COSMOS fields confirmed that our stellar-mass measurements are in excellent agreement with those derived by the CANDELS team, following a tight 1:1
relation with a typical scatter of $\pm 0.05$ dex.

\section{GSMF determination}
In this section we present our basic determination of the evolving GSMF, taking into account the
effects of stellar-mass completeness and an assessment of the impact of cosmic variance. We also provide a full description
of how we fit the observed and intrinsic GSMF, after accounting for the effects of Eddington bias.

\subsection{Number densities}
After first splitting the data into six redshift bins, we employed the $\frac{1}{V_{\rm max}}$ estimator \citep{Schmidt1968} to
determine the number densities:
\begin{equation}
\phi(\mathcal{M}) \Delta\mathcal{M} = \sum\limits_{i=1}^{N_{\rm gal}} \frac{1}{C_{i}(\mathcal{M},z)V_{\rm max,i}},
\end{equation}
where $\phi(\mathcal{M})$ [dex$^{-1}$ Mpc$^{-3}$] is the number density of galaxies per dex, per unit comoving volume, $\Delta\mathcal{M}$ is the logarithmic stellar mass bin and
$C_{i}(\mathcal{M},z)$ is the completeness calculated for each galaxy. In Eqn.~1 we have defined $\mathcal{M}~\equiv~\log_{10}(M_{\star}/\Msun)$ and will repeatedly adopt this
shorthand throughout the rest of the paper.

Incompleteness in each of the three ground-based surveys was accounted for separately via simulations, in which artificial galaxies with a wide range of
physical properties (z, $M_{K}$, $\mathcal{M}$, $r_{e}$) were injected into the $K-$band imaging and recovered using the same {\sc sextractor} set-up used to construct the original photometry catalogues.
In performing the simulations we converted between $M_{K}$ and stellar mass by drawing randomly from the SED templates fitted to the real galaxies and adopted half-light radii predicted by the
size-mass-redshift distributions derived by \cite{Shibuya2015}. To calculate the effective stellar-mass limit of our ground-based survey data, we followed the procedure proposed by
\cite{Pozzetti2010} and calculated the distribution of limiting masses for the galaxies at each redshift, where the limited mass is defined as:
\begin{equation}
\mathcal{M}_{\rm lim}= \mathcal{M}+0.4(K-K_{\rm lim}),
\end{equation}
and $(K-K_{\rm lim})$ is the difference between the apparent $K-$band magnitude of a galaxy and the $5\sigma$ magnitude limit. Using this method we define the 90\% mass-completeness limit at each redshift as the stellar mass below which 90\% of the limiting stellar masses lie. Unlike \cite{Pozzetti2010}, at each redshift we calculate a more conservative limiting-mass based on the full galaxy sample (i.e. using the full range of mass-to-light ratios), rather than adopting the faintest 20\% of galaxies. In Fig. \ref{fig:mass_completeness} we plot the 90\% mass-completeness limit versus redshift for
both the ground-based and CANDELS data. All of the GSMF plots shown in this paper include only those galaxies that lie above the appropriate 90\% mass-completeness limit. 

\begin{figure*}
\begin{center}
\begin{tabular}{ccc}
\includegraphics[width=0.6\columnwidth]{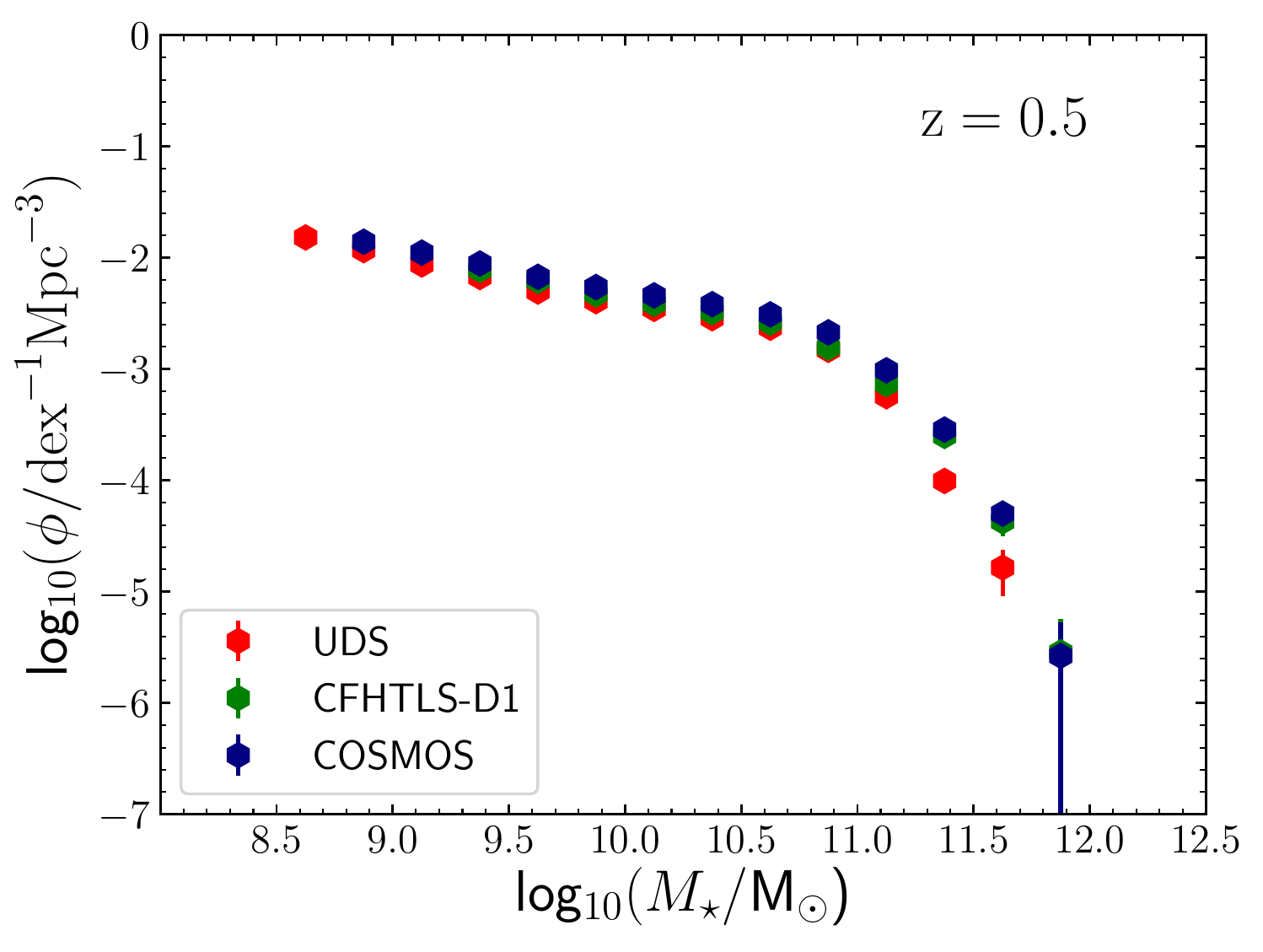} &
\includegraphics[width=0.6\columnwidth]{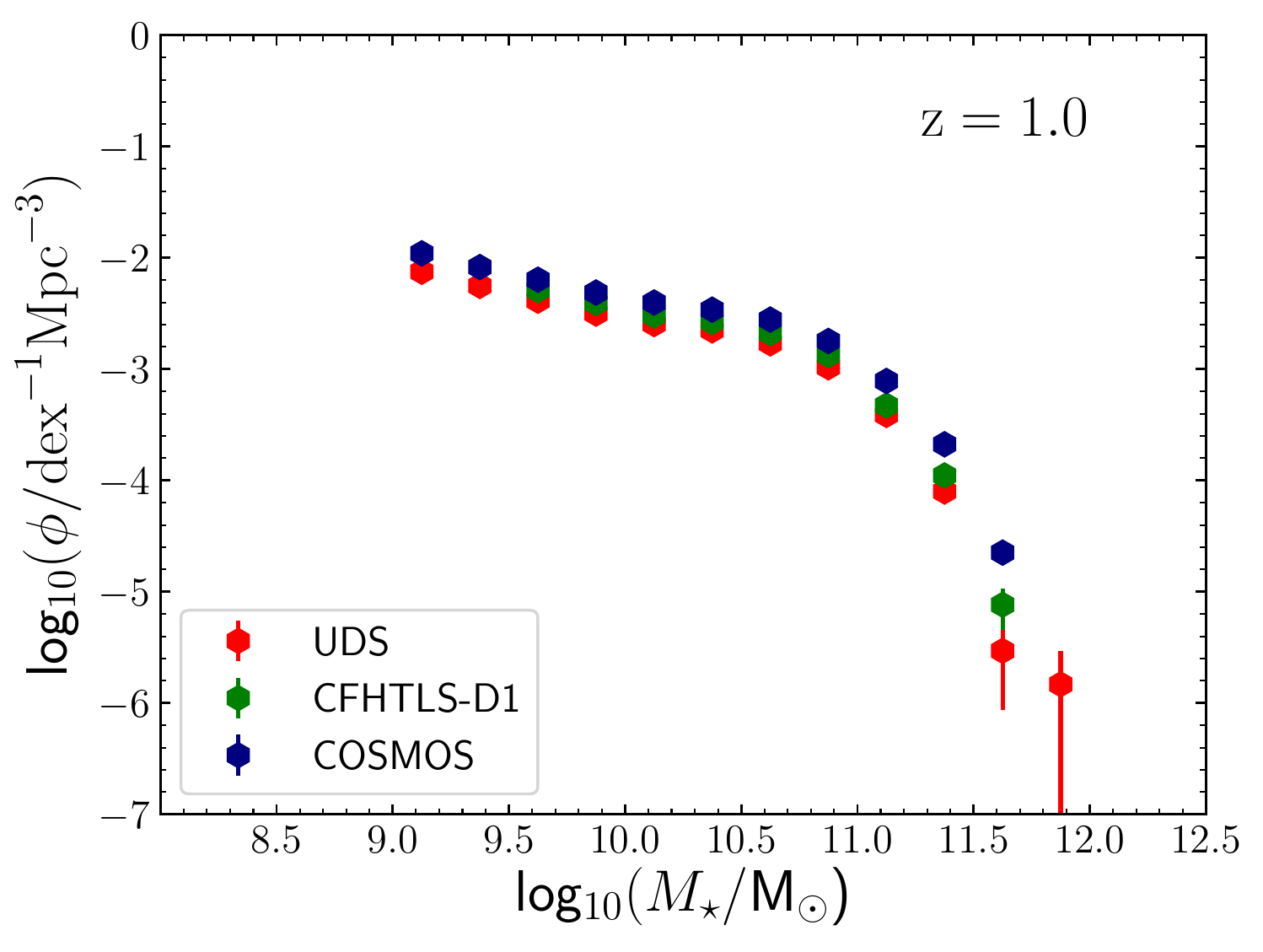} &
\includegraphics[width=0.6\columnwidth]{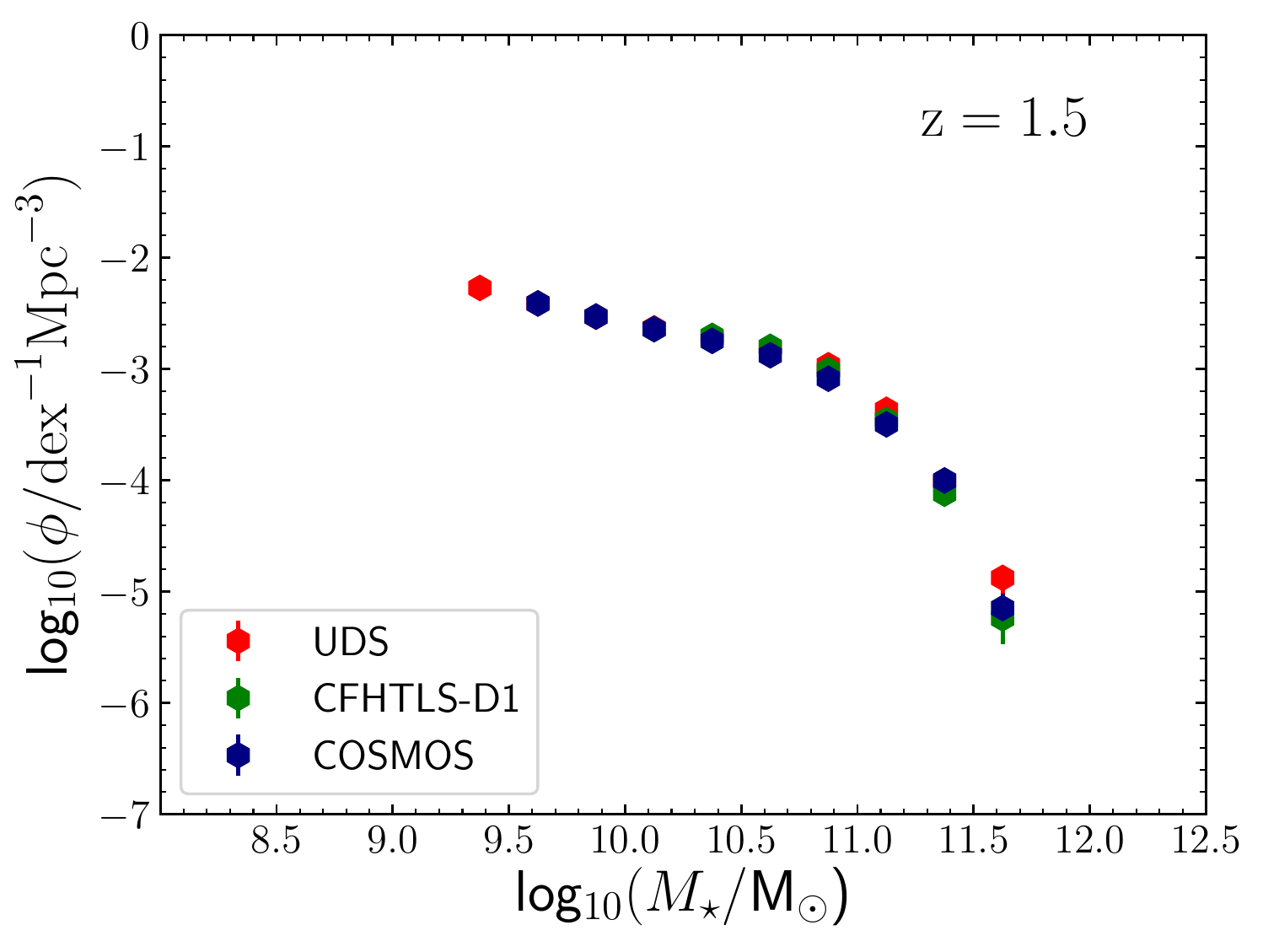} \\
\includegraphics[width=0.6\columnwidth]{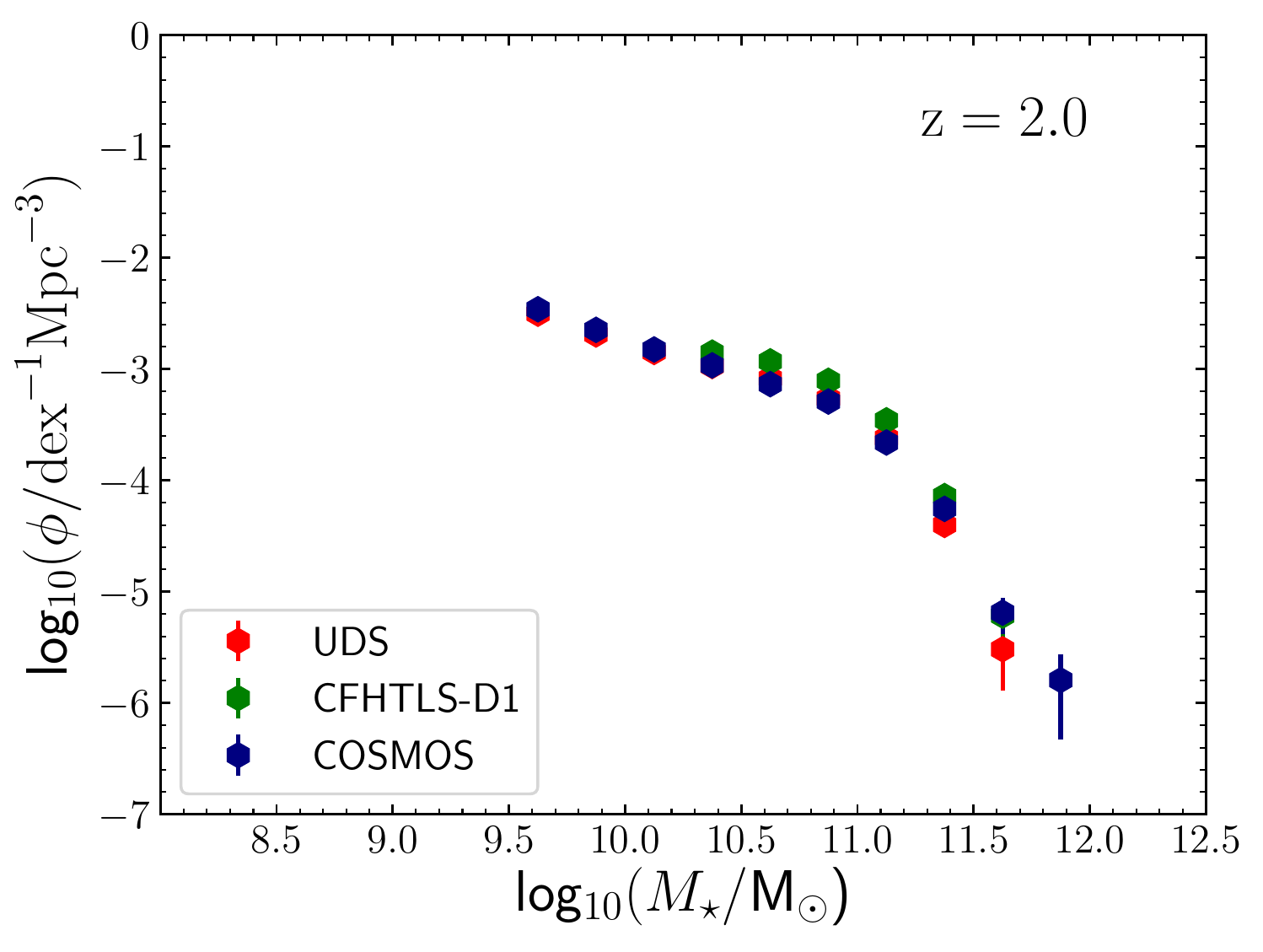} &
\includegraphics[width=0.6\columnwidth]{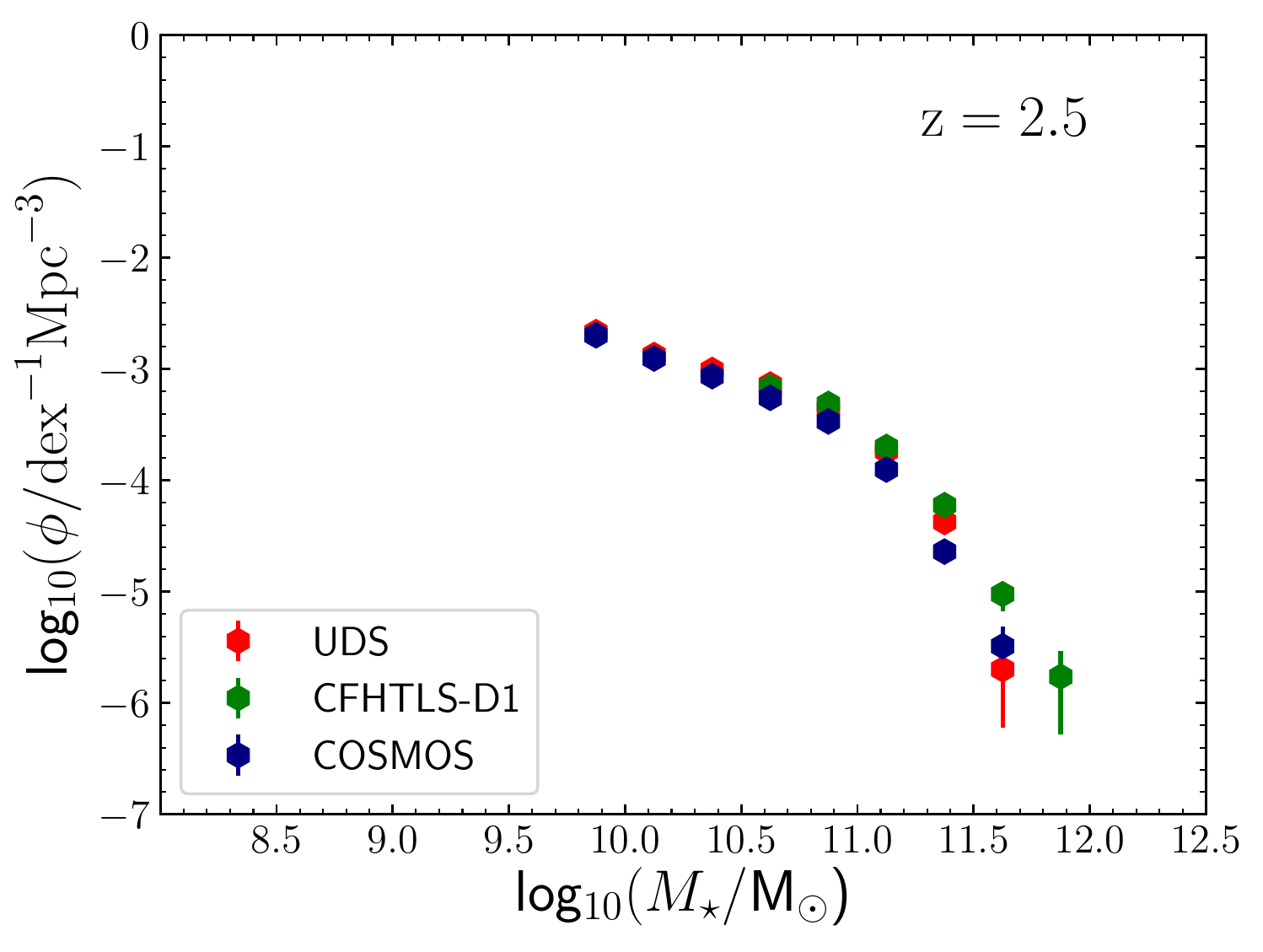} &
\includegraphics[width=0.6\columnwidth]{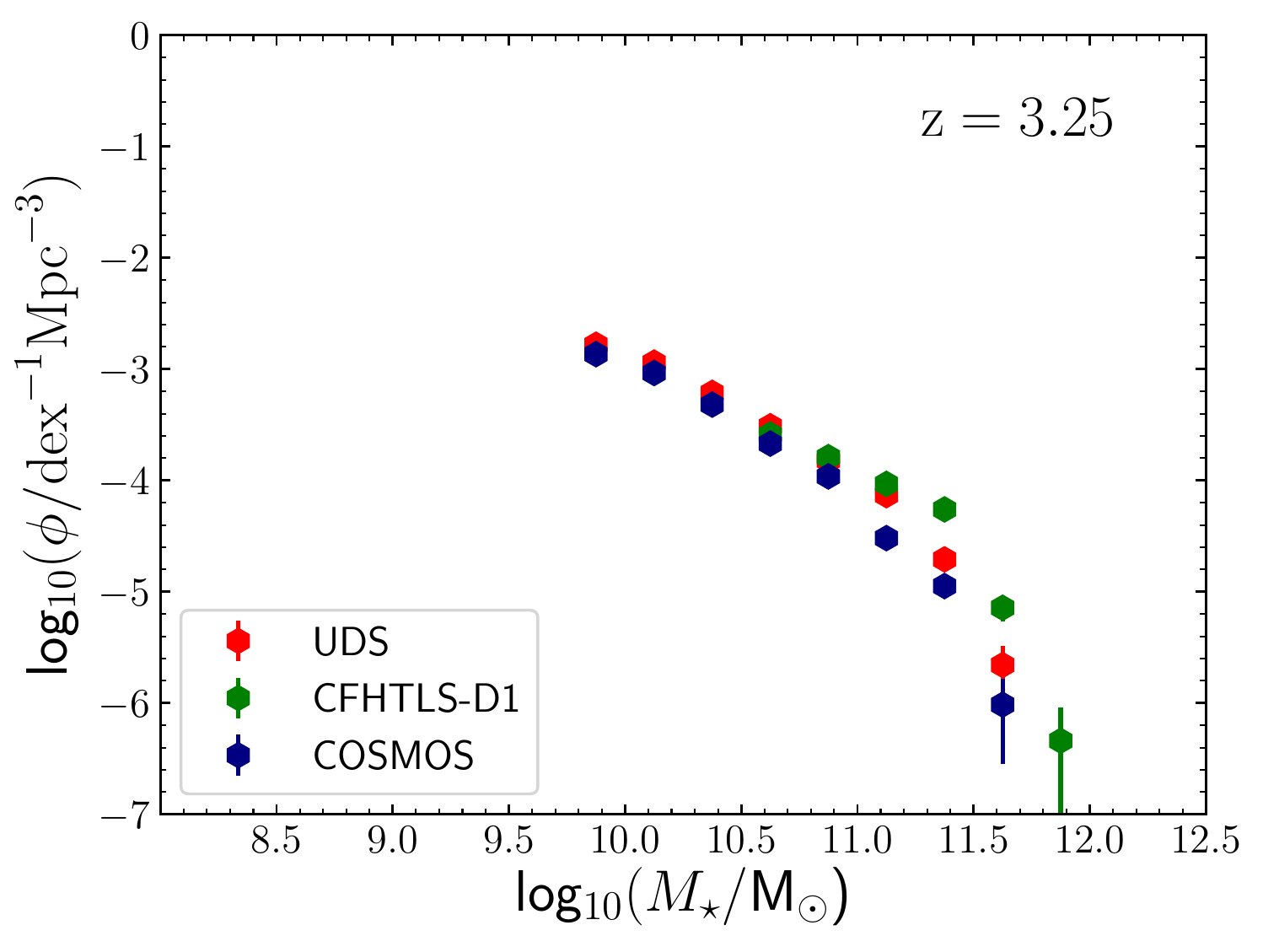}
\end{tabular}
\end{center}
\caption{A comparison of how the observed GSMF evolves as a function of redshift in our three, degree-scale, survey fields. In this plot the number density uncertainties are simply the Poissonian
counting errors. The availability of three, non-contiguous, degree-scale survey fields allows an empirical measurement of the level of cosmic variance in the high-mass end of the GSMF (see text for discussion).}
\label{fig:GSMF_each_field_separately}
\end{figure*}

\subsubsection{Individual ground-based GSMF determinations}
In Fig. \ref{fig:GSMF_each_field_separately} we show our determinations of the GSMF, based on the ground-based data in the UDS, COSMOS and CFHTLS-D1 and
fields alone. The first five redshift bins all have a width of $\Delta z=0.5$ and are centred on $z=0.5,1.0,1.5,2.0$ and 2.5, whereas the
final redshift bin spans the range $2.75 \leq z < 3.75$ in order to maintain the statistics at a similar level to the lower-redshift bins.

It can be seen that the independent GSMF determinations for the three, degree-scale, fields are generally in good agreement, although significant
differences are present, most noticeably at the high-mass end. For example, at $z=0.5-1.0$ the COSMOS field can be seen to be somewhat overdense compared to the other two fields, a fact which was also noted by \cite{Moustakas2013}. In terms of cosmic variance, the advantage of determining the high-mass end of the GSMF
from three independent degree-scale fields is therefore clear. Indeed, a further advantage of having three non-contiguous fields is that we are able to
empirically quantify the level of cosmic variance in the GSMF.

The number density uncertainties
plotted in Fig. \ref{fig:GSMF_each_field_separately} are simply the Poissonian counting errors. However, in all subsequent plots and tables, the quoted uncertainties are based on the
quadrature addition of $\sigma_{\rm poisson}$, $\sigma_{\rm boot}$ and $\sigma_{\rm cv}$, where $\sigma_{\rm boot}$ is the error contribution calculated from a bootstrap analysis based on
many thousands of GSMF realisations. For $\sigma_{\rm cv}$, we consider both an empirical estimate based on the field-to-field variance and that estimated using \cite{Moster2011}. The cosmic variance contribution ($\sigma_{\rm cv}$) to each bin is taken as the greater of the \cite{Moster2011} estimate and the measured field-to-field variance of those
fields contributing to the bin. The number density uncertainties for the {\it HST} CANDELS data were calculated in an identical fashion.

\subsubsection{The combined HST and ground-based GSMF}
\begin{figure*}
\begin{tabular}{cc}
\includegraphics[width=0.8\columnwidth]{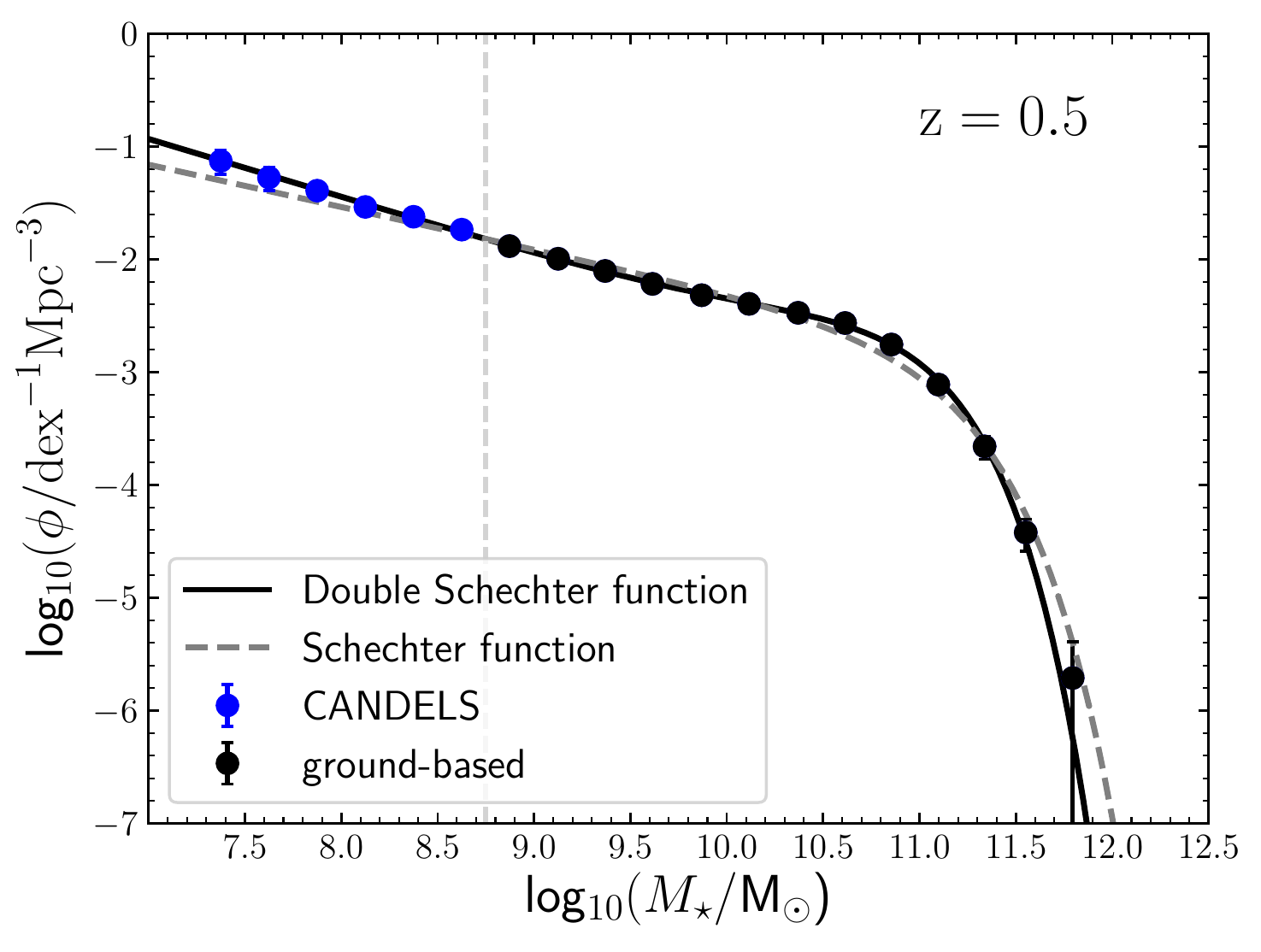} &
\includegraphics[width=0.8\columnwidth]{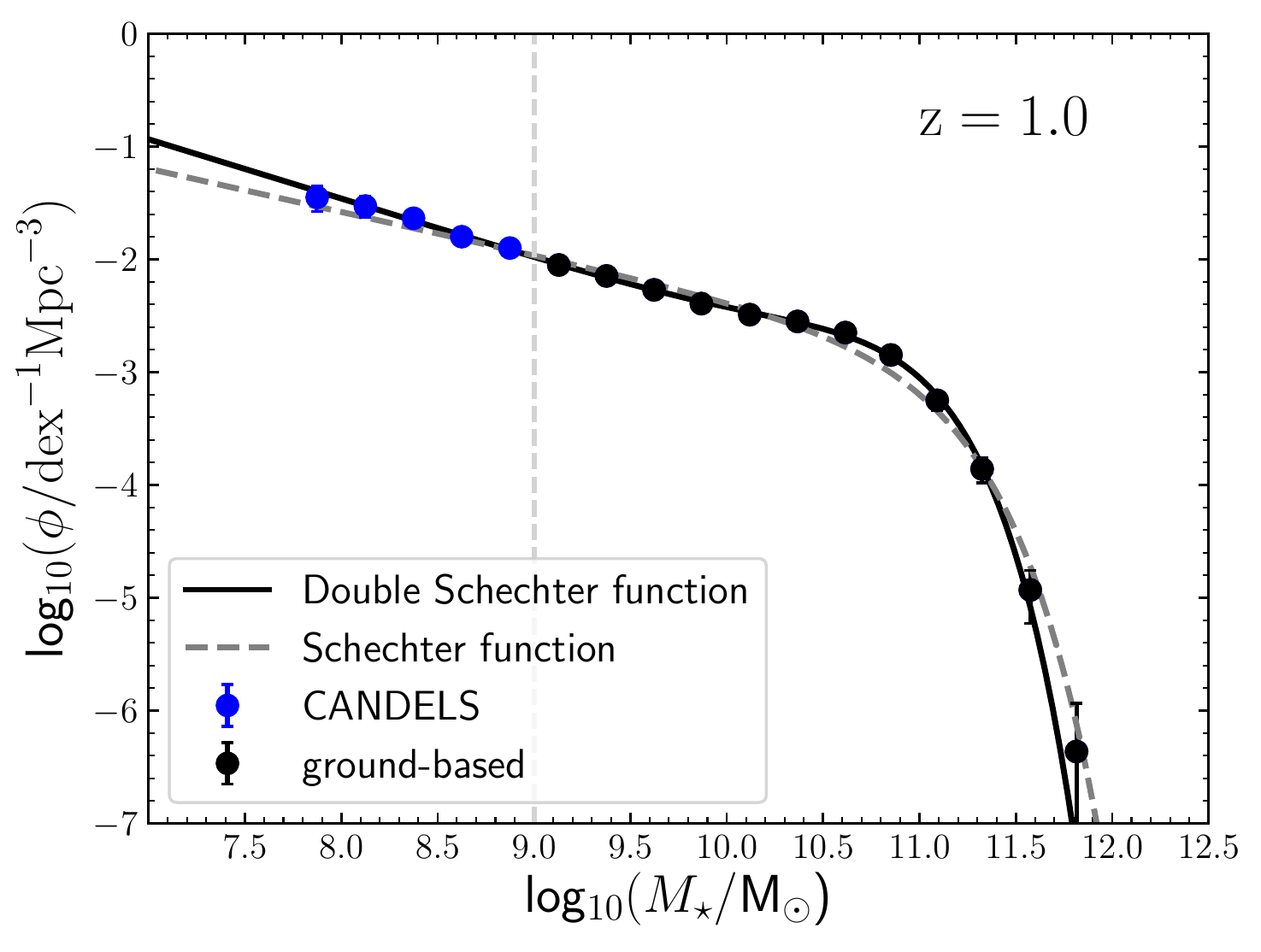} \\
\includegraphics[width=0.8\columnwidth]{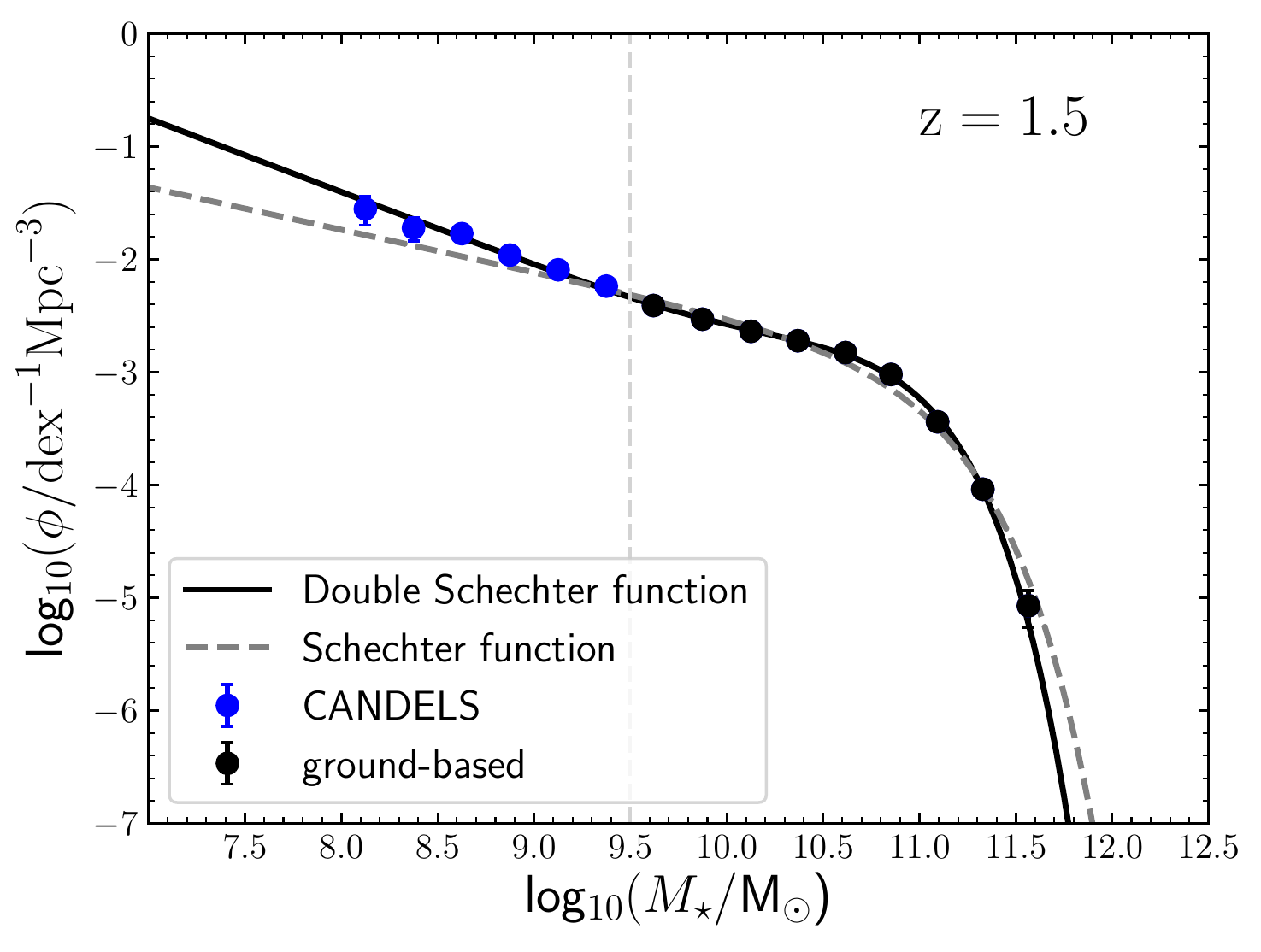} &
\includegraphics[width=0.8\columnwidth]{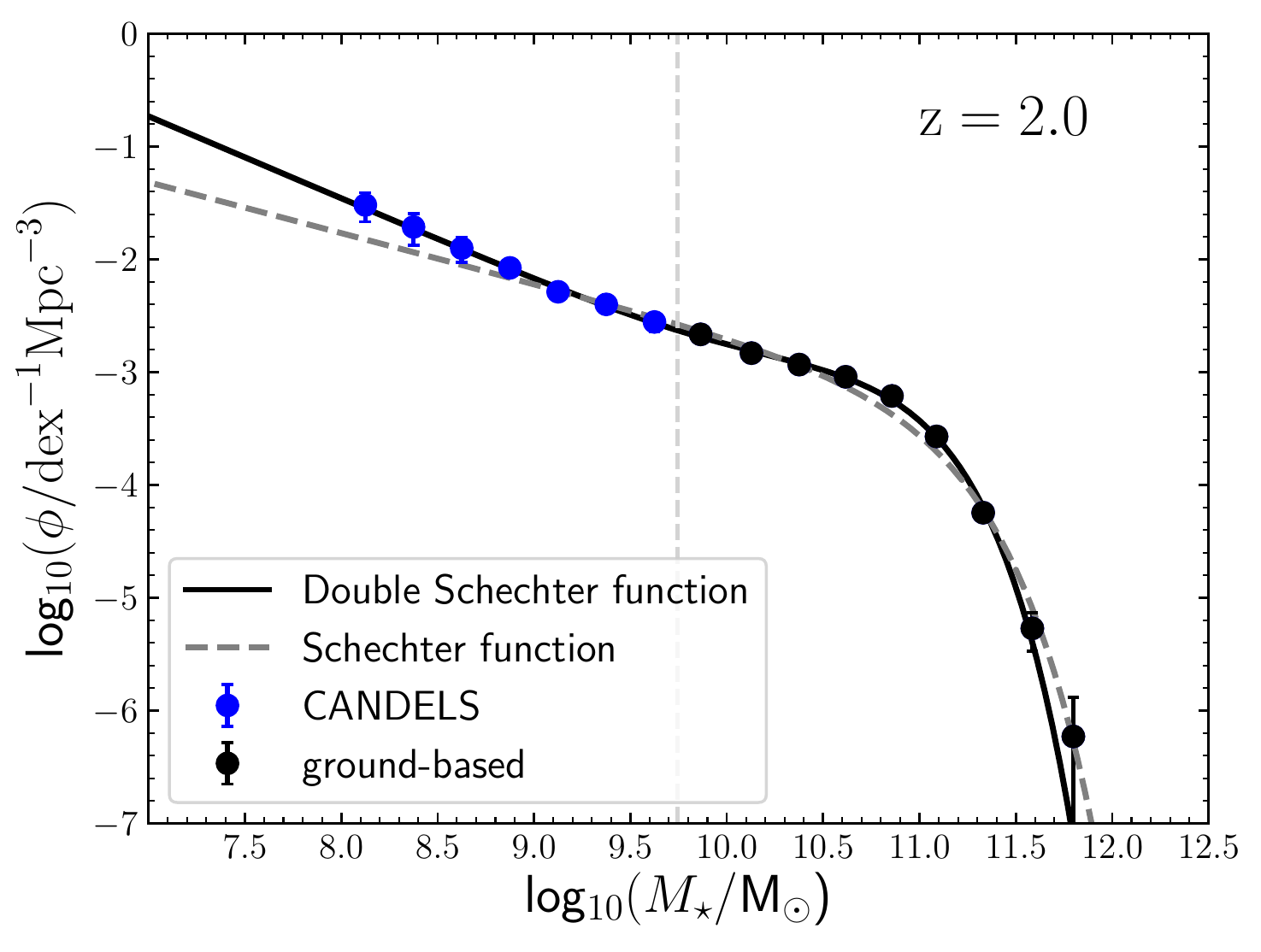} \\
\includegraphics[width=0.8\columnwidth]{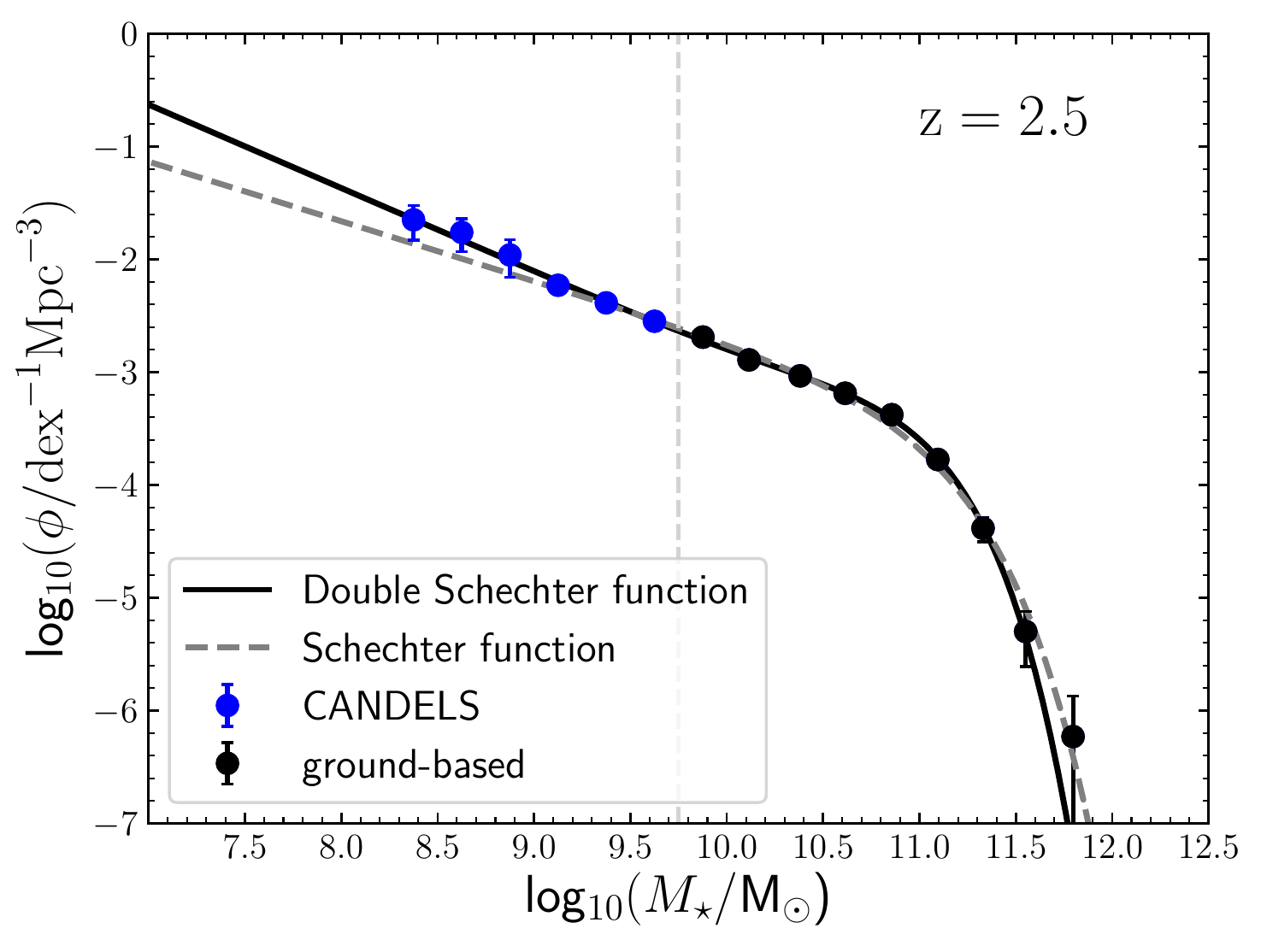} &
\includegraphics[width=0.8\columnwidth]{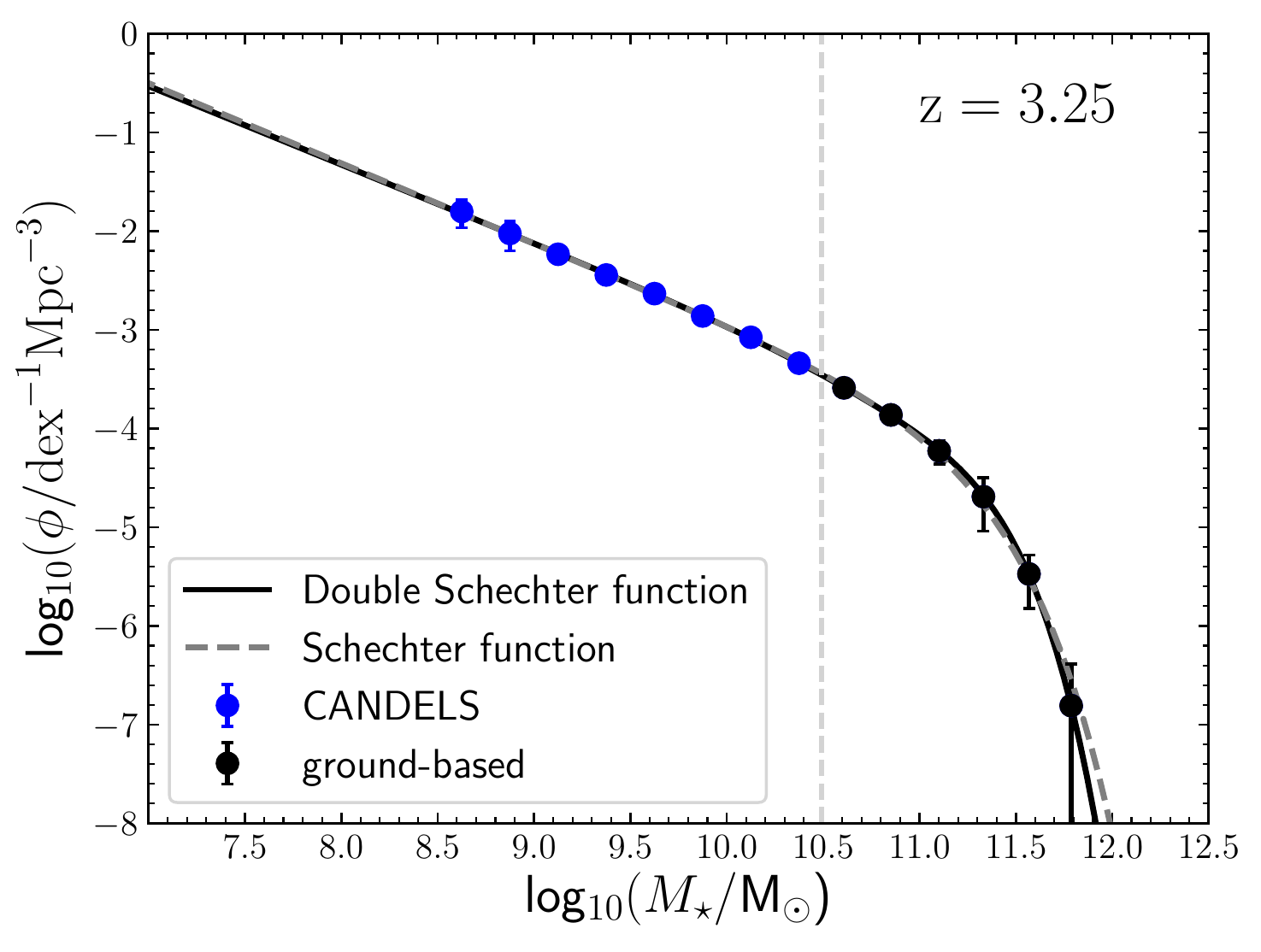}
\end{tabular}
\caption{The observed GSMF as a function of redshift, based on the combined ground-based and {\it HST} data set. Also plotted are the best-fitting single (dashed grey) and double (solid black) Schechter function fits. The black data points are derived from the ground-based data set alone, while the blue data points are derived from the {\it HST} data set alone. The split between the ground-based and {\it HST} data is highlighted by the dashed grey vertical line. Over the redshift range $0.25\leq z\leq 2.75$, the double Schechter fit is seen to be a better representation of the observed GSMF than the single Schechter fit. However, in the final redshift bin centred on $z=3.25$, the single and double Schechter function fits are basically indistinguishable.}
\label{fig:observed_GSMFs}
\end{figure*}

In Table \ref{tab:number_densities_total} and Fig. \ref{fig:observed_GSMFs} we present our determination of the observed GSMF over the redshift range $0.25 \leq z < 3.75$, based on a combination of
the full ground-based and {\it HST} data set. The process adopted for producing the combined GSMF determination was as follows. Firstly, we produced a combined ground-based GSMF by merging the three
ground-based catalogues into a single catalogue, calculating the numbers of objects and the cosmological volume contributing to each redshift-mass bin based on the 90\% mass-completeness limits for each field.
Secondly, we produced a combined {\it HST}-based GSMF by applying an identical methodology to the data for the five CANDELS fields. The third step in the process was to match the ground-based
and {\it HST}-based
GSMFs in each redshift bin, but adjusting the normalization of the {\it HST}-based GSMF to match that of the ground-based GSMF in the overlap region between the two. The typical adjustment required was at the $\simeq \pm 0.03$ dex level. In each redshift bin, the final split between the ground-based (black data points) and {\it HST}-based (blue data points) shown in Fig. \ref{fig:observed_GSMFs}
is based on the 90\% completeness limit of the deepest ground-based survey field (typically the UDS). For clarity, the blue data points shown in Fig. \ref{fig:observed_GSMFs} are entirely based on
{\it HST} data, whereas the black points are entirely based on ground-based data.

\begin{table*}
\caption{The observed GSMF as a function of redshift, based on the combined ground-based and {\it HST} data set. The first column lists the adopted stellar mass bins, where $\mathcal{M}~\equiv~\log_{10}(M_{\star}/\Msun)$, while columns $2-7$ list the logarithm of the number densities ($\phi_{k}$) within six redshift bins. The units of $\phi_{k}$ are dex$^{-1}$ Mpc$^{-3}$. The data presented in
this table are plotted in Fig. \ref{fig:observed_GSMFs}.}
\begin{tabular}{ | c | c | c | c | c | c | c |}
\hline
               & $0.25\leq z <0.75$ & $0.75\leq z <1.25$ & $1.25\leq z <1.75$ & $1.75\leq z <2.25$ & $2.25\leq z <2.75$ & $2.75\leq z <3.75$ \\ 
 & & & & & & \\
$\mathcal{M}$ & $\log_{10}(\phi_{k})$ & $\log_{10}(\phi_{k})$ &$\log_{10}(\phi_{k})$&$\log_{10}(\phi_{k})$ &$\log_{10}(\phi_{k})$ & $\log_{10}(\phi_{k})$ \\
\hline
$\phantom{0}7.25\leq\mathcal{M}<\phantom{0}7.50$ & $-1.13^{+0.09}_{-0.12}$ &  &  &  &  &  \\[0.15cm] 
$\phantom{0}7.50\leq\mathcal{M}<\phantom{0}7.75$ & $-1.28^{+0.09}_{-0.11}$ &  &  &  &  &  \\[0.15cm]
$\phantom{0}7.75\leq\mathcal{M}<\phantom{0}8.00$ & $-1.39^{+0.06}_{-0.07}$ & $-1.45^{+0.10}_{-0.13}$ &  &  &  &  \\ [0.15cm]
$\phantom{0}8.00\leq\mathcal{M}<\phantom{0}8.25$ &  $-1.54^{+0.05}_{-0.05}$ & $-1.53^{+0.08}_{-0.10}$ & $-1.55^{+0.11}_{-0.14}$ & $-1.52^{+0.11}_{-0.15}$ &  &  \\[0.15cm] 
$\phantom{0}8.25\leq\mathcal{M}<\phantom{0}8.50$ &  $-1.62^{+0.03}_{-0.03}$ & $-1.64^{+0.04}_{-0.05}$ & $-1.72^{+0.09}_{-0.12}$ & $-1.71^{+0.12}_{-0.16}$ & $-1.65^{+0.13}_{-0.18}$ &  \\[0.15cm] 
$\phantom{0}8.50\leq\mathcal{M}<\phantom{0}8.75$ &  $-1.74^{+0.04}_{-0.04}$ & $-1.80^{+0.04}_{-0.05}$ & $-1.77^{+0.05}_{-0.05}$ & $-1.90^{+0.10}_{-0.13}$ & $-1.76^{+0.12}_{-0.17}$ & $-1.80^{+0.12}_{-0.16}$ \\[0.15cm] 
$\phantom{0}8.75\leq\mathcal{M}<\phantom{0}9.00$ &  $-1.88^{+0.03}_{-0.03}$ & $-1.90^{+0.05}_{-0.06}$ & $-1.96^{+0.05}_{-0.05}$ & $-2.08^{+0.06}_{-0.06}$ & $-1.96^{+0.14}_{-0.20}$ & $-2.02^{+0.12}_{-0.17}$ \\[0.15cm] 
$\phantom{0}9.00\leq\mathcal{M}<\phantom{0}9.25$ &  $-2.00^{+0.03}_{-0.03}$ & $-2.05^{+0.04}_{-0.04}$ & $-2.09^{+0.06}_{-0.08}$ & $-2.29^{+0.04}_{-0.05}$ & $-2.23^{+0.06}_{-0.07}$ & $-2.23^{+0.07}_{-0.08}$ \\[0.15cm] 
$\phantom{0}9.25\leq\mathcal{M}<\phantom{0}9.50$ &  $-2.10^{+0.03}_{-0.03}$ & $-2.15^{+0.04}_{-0.04}$ & $-2.24^{+0.04}_{-0.04}$ & $-2.40^{+0.06}_{-0.07}$ & $-2.39^{+0.06}_{-0.07}$ & $-2.44^{+0.07}_{-0.09}$ \\[0.15cm] 
$\phantom{0}9.50\leq\mathcal{M}<\phantom{0}9.75$ &  $-2.22^{+0.03}_{-0.03}$ & $-2.27^{+0.04}_{-0.05}$ & $-2.41^{+0.02}_{-0.02}$ & $-2.56^{+0.07}_{-0.09}$ & $-2.55^{+0.05}_{-0.05}$ & $-2.63^{+0.05}_{-0.05}$ \\[0.15cm] 
$\phantom{0}9.75\leq\mathcal{M}<10.00$ &  $-2.32^{+0.03}_{-0.03}$ & $-2.39^{+0.04}_{-0.05}$ & $-2.53^{+0.03}_{-0.03}$ & $-2.67^{+0.03}_{-0.03}$ & $-2.69^{+0.03}_{-0.03}$ & $-2.86^{+0.07}_{-0.08}$ \\[0.15cm] 
$10.00\leq\mathcal{M}<10.25$ &  $-2.40^{+0.03}_{-0.03}$ & $-2.49^{+0.04}_{-0.05}$ & $-2.64^{+0.02}_{-0.02}$ & $-2.83^{+0.03}_{-0.03}$ & $-2.89^{+0.03}_{-0.03}$ & $-3.08^{+0.06}_{-0.07}$ \\[0.15cm] 
$10.25\leq\mathcal{M}<10.50$ &  $-2.47^{+0.03}_{-0.03}$ & $-2.55^{+0.04}_{-0.05}$ & $-2.72^{+0.02}_{-0.03}$ & $-2.93^{+0.03}_{-0.04}$ & $-3.03^{+0.04}_{-0.05}$ & $-3.34^{+0.06}_{-0.08}$ \\[0.15cm] 
$10.50\leq\mathcal{M}<10.75$ &  $-2.57^{+0.03}_{-0.03}$ & $-2.65^{+0.05}_{-0.05}$ & $-2.83^{+0.02}_{-0.03}$ & $-3.04^{+0.05}_{-0.06}$ & $-3.19^{+0.04}_{-0.04}$ & $-3.59^{+0.04}_{-0.05}$ \\[0.15cm] 
$10.75\leq\mathcal{M}<11.00$ &  $-2.76^{+0.04}_{-0.05}$ & $-2.85^{+0.05}_{-0.06}$ & $-3.02^{+0.04}_{-0.04}$ & $-3.21^{+0.05}_{-0.06}$ & $-3.38^{+0.05}_{-0.06}$ & $-3.86^{+0.05}_{-0.06}$ \\[0.15cm] 
$11.00\leq\mathcal{M}<11.25$ &  $-3.11^{+0.05}_{-0.06}$ & $-3.25^{+0.07}_{-0.09}$ & $-3.44^{+0.04}_{-0.04}$ & $-3.57^{+0.05}_{-0.06}$ & $-3.77^{+0.06}_{-0.06}$ & $-4.23^{+0.10}_{-0.14}$ \\[0.15cm] 
$11.25\leq\mathcal{M}<11.50$ &  $-3.66^{+0.09}_{-0.12}$ & $-3.86^{+0.10}_{-0.12}$ & $-4.04^{+0.05}_{-0.06}$ & $-4.25^{+0.07}_{-0.08}$ & $-4.38^{+0.09}_{-0.12}$ & $-4.69^{+0.19}_{-0.35}$ \\[0.15cm] 
$11.50\leq\mathcal{M}<11.75$ &  $-4.42^{+0.12}_{-0.16}$ & $-4.93^{+0.17}_{-0.29}$ & $-5.07^{+0.13}_{-0.19}$ & $-5.27^{+0.14}_{-0.20}$ & $-5.30^{+0.18}_{-0.31}$ & $-5.47^{+0.19}_{-0.35}$ \\[0.15cm] 
$11.75\leq\mathcal{M}<12.00$ &  $-5.71^{+0.32}_{-\inf}$ & $-6.36^{+0.43}_{-\inf}$ & & $-6.23^{+0.35}_{-\inf}$ & $-6.23^{+0.36}_{-\inf}$ & $-6.81^{+0.42}_{-\inf}$ \\[0.15cm] 
\hline
\end{tabular}
\label{tab:number_densities_total}
\end{table*}

\begin{table*}
\caption{The best-fitting single (upper section) and double (lower section) Schechter function parameters to the observed GSMF, where $\mathcal{M}^{\star} \equiv \log_{10}(M^{\star}/\Msun)$ and the units of $\phi^{\star}$, $\phi_{1}^{\star}$ and $\phi_{2}^{\star}$ are dex$^{-1}$ Mpc$^{-3}$. The final two columns list the $\chi^2$ and $\chi^2_{\nu}$ values of the fits, respectively. The best-fitting double Schechter function parameters
for the local observed GSMF derived by \protect\cite{Baldry2012} are provided for comparison.}
\begin{tabular}{ | c | c | c | c | c | c |c|c|}
\hline
Redshift & $\mathcal{M}^{\star}$ & & & $\log_{10} (\phi^{\star})$ &  
$\alpha$ & $\chi^{2}$ & $\chi^{2}_{\nu}$ \\ 
\hline
$0.25 \leq z < 0.75$ & 11.05 $\pm$ 0.03 & & &$-$3.04 $^{+0.03}_{-0.03}$ &  $-$1.38 $\pm$ 0.01 & 44.60 & 2.79 \\[0.15cm]
$0.75 \leq z < 1.25$ & 10.96 $\pm$ 0.03 & & &$-$3.08 $^{+0.04}_{-0.05}$ &  $-$1.38 $\pm$ 0.02 & 24.85 & 1.77 \\[0.15cm]
$1.25 \leq z < 1.75$ & 10.96 $\pm$ 0.02 & & &$-$3.21 $^{+0.03}_{-0.04}$ &  $-$1.37 $\pm$ 0.02 & 57.19 & 4.77 \\[0.15cm]
$1.75 \leq z < 2.25$ & 10.99 $\pm$ 0.03 & & &$-$3.48 $^{+0.05}_{-0.05}$ &  $-$1.45 $\pm$ 0.03 & 27.17 & 2.09 \\[0.15cm]
$2.25 \leq z < 2.75$ & 11.00 $\pm$ 0.04 & & &$-$3.61 $^{+0.06}_{-0.07}$ &  $-$1.53 $\pm$ 0.04 & 13.48 & 1.12 \\[0.15cm]
$2.75 \leq z < 3.75$ & 11.09 $\pm$ 0.07 & & &$-$4.18 $^{+0.09}_{-0.12}$ &  $-$1.81 $\pm$ 0.04 & \phantom{1}0.89 & 0.08 \\[0.15cm]
\hline
Redshift & $\mathcal{M}^{\star}$ & $\log_{10} (\phi_{1}^{\star})$ &  
$\alpha_{1}$ &  $\log_{10}(\phi_{2}^{\star})$ & $\alpha_{2}$ & $\chi^{2}$ & $\chi^{2}_{\nu}$ \\ 
\hline
$z < 0.06$      & 10.66 $\pm$ 0.05 & $-$2.40 $^{+0.04}_{-0.04}$ &  $-$0.35 $\pm$ 0.18 & $-$3.10 $^{+0.11}_{-0.15}$ & $-$1.47 $\pm$ 0.05 & \multicolumn{2}{c}{Baldry et al. (2012)} \\[0.15cm]
$0.25 \leq z < 0.75$ & 10.80 $\pm$ 0.06 & $-$2.77 $^{+0.06}_{-0.07}$ &  $-$0.61 $\pm$ 0.23 & $-$3.26 $^{+0.12}_{-0.17}$ & $-$1.52 $\pm$ 0.05 & 3.03 & 0.22  \\[0.15cm]
$0.75 \leq z < 1.25$ & 10.72 $\pm$ 0.07 & $-$2.80 $^{+0.07}_{-0.09}$ &  $-$0.46 $\pm$ 0.34 & $-$3.26 $^{+0.15}_{-0.23}$ & $-$1.53 $\pm$ 0.07 & 2.54 & 0.21  \\[0.15cm]
$1.25 \leq z < 1.75$ & 10.72 $\pm$ 0.05 & $-$2.94 $^{+0.04}_{-0.05}$ &  $-$0.55 $\pm$ 0.22 & $-$3.54 $^{+0.14}_{-0.22}$ & $-$1.65 $\pm$ 0.07 & 5.36 & 0.54  \\[0.15cm]
$1.75 \leq z < 2.25$ & 10.77 $\pm$ 0.06 & $-$3.18 $^{+0.07}_{-0.08}$ &  $-$0.68 $\pm$ 0.29 & $-$3.84 $^{+0.22}_{-0.46}$ & $-$1.73 $\pm$ 0.12 & 4.02 & 0.37  \\[0.15cm]
$2.25 \leq z < 2.75$ & 10.77 $\pm$ 0.10 & $-$3.39 $^{+0.09}_{-0.11}$ &  $-$0.62 $\pm$ 0.50 & $-$3.78 $^{+0.23}_{-0.50}$ & $-$1.74 $\pm$ 0.13 & 2.73 & 0.27 \\[0.15cm]
$2.75 \leq z < 3.75$ & 10.84 $\pm$ 0.18 &$-$4.30 $^{+0.23}_{-0.52}$ & $-$0.00 $\pm$ 1.03 & $-$3.94 $^{+0.20}_{-0.37}$   & $-$1.79 $\pm$ 0.09 & 0.16 & 0.02  \\[0.15cm]
\hline
\end{tabular}
\label{tab:observed_Schechter_parameters}
\end{table*}

\subsection{Fitting the observed GSMF}
In each redshift range we derive maximum likelihood fits to the binned GSMF data shown in Fig. \ref{fig:observed_GSMFs} using both a single and double Schechter function parameterization \citep{Schechter1976}.
The single Schechter function has the following functional form:
\begin{equation}
\phi(\mathcal{M}) = \phi^{\star}\cdot \rm \ln(10)\cdot[10^{(\mathcal{M}-\mathcal{M}^{\star})}]^{(1+\alpha)}\cdot \rm \exp[-10^{(\mathcal{M}-\mathcal{M}^{\star})}],
\end{equation}
where  $\phi(\mathcal{M})$ is the number density of galaxies per Mpc$^{3}$ per dex stellar mass, $\mathcal{M^{\star}}\equiv \log(M^{\star}/\Msun)$,
where $M^{\star}$ is the characteristic stellar mass and $\alpha$ is the low-mass slope. The double Schechter function has the form:
\begin{multline}
\phi(\mathcal{M}) = \ln(10)\cdot \exp[-10^{(\mathcal{M}-\mathcal{M}^{\star})}]\cdot 10^{(\mathcal{M}-\mathcal{M}^{\star})}\\
\cdot [\phi_{1}^{\star}\cdot 10^{(\mathcal{M}-\mathcal{M}^{\star})\alpha_{1}}+\phi_{2}^{\star}\cdot 10^{(\mathcal{M}-\mathcal{M}^{\star})\alpha_{2}}],  
\end{multline}
where both components have the same characteristic mass and we define $\alpha_{1}$ to be the high-mass slope and $\alpha_{2}$ to be the low-mass slope.
The best-fitting parameters and their corresponding uncertainties are presented in Table \ref{tab:observed_Schechter_parameters}, which also includes, for comparison, the double Schechter function parameters
for the local GSMF derived by \cite{Baldry2012}. The best-fitting single and double Schechter functions are plotted as the
dashed grey curves and solid black curves in Fig. \ref{fig:observed_GSMFs}, respectively.

It can be seen from Fig. \ref{fig:observed_GSMFs} that a double Schechter function appears to provide a better description of the data in all redshift bins out to $z=2.5$, whereas in the final
redshift bin at $z=3.25$ the difference between the single and double Schechter function fit is negligible. This impression is confirmed by the information displayed in Table \ref{tab:observed_Schechter_parameters}, which shows that the double Schechter function provides a better statistical description of the data in all six redshifts bins. Notably, in the first four redshift bins, covering the redshift range $0.25 \leq z < 2.25$, the single Schechter function does not provide a statistically acceptable fit to the data, whereas the double Schechter function fit is statistically acceptable at all redshifts. However, it is also worth noting that in the last two redshift bins, covering the range $2.25 \leq z < 3.75$, the single and double Schechter function fits are both statistically acceptable.

Overall, it can be seen from Table \ref{tab:observed_Schechter_parameters} that the best-fitting Schechter function parameters exhibit remarkably smooth and modest evolution over the redshift range
studied here, a subject we will return to in Section 4.6. Focusing on the double Schechter function parameters, it can be seen that the characteristic stellar mass, in particular, remains remarkably constant, lying
within the range $\mathcal{M^{\star}}=10.75\pm0.1$ at all redshifts. Moreover, it is noteworthy that the difference in the fitted Schechter function power-law indices (i.e. $\Delta\alpha = \alpha_{1}-\alpha_{2}$) is
consistent with unity
at all redshifts, with a variance weighted mean difference of $\Delta \alpha = 1.09\pm0.21$. This result is in good agreement with the phenomenological model of
\cite{Peng2010}, and is a subject we will return to when we investigate the individual star-forming and passive GSMFs in Section 5.

\begin{figure}
\includegraphics[width=\columnwidth]{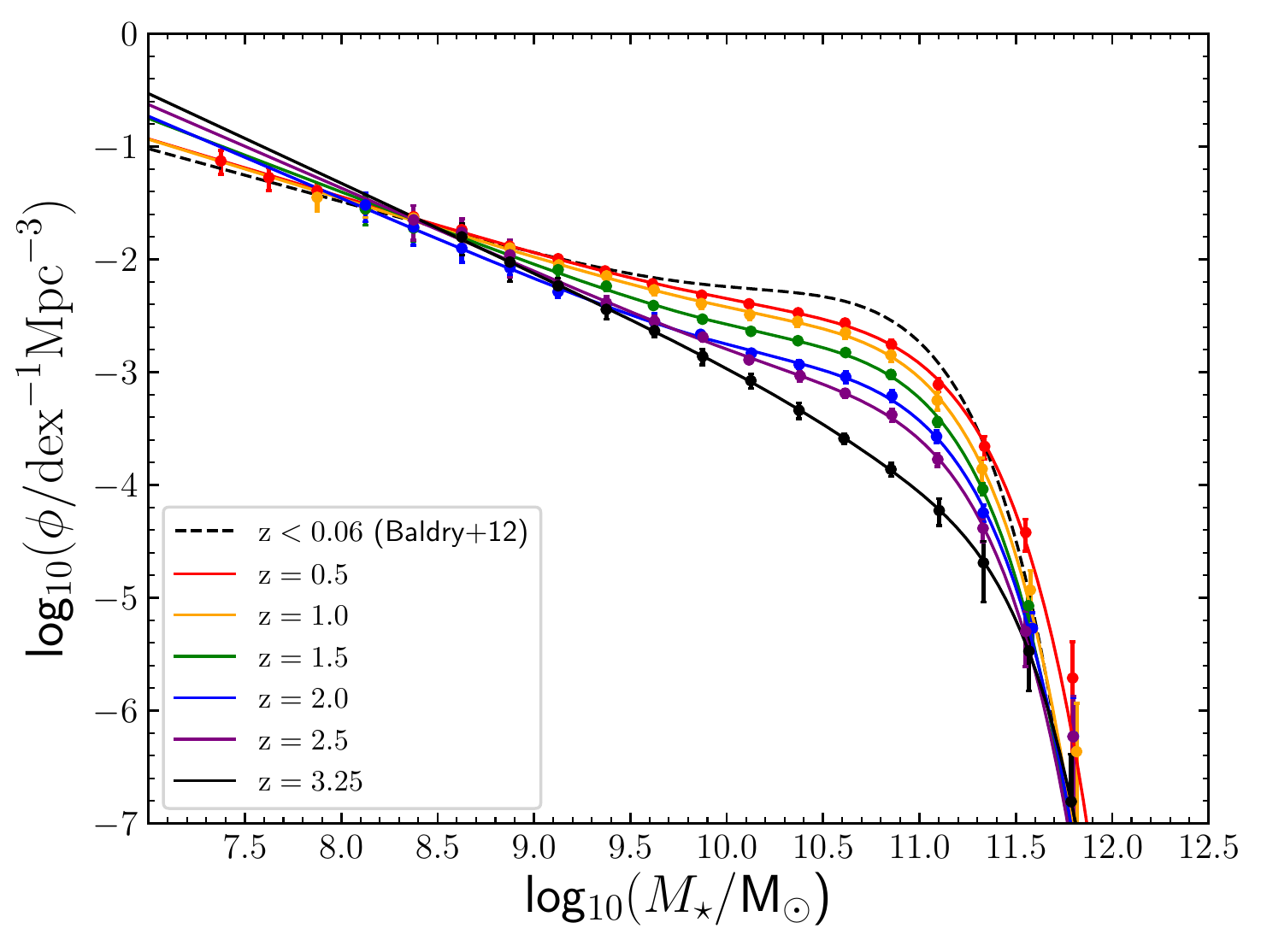}
\caption{An overlay of the observed GSMF data in all six redshift bins, together with the corresponding best-fitting double Schechter functions. The best-fitting double
Schechter function to the local observed GSMF derived by \protect\cite{Baldry2012} is included for comparison. This plot highlights the
substantial redshift evolution around the characteristic stellar mass (i.e. $\mathcal{M}\simeq 10.75$), in contrast to the lack of evolution at either low or high stellar masses.}
\label{fig:all_z_observed_GSMFs}
\end{figure}

The evolution of the observed GSMF can be seen more clearly in Fig. \ref{fig:all_z_observed_GSMFs}, which shows an overlay of both the data and the corresponding best-fitting double Schechter function fits
over the full $0.25 \leq z < 3.75$ redshift range. To allow comparison with the local GSMF, we have also included the double Schechter function fit from \cite{Baldry2012}. This plot very clearly illustrates that
there is very little evolution in the observed GSMF at either the low-mass ($8.0<\mathcal{M}<9.0$) or high-mass ($\mathcal{M}>11.5$) end. In contrast, substantial evolution is apparent in the number density of
galaxies close to the characteristic stellar mass (i.e. $\mathcal{M}\simeq 10.75$).

\subsection{Eddington bias}
\begin{table*}
\caption{The best-fitting single (upper section) and double (lower section) Schechter function parameters to the intrinsic GSMF, where $\mathcal{M}^{\star} \equiv \log_{10}(M^{\star}/\Msun)$ and the units of $\phi^{\star}$, $\phi_{1}^{\star}$ and $\phi_{2}^{\star}$ are dex$^{-1}$ Mpc$^{-3}$. The final two columns list the $\chi^2$ and $\chi^2_{\nu}$ values of the fits, respectively. As described in the text, the best-fitting intrinsic Schechter function parameters have been derived by incorporating a convolution of $\sigma_{\mathcal{M}}=0.15$ dex when fitting the observed GSMF data. For comparison, we also include our estimate of the intrinsic local GSMF, derived by fitting the data from \protect\cite{Baldry2012} assuming that $\sigma_{\mathcal{M}}=0.1$ dex \citep{Wright2018}.}
\begin{tabular}{ | c | c | c | c | c | c | c | c |}
\hline
Redshift & $\mathcal{M}^{\star}$ & & & $\log_{10} (\phi^{\star})$ &  
$\alpha$ & $\chi^{2}$ & $\chi^{2}_{\nu}$ \\ 
\hline
$0.25 \leq z < 0.75$ & 10.96 $\pm$ 0.03& & &$-$2.99 $^{+0.03}_{-0.03}$ &  $-$1.37 $\pm$ 0.01 & 59.50 & 3.72 \\[0.15cm]
$0.75 \leq z < 1.25$ & 10.86 $\pm$ 0.03& & &$-$3.01 $^{+0.04}_{-0.05}$ &  $-$1.37 $\pm$ 0.02 & 36.83 & 2.63 \\[0.15cm]
$1.25 \leq z < 1.75$ & 10.88 $\pm$ 0.02& & &$-$3.15 $^{+0.04}_{-0.04}$ &  $-$1.36 $\pm$ 0.03 & 72.45 & 6.04 \\[0.15cm]
$1.75 \leq z < 2.25$ & 10.90 $\pm$ 0.03& & &$-$3.41 $^{+0.05}_{-0.06}$ &  $-$1.43 $\pm$ 0.03 & 35.91 & 2.76 \\[0.15cm]
$2.25 \leq z < 2.75$ & 10.91 $\pm$ 0.04& & &$-$3.54 $^{+0.06}_{-0.08}$ &  $-$1.51 $\pm$ 0.04 & 17.42 & 1.45 \\[0.15cm]
$2.75 \leq z < 3.75$ & 10.97 $\pm$ 0.07& & &$-$4.05 $^{+0.10}_{-0.13}$ &  $-$1.79 $\pm$ 0.05 & \phantom{1}1.82 & 0.17  \\[0.15cm]
\hline
Redshift & $\mathcal{M}^{\star}$ & $\log_{10} (\phi_{1}^{\star})$ &  
$\alpha_{1}$ &  $\log_{10}(\phi_{2}^{\star})$ & $\alpha_{2}$ & $\chi^{2}$ & $\chi^{2}_{\nu}$ \\ 
\hline
$z < 0.06$      & 10.60 $\pm$ 0.05 &$-$2.37 $^{+0.03}_{-0.04}$  & $-$0.20 $\pm$ 0.20            & $-$3.03 $^{+0.10}_{-0.13}$ & $-$1.45 $\pm$ 0.04 & 9.85 & 0.62\\[0.15cm]
$0.25 \leq z < 0.75$ & 10.64 $\pm$ 0.06 &$-$2.63 $^{+0.05}_{-0.05}$  & $-$0.25 $\pm$ 0.25            & $-$3.11 $^{+0.09}_{-0.11}$ & $-$1.49 $\pm$ 0.03 & 1.00 & 0.07  \\[0.15cm]
$0.75 \leq z < 1.25$ & 10.51 $\pm$ 0.07 &$-$2.67 $^{+0.06}_{-0.07}$  & \phantom{$-$}0.08 $\pm$ 0.37 & $-$3.07 $^{+0.11}_{-0.14}$ & $-$1.49 $\pm$ 0.05 & 1.01 & 0.08  \\[0.15cm]
$1.25 \leq z < 1.75$ & 10.54 $\pm$ 0.05 &$-$2.83 $^{+0.04}_{-0.04}$  & $-$0.07 $\pm$ 0.26            & $-$3.32 $^{+0.10}_{-0.14}$ & $-$1.60 $\pm$ 0.06 & 2.79 & 0.28  \\[0.15cm]
$1.75 \leq z < 2.25$ & 10.56 $\pm$ 0.07 &$-$3.05 $^{+0.06}_{-0.07}$  & $-$0.06 $\pm$ 0.39            & $-$3.51 $^{+0.15}_{-0.22}$ & $-$1.63 $\pm$ 0.09 & 3.27 & 0.30  \\[0.15cm]
$2.25 \leq z < 2.75$ & 10.55 $\pm$ 0.11 &$-$3.28 $^{+0.08}_{-0.10}$  & \phantom{$-$}0.02 $\pm$ 0.59 & $-$3.50 $^{+0.17}_{-0.28}$ & $-$1.66 $\pm$ 0.10 & 2.45 & 0.25  \\[0.15cm]
$2.75 \leq z < 3.75$ & 10.64 $\pm$ 0.17 &$-$4.08 $^{+0.18}_{-0.33}$  & \phantom{$-$}0.35 $\pm$ 1.06 & $-$3.74 $^{+0.20}_{-0.38}$ & $-$1.76 $\pm$ 0.10 & 0.32 & 0.04  \\[0.15cm]
\hline
\end{tabular}
\label{tab:intrinsic_Schechter_parameters}
\end{table*}

The observed GSMF results presented in the previous section will inevitably be subject to Eddington bias, whereby the combination of stellar-mass uncertainties and the
steep exponential fall-off of the GSMF leads to a net bias towards higher number densities at the high-mass end of the GSMF. As a consequence, the Schechter function parameters derived from a direct
fit to the observed GSMF data points will be biased with respect to the intrinsic Schechter function parameters. Recovering the intrinsic form of the GSMF is of particular interest for direct comparison
to galaxy simulation results (see Section 4.5) and for deriving accurate measurements of the integrated stellar-mass density (see Section 6).

In order to recover the intrinsic form of the GSMF it is necessary to determine the effective uncertainties in the stellar-mass measurements, which we model as a log-normal distribution with $\sigma_{\mathcal{M}}$.
We adopted two approaches to quantifying $\sigma_{\mathcal{M}}$. The first approach was to run a series of simulations in which the photometry for each galaxy was scattered according to its errors, before the
photometric redshifts and stellar masses were re-calculated. The results of these simulations indicated that $\sigma_{\mathcal{M}}\simeq 0.2$ dex and
was not a strong function of either redshift or stellar mass.

\begin{figure}
\includegraphics[width=\columnwidth]{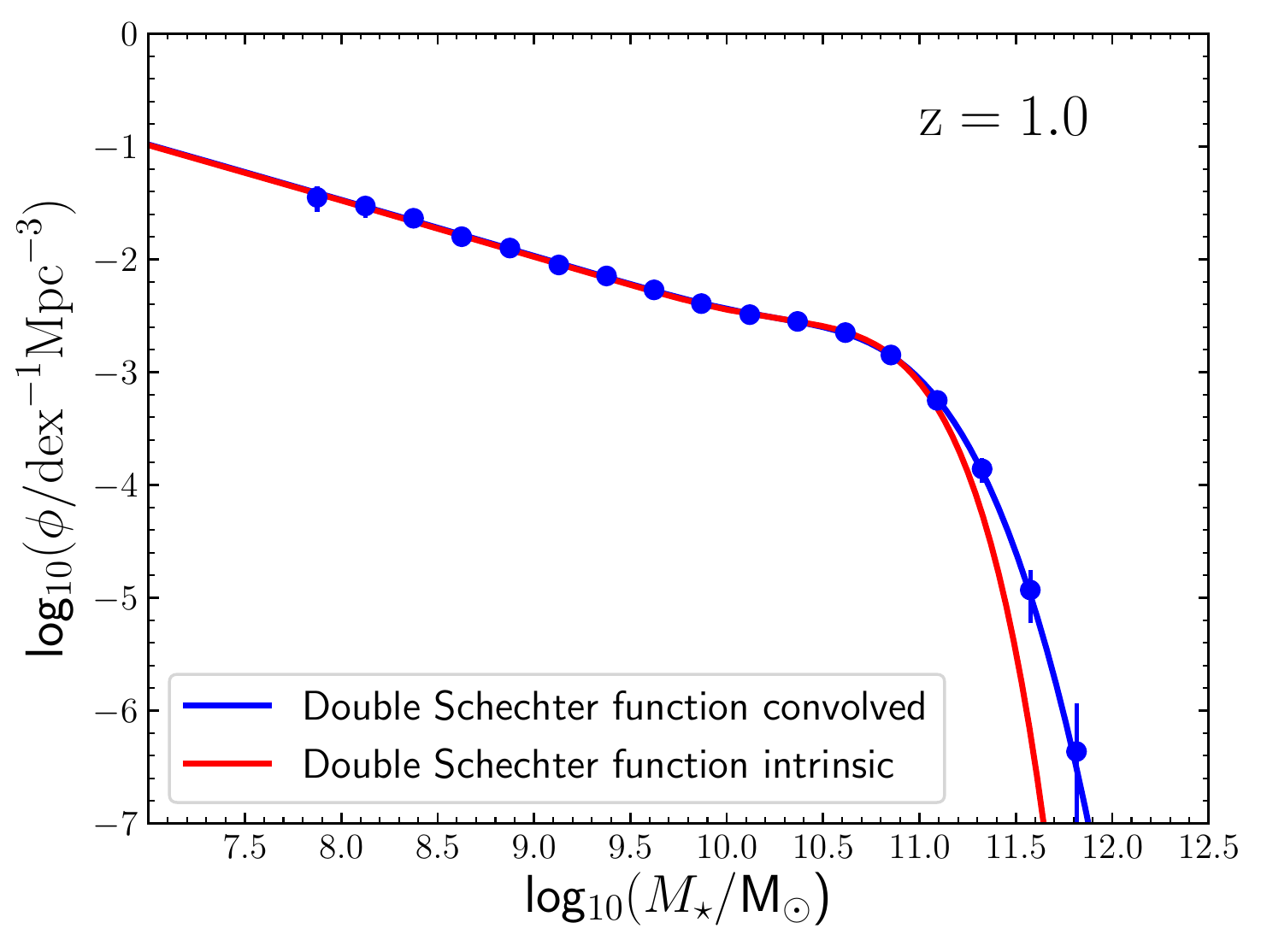}
\caption{A demonstration of the effect of Eddington bias on GSMF Schechter function parameter estimation. The blue data points are the observed GSMF at $z=1.0$. The red curve is our best estimate of
the {\it intrinsic} GSMF, which is obtained by fitting to the blue data points with a double Schechter function, assuming that $\sigma_{\mathcal{M}}=0.15$. Convolving the red curve with a log-normal distribution
with $\sigma_{\mathcal{M}}=0.15$ produces the blue curve, which is an excellent fit to the observed data points.}
\label{fig:edd_bias_demo}
\end{figure}

The second approach was to use the binned GSMF data itself to constrain the value of $\sigma_{\mathcal{M}}$. This process involved re-fitting the observed GSMF data with a double Schechter function as before,
but convolving the intrinsic Schechter function with a log-normal distribution with $\sigma_{\mathcal{M}}$, where $\sigma_{\mathcal{M}}$ is a free parameter in the fit. The results of this fitting process demonstrated
that $\sigma_{\mathcal{M}}$ had a mean value of 0.15 dex and lay within the range $0.15\pm0.04$ dex in all six redshift bins. As a result,
we adopted a value of $\sigma_{\mathcal{M}}=0.15$ dex and re-ran fits to the
observed GSMF data including the convolution due to stellar-mass uncertainties. The best-fitting Schechter function parameters from this fitting process represent our best estimates of the {\it intrinsic}
form of the GSMF. The typical impact of Eddington bias on the form of the best-fitting Schechter function is illustrated in Fig. \ref{fig:edd_bias_demo}.

\subsection{The intrinsic GSMF}
Our determination of the best-fitting intrinsic Schechter function parameters is presented in Table \ref{tab:intrinsic_Schechter_parameters}. As with the best-fitting observed Schechter function parameters
shown in Table \ref{tab:observed_Schechter_parameters}, it can be seen that the double Schechter function provides a better description of the data at all redshifts, and that the single Schechter function fits are statistically unacceptable at $z < 2.25$. Once again, the single and double Schechter function fits to the highest redshift bin are basically indistinguishable.

\begin{figure*}
\includegraphics[width=1.5\columnwidth]{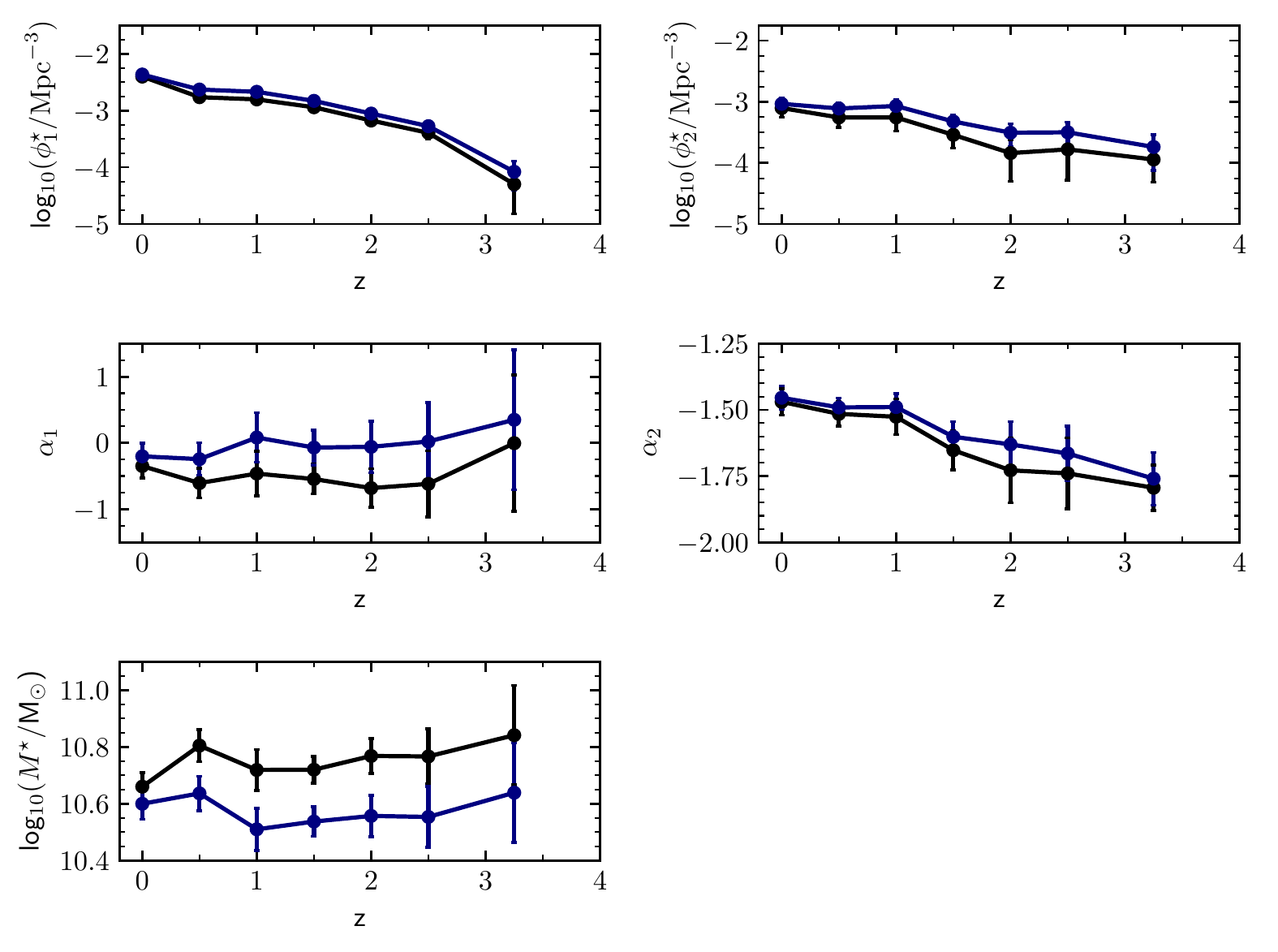}
\caption{The redshift evolution of the best-fitting observed (black) and intrinsic (blue) GSMF double Schechter function parameters. It can be seen that, following what is essentially a
deconvolution process, the normalizations of the intrinsic double Schechter function shift to slightly higher values and the slopes shift to slightly shallower values. The largest difference between the
observed and intrinsic parameters is $\mathcal{M}^{\star}$, which shifts $\simeq 0.2$ dex lower after Eddington bias has been accounted for.}
\label{fig:parameter_evolution_total_GSMF}
\end{figure*}

The evolution of the best-fitting observed and intrinsic double Schechter function parameters is shown in Fig. \ref{fig:parameter_evolution_total_GSMF}. As is to be expected, following what is essentially a
deconvolution process, the normalizations of the intrinsic double Schechter function shift to slightly higher values, and the slopes shift to slightly shallower values. Unsurprisingly, the largest difference between the observed and intrinsic Schechter function parameters
is the best-fitting value of the characteristic stellar mass. In both cases the characteristic stellar mass displays remarkably little redshift evolution but, after accounting for the effect of Eddington bias,
the best-fitting intrinsic value of $\mathcal{M^{\star}}$ is shifted $\simeq 0.2$ dex lower, lying within the range $\mathcal{M^{\star}}\simeq 10.55\pm{0.1}$ at all redshifts.

\begin{figure*}
\includegraphics[width=1.5\columnwidth]{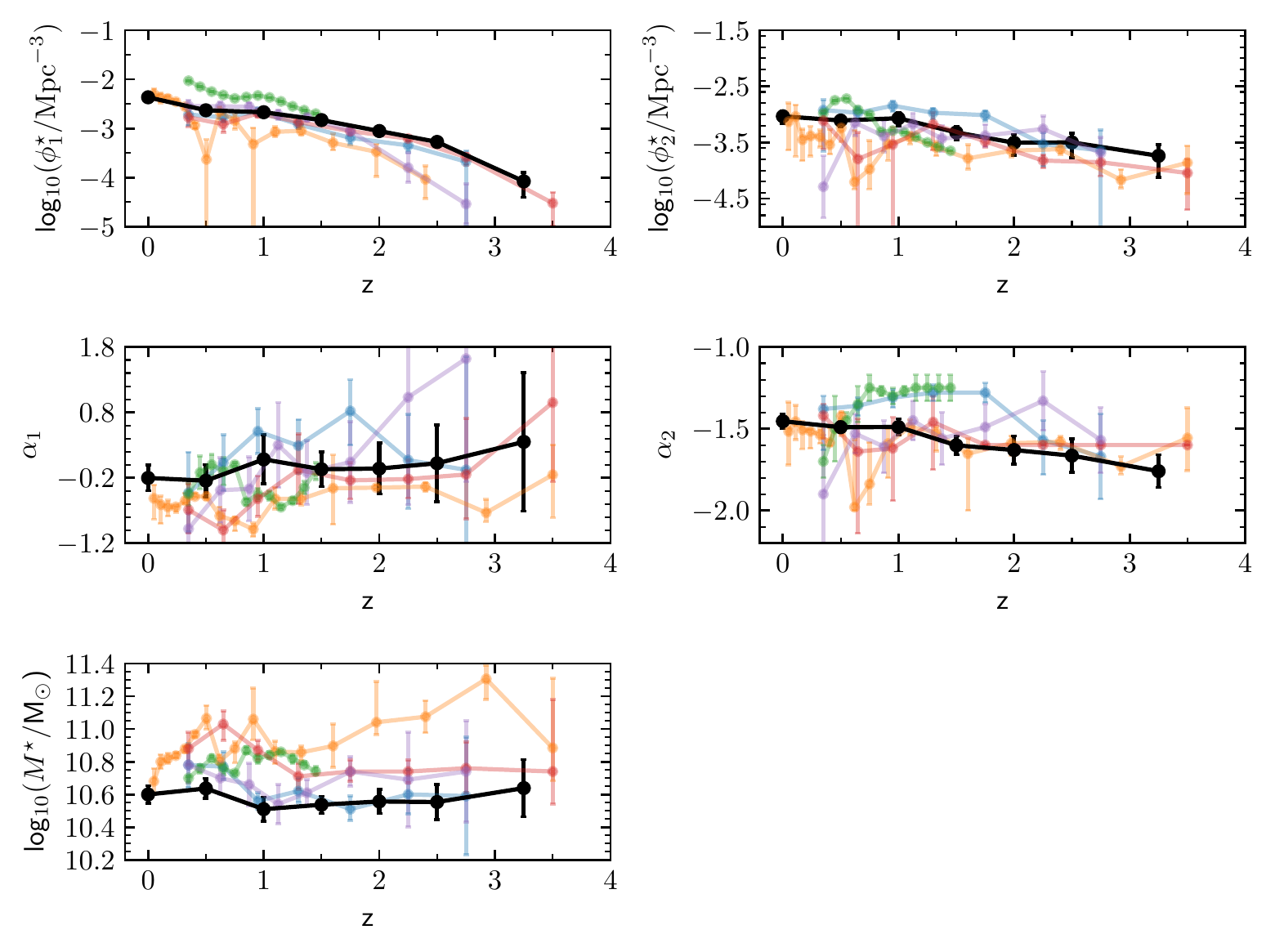}
\caption{A comparison between the best-fitting intrinsic GSMF double Schechter parameters derived in this study (black) with those from previous literature studies. Plotted for comparison are the results from
\protect\citet[red]{Ilbert2013}; \protect\citet[purple]{Tomczak2014}; \protect\citet[blue]{Davidzon2017}; \protect\citet[orange]{Wright2018} and \protect\citet[green]{Kawinwanichakij2020}. Note that the parameters derived by \protect\cite{Tomczak2014} are observed, rather than intrinsic.}
\label{fig:parameter_evolution_total_vs_literature}
\end{figure*}

In Fig. \ref{fig:parameter_evolution_total_vs_literature} we show a comparison between our intrinsic double Schechter function parameters and those derived by recent studies in the literature
\citep{Ilbert2013, Tomczak2014, Davidzon2017, Wright2018, Kawinwanichakij2020}. We note that all of the studies we compare to in
Fig. \ref{fig:parameter_evolution_total_vs_literature} quote intrinsic Schechter function parameters, with the exception of \cite{Tomczak2014} who quote observed parameters.
It can be seen from Fig. \ref{fig:parameter_evolution_total_vs_literature} that our intrinsic Schechter function parameters are in reasonable agreement with previous
determinations, although the parameter estimates in the literature span a significant range. It is also clear from Fig. \ref{fig:parameter_evolution_total_vs_literature} that
the unique combination of the dynamic range and cosmological volume sampled by this work has led to significantly improved parameter constraints.

It is noteworthy that our determination of the evolving characteristic stellar mass is significantly lower than most previous studies. However, it
can be seen from Fig. \ref{fig:parameter_evolution_total_vs_literature} that our determination of $\mathcal{M^{\star}}$ is in good agreement with \cite{Davidzon2017}, and is also likely to be in good agreement with \cite{Tomczak2014}, assuming that the Eddington bias correction for
their data set is similar to our estimate.

\subsection{Comparison to simulations}
\begin{figure}
\begin{center}
\begin{tabular}{c}
\includegraphics[width=0.8\columnwidth]{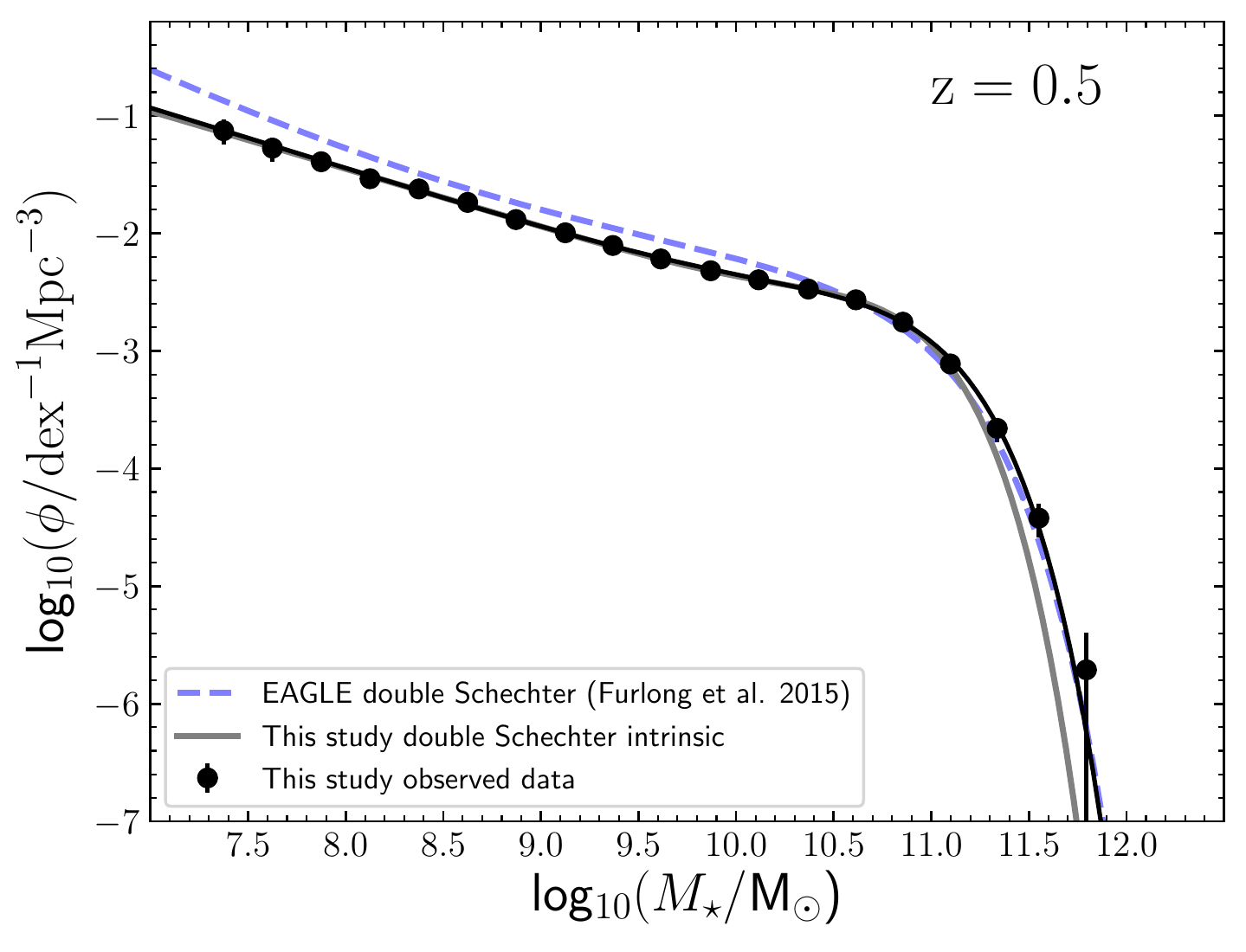}\\
\includegraphics[width=0.8\columnwidth]{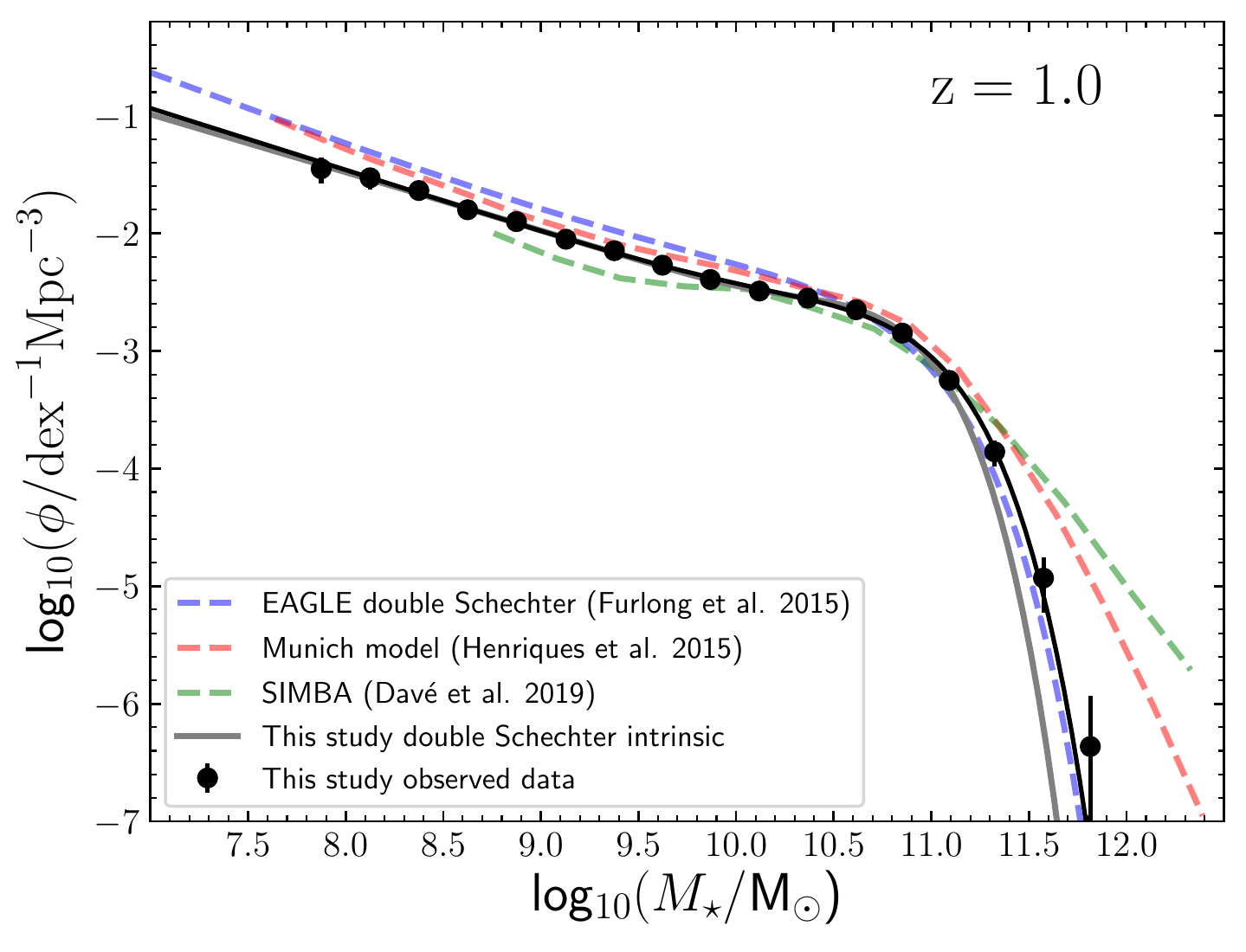} \\
\includegraphics[width=0.8\columnwidth]{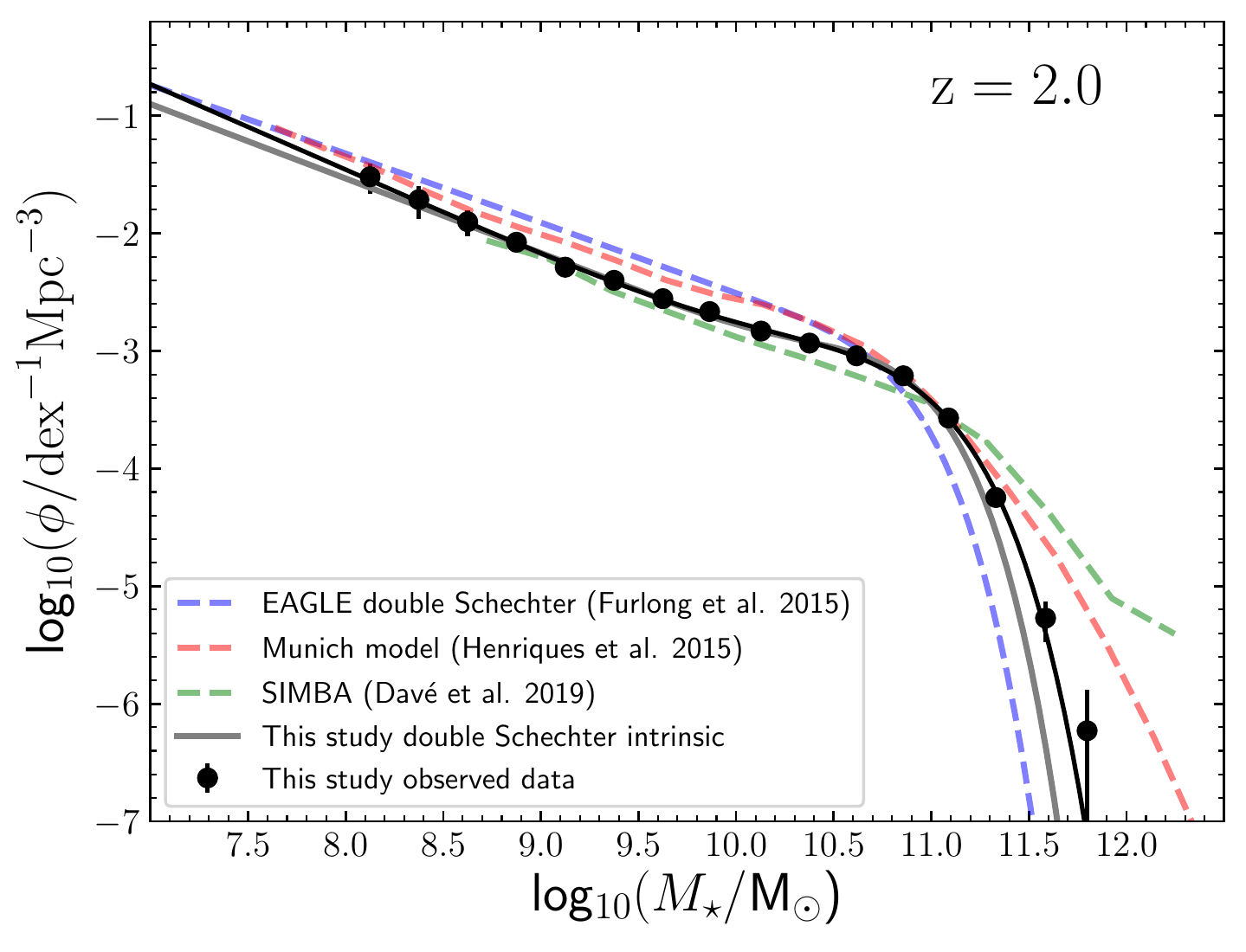} \\
\includegraphics[width=0.8\columnwidth]{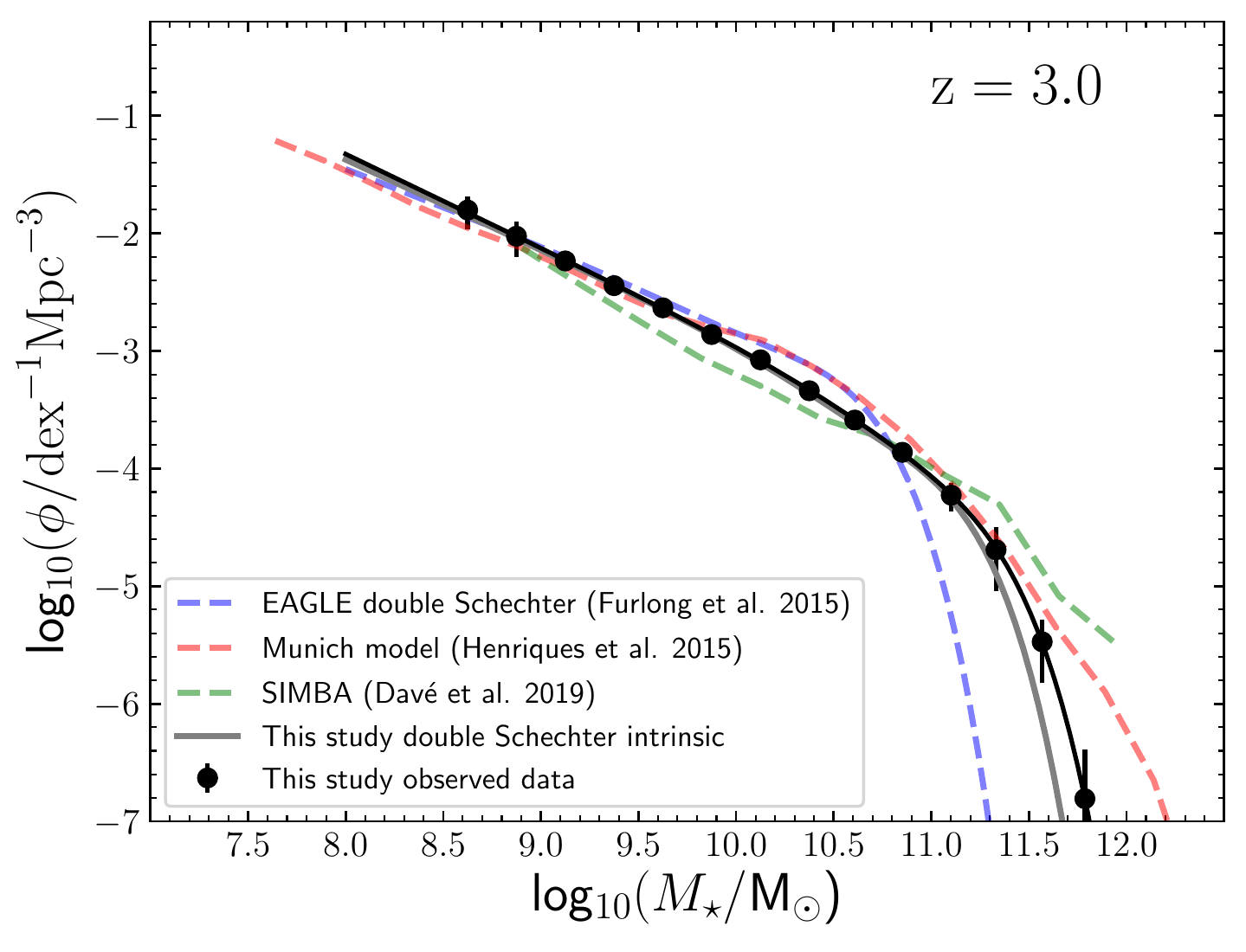} 
\end{tabular}
\end{center}
\caption{A comparison between our determination of the observed (black data points and curve) and intrinsic (grey curve) GSMF and the predictions of
the EAGLE \protect\citep{Furlong2015}, SIMBA \protect\citep{Dave2019} and Munich \protect\citep{Henriques2015} galaxy formation models.}
\label{fig:GSMF_vs_models}
\end{figure}

In Fig. \ref{fig:GSMF_vs_models} we show a comparison between our observed and
intrinsic GSMFs and the results of the EAGLE \citep{Furlong2015}, SIMBA \citep{Dave2019} and Munich galaxy formation models \citep{Henriques2015}.

At $z=0.5$ and $z=1.0$ the EAGLE hydrodynamical simulation is in generally good agreement with our GSMF determinations, particularly at $\mathcal{M}\geq\mathcal{M}^{\star}$. However, it
can also be seen that EAGLE systematically over-predicts the numbers of $\mathcal{M}\leq\mathcal{M}^{\star}$ galaxies in the redshift
range $0.5 \leq z \leq 2.0$, by a factor of $1.5-2.0$. By $z=3$, it is notable that EAGLE predicts a significantly lower value of $\mathcal{M^{\star}}$ than we observe.

The Munich semi-analytic model is in good agreement with our determination of the $\mathcal{M}\leq\mathcal{M}^{\star}$ number densities at $z=1.0$ and $z=3.0$, but over-predicts the $\mathcal{M}\leq\mathcal{M}^{\star}$ number densities at $z=2.0$ by a factor of $\simeq 1.5$. It is notable that the Munich model systematically over-predicts the number of galaxies with the highest stellar masses (i.e. $\mathcal{M}\geq 11.5$).

The systematic over-prediction of the high-mass end of the GSMF is a problem that is shared by the SIMBA hydrodynamical simulation. That said,
the SIMBA results at $z=1.0$ and $z=2.0$ are in generally good agreement with our observational results at $\mathcal{M}\leq\mathcal{M}^{\star}$. 

In Section 6, we compare the redshift evolution of the integrated stellar-mass density predicted
by the three theoretical models with our observational results.

\subsection{An evolving fit to the galaxy stellar mass function}
For the purposes of comparing to a variety of different theoretical and observational results, 
it is clearly desirable to be able to derive an accurate estimate of the total GSMF at any redshift. Guided by the smoothly evolving double Schechter function
parameters produced by our maximum likelihood fitting (see Fig. \ref{fig:parameter_evolution_total_GSMF}), we
adopt the following functional forms:
\begin{align}
\mathcal{M^{\star}} &=a_1 + a_2z\\
\alpha_1    &=a_3 + a_4z\\
\alpha_2    &=a_5 + a_6z\\
\mathrm{log}(\phi_{1}^{\star}) &=a_7 + a_8z + a_9z^2\\
\mathrm{log}(\phi_{2}^{\star}) &=a_{10} + a_{11}z.
\end{align}
It can be seen that all of the parameters follow a simple linear evolution with $z$, with the exception of $\mathrm{log}(\phi_{1}^{\star})$, which includes an additional $z^2$ term to account for
the steep decline at $z>2.5$.

To fit this 11-parameter model to the data we used \textsc{dynesty} \citep{Speagle2020}, a python implementation of the Bayesian Nested Sampling algorithm \citep{Skilling2006}, that can be used to estimate
posteriors on model parameters after specifying an appropriate likelihood function and prior. We fitted the
model across all $96$ ($z, \log_{10}(\phi)$, $\mathcal{M}$) values in our data set (Table \ref{tab:number_densities_total}), assuming a standard Gaussian likelihood of the form:
\begin{equation}
\mathrm{ln}(L) = -\frac{1}{2}\sum_{i}\bigg[\frac{(\phi_{i}-\phi_{i}^{\prime}(\theta))^2}{\sigma_{\phi_{i}}^2} + \mathrm{ln}(2\pi\sigma_{\phi_{i}}^2)\bigg],
\end{equation}
where $\phi_{i}$ and $\sigma_{\phi_{i}}$ are the observed number densities and their corresponding errors at a given redshift and stellar mass, and $\phi_{i}^{\prime}(\theta)$ are the corresponding model
number densities, based on the parameter set $\theta=(a_1,\ldots, a_{11})$. We assumed flat priors on all parameters, within the ranges specified in Table \ref{table:best_fit_params}.

We ran fits to both the observed and the intrinsic GSMF.
For the intrinsic GSMF we assumed a constant $\sigma_{\mathcal{M}}=0.15$ dex based on our maximum likelihood analysis.
The median posterior parameter values and their corresponding $68\%$ confidence intervals for both fits are
listed in Table \ref{table:best_fit_params}, and a corner plot showing the 1-D and 2-D marginalized posteriors for the
fit to the intrinsic GSMF is shown in Fig.~\ref{fig:corner}.

\begin{figure*}
\begin{tabular}{cc}
\includegraphics[width=\columnwidth]{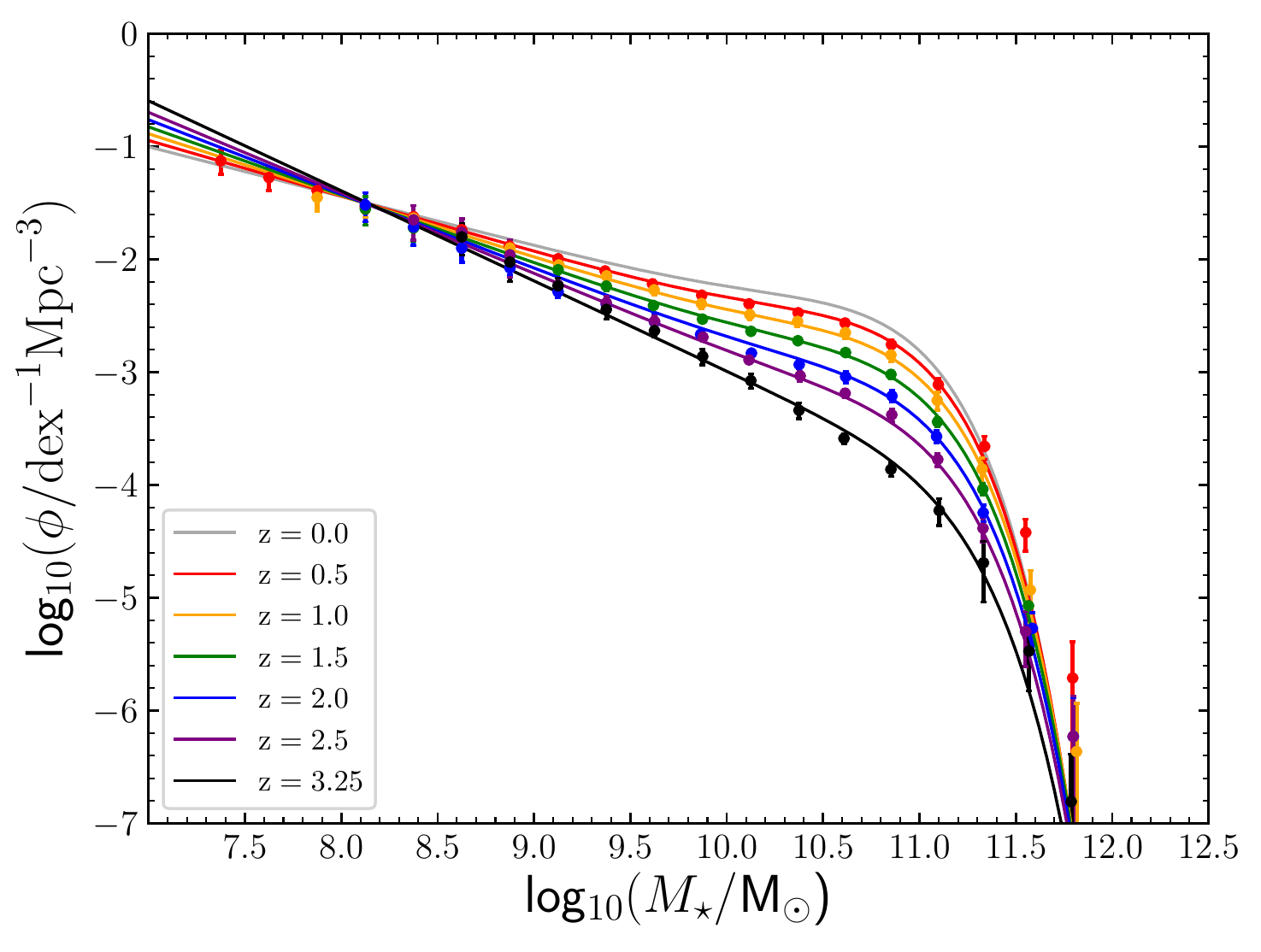} &
\includegraphics[width=\columnwidth]{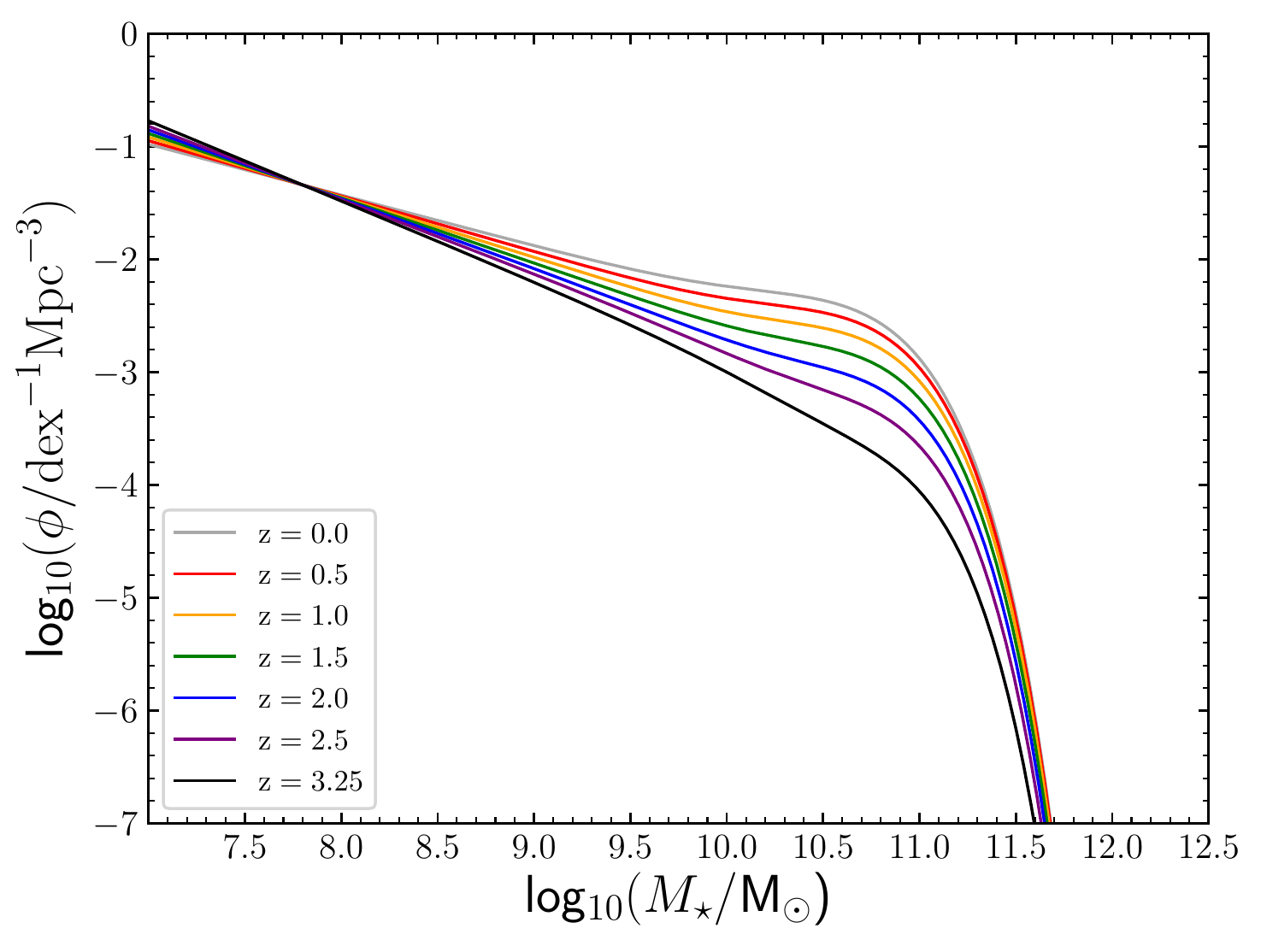}
\end{tabular}
\caption{The left-hand panel shows a comparison between the observed GSMF data and the fit produced using our simple evolving parameterization (Eqns. $5-9$, Table \ref{table:best_fit_params}) over the full redshift range of this study. The right-hand panel shows our simple parameterization of the evolving intrinsic GSMF, after accounting for Eddington bias (see text for details.)}
\label{fig:simple_model_GSMFs}
\end{figure*}

The best-fitting evolving model is shown in Fig. \ref{fig:simple_model_GSMFs} and provides an excellent fit to the data, with $\chi^2_{\nu}=1.00$ and $0.85$ for the observed and intrinsic fits, respectively. As a result, this simplified evolving prescription can be used to provide an
accurate estimate of the GSMF (both observed and intrinsic) at any desired redshift within the range $0.0 \leq z \leq 3.75$. In fact, comparison with GSMF constraints at higher
redshift (e.g. \citealt{Duncan2014,Grazian2015,Song2016}) suggests that the parameterization presented here remains in reasonable agreement with observational constraints out to $z\simeq 5$.

\begin{table}
    \centering
\caption{Details of our simple parameterization of the evolving GSMF. The first column lists the parameters (see Eqns. $5-9$) and the second column lists the range of
the corresponding flat priors adopted during the fitting process. The final two columns list the best-fitting parameter values from the fit to the observed and intrinsic GSMF, respectively.}
\label{table:best_fit_params}

\begin{tabular}{crrr}
        \hline
        Parameter & \multicolumn{1}{c}{Prior range} & \multicolumn{1}{c}{GSMF} & \multicolumn{1}{c}{GSMF}\\
                  &  & \multicolumn{1}{c}{observed} & \multicolumn{1}{c}{intrinsic}\\
        \hline
        $a_1$ & $(10.4,10.8)$    &$10.67^{+0.03}_{-0.03}$                 &$10.55^{+0.03}_{-0.03}$\\[0.15cm]
        $a_2$ & $(-1.0,1.0)$     &\phantom{$-$}$0.04^{+0.02}_{-0.02}$     &\phantom{$-$}$0.00^{+0.02}_{-0.02}$\\[0.15cm]
        $a_3$ & $(-0.5,0.5)$     &$-0.32^{+0.12}_{-0.11}$                 &$-0.16^{+0.10}_{-0.15}$\\[0.15cm]
        $a_4$ & $(-1.0,1.0)$     &$-0.12^{+0.10}_{-0.12}$                 &\phantom{$-$} $0.12^{+0.12}_{-0.11}$\\[0.15cm]
        $a_5$ & $(-1.2,-1.7)$    &$-1.44^{+0.03}_{-0.03}$                 &$-1.45^{+0.02}_{-0.02}$\\[0.15cm]
        $a_6$ & $(-1.0,1.0)$     &$-0.11^{+0.02}_{-0.03}$                 &$-0.08^{+0.02}_{-0.02}$\\[0.15cm]
        $a_7$ & $(-2.2,-2.9)$    &$-2.53^{+0.04}_{-0.05}$                 &$-2.43^{+0.04}_{-0.04}$\\[0.15cm]
        $a_8$ & $(-1.0,1.0)$     &$-0.20^{+0.05}_{-0.05}$                 &$-0.17^{+0.05}_{-0.05}$\\[0.15cm]
        $a_9$ & $(-1.0,1.0)$     &$-0.07^{+0.02}_{-0.02}$                 &$-0.08^{+0.02}_{-0.02}$\\[0.15cm]
        $a_{10}$ & $(-2.8,-3.2)$  &$-2.98^{+0.06}_{-0.06}$                &$-2.94^{+0.05}_{-0.05}$\\[0.15cm]
        $a_{11}$ & $(-1.0,1.0)$   &$-0.31^{+0.05}_{-0.07}$                 &$-0.22^{+0.04}_{-0.04}$\\[0.15cm]
        \hline
    \end{tabular}
\end{table}

\section{The passive and star-forming galaxy stellar mass functions}
In this section we proceed to split our galaxy sample into its
star-forming and passive components, in order to explore how the
GSMF and integrated stellar-mass density of each component evolves with redshift. Fundamentally,
the differential evolution of the star-forming and passive GSMFs provides crucial constraints
on the impact of mass and environmental quenching as a function of redshift and stellar/halo mass.

\subsection{UVJ selection}
\begin{figure}
\includegraphics[width=\columnwidth]{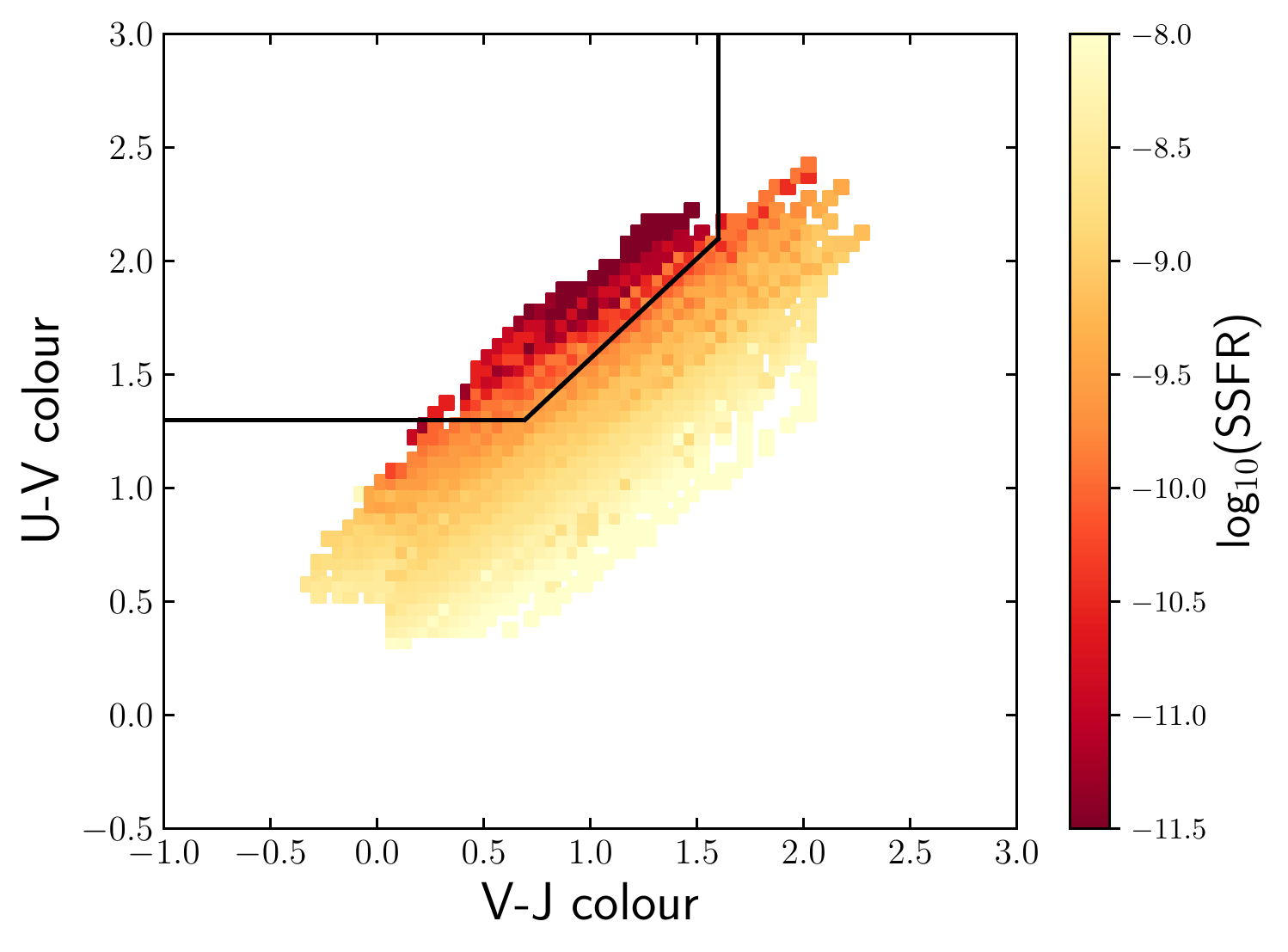}
\caption{The (U-V) colour versus (V-J) colour for our sample of galaxies over $0.25\leq z \leq 3.75$, demonstrating our UVJ selection between passive and star-forming galaxies. Objects are binned by their (U-V), (V-J) colours. The minimum number of objects in a bin is 20. A colour map of the median $\log_{10} SSFR$ in each bin is also shown. The boundaries of our selection criteria for splitting our sample into passive and star-forming galaxies are displayed, following Eqns 11-13.}
\label{fig:UVJ_selection}
\end{figure}

We separate our galaxy sample into its star-forming and passive components using the UVJ colour-colour
criteria proposed by \cite{Williams2009}. Specifically, we follow the results of \cite{Carnall2018, Carnall2020} and apply the following criteria:
\begin{align}
(U-V)> &\, 0.88\times(V-J) + 0.69;\\
(U-V)> &\, 1.3;\\
(V-J)< &\, 1.6,
\end{align}
at all redshifts. Although not entirely model independent, applying this set of criteria is robust, and has
the advantage of being easy to apply to a wide variety of observed and simulated data sets. The rest-frame
colours for the ground-based component of our final galaxy sample were generated from the SED fitting described in
Section 3.4. For the {\it HST} CANDELS component, we adopted the rest-frame colours from the relevant publicly
available catalogue (see Table \ref{tab:CANDELS_cats}), and used the overlap with our ground-based data to adjust for off-sets due to
different filter definitions (typically at the $\pm 0.1$ mag level). We include an illustration of our UVJ selection in Fig. \ref{fig:UVJ_selection}.

The number densities of the passive and star-forming populations were calculated according to the description provided in Section 4.1.
However, when calculating the uncertainties associated with the number densities in each ($\mathcal{M}$, $z$) bin, an additional
contribution was included to account for objects scattering into and out of the UVJ selection box due
to photometric uncertainties. This additional contribution was calculated as part of the bootstrap simulations described previously,
by scattering the $U-V$ and $V-J$ colours according to their uncertainties. The number densities for the star-forming and passive galaxies
are presented in Table \ref{tab:blue_number_densities} and Table \ref{tab:red_number_densities}, respectively.

\begin{table*}
\caption{The observed GSMF for UVJ--selected star-forming galaxies (see text for details). The first column lists the adopted stellar mass bins, where $\mathcal{M}~\equiv~\log_{10}(M_{\star}/\Msun)$, while columns $2-7$ list the logarithm of the number densities ($\phi_{k}$) within six redshift bins. The units of $\phi_{k}$ are dex$^{-1}$~Mpc$^{-3}$. The data presented in this table are plotted in Fig. \ref{fig:UVJ_GSMFs}.}
\begin{tabular}{ | c | c | c | c | c | c | c |}
\hline
               & $0.25\leq z <0.75$ & $0.75\leq z <1.25$ &
 $1.25\leq z <1.75$ & $1.75\leq z <2.25$ & $2.25\leq z <2.75$
 & $2.75\leq z <3.75$ \\  & & & & & & \\ $\mathcal{M}$ &
 $\log_{10}(\phi_{k})$ & $\log_{10}(\phi_{k})$ & $\log_{10}(\phi_{k})$
 &$\log_{10}(\phi_{k})$ & $\log_{10}(\phi_{k})$ &
 $\log_{10}(\phi_{k})$\\
\hline
$\phantom{0}8.00\leq\mathcal{M}<\phantom{0}8.25$ & $-$1.60 $^{+0.05}_{-0.05}$ &  & &  &  &   \\[0.15cm] 
$\phantom{0}8.25\leq\mathcal{M}<\phantom{0}8.50$ & $-$1.69 $^{+0.03}_{-0.03}$ & $-$1.70 $^{+0.04}_{-0.05}$ & &  &  &   \\[0.15cm] 
$\phantom{0}8.50\leq\mathcal{M}<\phantom{0}8.75$ & $-$1.83 $^{+0.04}_{-0.04}$ & $-$1.82 $^{+0.04}_{-0.05}$ & $-$1.79 $^{+0.05}_{-0.05}$ & & & \\[0.15cm] 
$\phantom{0}8.75\leq\mathcal{M}<\phantom{0}9.00$ & $-$1.93 $^{+0.03}_{-0.03}$ & $-$1.94 $^{+0.03}_{-0.03}$ & $-$1.96 $^{+0.05}_{-0.05}$ &$-$2.08 $^{+0.06}_{-0.06}$ & &  \\[0.15cm] 
$\phantom{0}9.00\leq\mathcal{M}<\phantom{0}9.25$ & $-$2.06 $^{+0.03}_{-0.03}$ &$-$2.07 $^{+0.03}_{-0.03}$ & $-$2.08 $^{+0.03}_{-0.03}$ & $-$2.24 $^{+0.04}_{-0.05}$ & $-$2.24 $^{+0.06}_{-0.07}$ & $-$2.23 $^{+0.06}_{-0.07}$  \\[0.15cm] 
$\phantom{0}9.25\leq\mathcal{M}<\phantom{0}9.50$ & $-$2.16 $^{+0.02}_{-0.03}$ & $-$2.19 $^{+0.03}_{-0.03}$ & $-$2.24 $^{+0.03}_{-0.03}$ &$-$2.36 $^{+0.03}_{-0.04}$ & $-$2.40 $^{+0.04}_{-0.05}$ & $-$2.45 $^{+0.06}_{-0.07}$  \\[0.15cm] 
$\phantom{0}9.50\leq\mathcal{M}<\phantom{0}9.75$ & $-$2.29 $^{+0.02}_{-0.03}$ & $-$2.31 $^{+0.02}_{-0.02}$ & $-$2.40 $^{+0.03}_{-0.03}$ &$-$2.53 $^{+0.04}_{-0.04}$ & $-$2.56 $^{+0.04}_{-0.04}$ & $-$2.61 $^{+0.05}_{-0.05}$  \\[0.15cm] 
$\phantom{0}9.75\leq\mathcal{M}<10.00$ & $-$2.41 $^{+0.03}_{-0.03}$ & $-$2.45 $^{+0.02}_{-0.02}$ & $-$2.55 $^{+0.02}_{-0.03}$ & $-$2.70 $^{+0.04}_{-0.04}$ & $-$2.74 $^{+0.05}_{-0.05}$& $-$2.85 $^{+0.04}_{-0.05}$  \\[0.15cm] 
$10.00\leq\mathcal{M}<10.25$ & $-$2.54 $^{+0.03}_{-0.03}$ & $-$2.60 $^{+0.02}_{-0.02}$ & $-$2.71 $^{+0.02}_{-0.03}$ & $-$2.85 $^{+0.03}_{-0.03}$ & $-$2.89 $^{+0.03}_{-0.03}$ & $-$3.08 $^{+0.05}_{-0.05}$  \\[0.15cm] 
$10.25\leq\mathcal{M}<10.50$ & $-$2.70 $^{+0.03}_{-0.03}$ & $-$2.74 $^{+0.02}_{-0.03}$ & $-$2.86 $^{+0.02}_{-0.03}$ & $-$3.01 $^{+0.03}_{-0.04}$ &$-$3.09 $^{+0.04}_{-0.05}$ & $-$3.38 $^{+0.07}_{-0.08}$  \\[0.15cm] 
$10.50\leq\mathcal{M}<10.75$ & $-$2.88 $^{+0.03}_{-0.03}$ & $-$2.92 $^{+0.02}_{-0.03}$ & $-$3.04 $^{+0.03}_{-0.03}$ & $-$3.17 $^{+0.03}_{-0.03}$ & $-$3.31 $^{+0.04}_{-0.04}$ & $-$3.65 $^{+0.04}_{-0.04}$  \\[0.15cm] 
$10.75\leq\mathcal{M}<11.00$ & $-$3.16 $^{+0.04}_{-0.05}$ & $-$3.21 $^{+0.04}_{-0.04}$ & $-$3.33 $^{+0.04}_{-0.04}$ & $-$3.39 $^{+0.04}_{-0.05}$ & $-$3.55 $^{+0.05}_{-0.06}$ & $-$3.96 $^{+0.06}_{-0.06}$  \\[0.15cm] 
$11.00\leq\mathcal{M}<11.25$ & $-$3.63 $^{+0.05}_{-0.06}$ & $-$3.69 $^{+0.04}_{-0.05}$ & $-$3.81 $^{+0.04}_{-0.05}$ & $-$3.78 $^{+0.05}_{-0.05}$ & $-$3.94 $^{+0.06}_{-0.07}$& $-$4.29 $^{+0.06}_{-0.07}$  \\[0.15cm] 
$11.25\leq\mathcal{M}<11.50$ & $-$4.34 $^{+0.09}_{-0.11}$ & $-$4.34 $^{+0.06}_{-0.07}$ & $-$4.45 $^{+0.06}_{-0.08}$ & $-$4.47 $^{+0.07}_{-0.08}$ & $-$4.57 $^{+0.08}_{-0.09}$ & $-$4.74 $^{+0.07}_{-0.09}$  \\[0.15cm] 
$11.50\leq\mathcal{M}<11.75$ & $-$5.31 $^{+0.21}_{-0.44}$ & $-$5.52 $^{+0.19}_{-0.33}$ & $-$5.34 $^{+0.14}_{-0.21}$ & $-$5.48 $^{+0.16}_{-0.25}$ & $-$5.62 $^{+0.18}_{-0.32}$& $-$5.56 $^{+0.14}_{-0.21}$  \\[0.15cm] 
$11.75\leq\mathcal{M}<12.00$ &  & $-$6.36 $^{+0.38}_{-\inf}$ & & $-$6.53 $^{+0.38}_{-\inf}$ & $-$6.23 $^{+0.30}_{-\inf}$ & $-$6.81 $^{+0.39}_{-\inf}$   \\
\hline
\end{tabular}
\label{tab:blue_number_densities}
\end{table*}

\begin{table*}
\caption{The observed GSMF for UVJ--selected passive galaxies (see text for details). The first column lists the adopted stellar mass bins, where $\mathcal{M}~\equiv~\log_{10}(M_{\star}/\Msun)$, while columns $2-7$ list the logarithm of the number densities ($\phi_{k}$) within six redshift bins. The units of $\phi_{k}$ are dex$^{-1}$~Mpc$^{-3}$. The data presented in this table are plotted in Fig. \ref{fig:UVJ_GSMFs}.}
\begin{tabular}{ | c | c | c | c | c | c | c |}
\hline
               & $0.25\leq z <0.75$ & $0.75\leq z <1.25$ & $1.25\leq z <1.75$ & $1.75\leq z <2.25$ & $2.25\leq z <2.75$ & $2.75\leq z <3.75$ \\ 
 & & & & & & \\
$\mathcal{M}$ & $\log_{10}(\phi_{k})$ & $\log_{10}(\phi_{k})$ & $\log_{10}(\phi_{k})$ &$\log_{10}(\phi_{k})$ & $\log_{10}(\phi_{k})$ &  $\log_{10}(\phi_{k})$\\
\hline

$\phantom{0}8.25\leq\mathcal{M}<\phantom{0}8.50$ & $-$2.40 $^{+0.05}_{-0.06}$ &  &  &  &  &  \\[0.15cm] 
$\phantom{0}8.50\leq\mathcal{M}<\phantom{0}8.75$ & $-$2.47 $^{+0.05}_{-0.05}$ & $-$3.11 $^{+0.08}_{-0.10}$ & &  &  &  \\[0.15cm] 
$\phantom{0}8.75\leq\mathcal{M}<\phantom{0}9.00$ & $-$2.56 $^{+0.04}_{-0.04}$ & $-$3.00 $^{+0.06}_{-0.07}$ & $-$3.61 $^{+0.15}_{-0.22}$ & &  &  \\[0.15cm] 
$\phantom{0}9.00\leq\mathcal{M}<\phantom{0}9.25$ & $-$2.68 $^{+0.04}_{-0.05}$ & $-$3.12 $^{+0.05}_{-0.05}$ & $-$3.74 $^{+0.12}_{-0.16}$ & &  &  \\[0.15cm] 
$\phantom{0}9.25\leq\mathcal{M}<\phantom{0}9.50$ & $-$2.83 $^{+0.05}_{-0.05}$ & $-$3.21 $^{+0.05}_{-0.06}$ & $-$3.68 $^{+0.08}_{-0.09}$ & &  &  \\[0.15cm] 
$\phantom{0}9.50\leq\mathcal{M}<\phantom{0}9.75$ & $-$2.91 $^{+0.05}_{-0.06}$ & $-$3.26 $^{+0.03}_{-0.03}$ & $-$3.68 $^{+0.07}_{-0.09}$ & $-$4.31 $^{+0.13}_{-0.18}$ & &  \\[0.15cm] 
$\phantom{0}9.75\leq\mathcal{M}<10.00$ & $-$2.97 $^{+0.03}_{-0.03}$ & $-$3.20 $^{+0.03}_{-0.03}$ & $-$3.59 $^{+0.07}_{-0.08}$ & $-$4.03 $^{+0.09}_{-0.12}$ & $-$4.42 $^{+0.14}_{-0.22}$ &  \\[0.15cm] 
$10.00\leq\mathcal{M}<10.25$ & $-$2.92 $^{+0.03}_{-0.03}$ & $-$3.09 $^{+0.02}_{-0.03}$ & $-$3.40 $^{+0.03}_{-0.03}$ & $-$3.85 $^{+0.08}_{-0.09}$ & $-$4.24 $^{+0.12}_{-0.17}$ & $-$4.60 $^{+0.15}_{-0.23}$  \\[0.15cm] 
$10.25\leq\mathcal{M}<10.50$ & $-$2.84 $^{+0.03}_{-0.03}$ & $-$2.99 $^{+0.03}_{-0.03}$ & $-$3.25 $^{+0.03}_{-0.03}$ & $-$3.68 $^{+0.04}_{-0.04}$ & $-$3.87 $^{+0.05}_{-0.06}$ & $-$4.46 $^{+0.13}_{-0.18}$ \\[0.15cm] 
$10.50\leq\mathcal{M}<10.75$ & $-$2.83 $^{+0.03}_{-0.03}$ & $-$2.96 $^{+0.02}_{-0.03}$ & $-$3.21 $^{+0.03}_{-0.03}$ & $-$3.60 $^{+0.03}_{-0.04}$ & $-$3.77 $^{+0.05}_{-0.05}$ & $-$4.40 $^{+0.06}_{-0.07}$ \\[0.15cm] 
$10.75\leq\mathcal{M}<11.00$ & $-$2.95 $^{+0.04}_{-0.04}$ & $-$3.08 $^{+0.03}_{-0.04}$ & $-$3.29 $^{+0.04}_{-0.04}$ & $-$3.65 $^{+0.05}_{-0.05}$ & $-$3.83 $^{+0.06}_{-0.06}$ & $-$4.50 $^{+0.06}_{-0.07}$ \\[0.15cm] 
$11.00\leq\mathcal{M}<11.25$ & $-$3.24 $^{+0.04}_{-0.05}$ & $-$3.43 $^{+0.04}_{-0.04}$ & $-$3.66 $^{+0.04}_{-0.05}$ & $-$3.95 $^{+0.05}_{-0.06}$ & $-$4.23 $^{+0.06}_{-0.07}$ & $-$5.03 $^{+0.09}_{-0.11}$ \\[0.15cm] 
$11.25\leq\mathcal{M}<11.50$ & $-$3.74 $^{+0.05}_{-0.06}$ & $-$4.02 $^{+0.05}_{-0.06}$ & $-$4.23 $^{+0.05}_{-0.06}$ & $-$4.61 $^{+0.07}_{-0.09}$ & $-$4.77 $^{+0.09}_{-0.11}$ & $-$5.53 $^{+0.13}_{-0.18}$ \\[0.15cm] 
$11.50\leq\mathcal{M}<11.75$ & $-$4.48 $^{+0.10}_{-0.13}$ & $-$5.04 $^{+0.12}_{-0.17}$ & $-$5.41 $^{+0.15}_{-0.23}$ & $-$5.63 $^{+0.18}_{-0.31}$ & $-$5.58 $^{+0.17}_{-0.30}$ & $-$6.22 $^{+0.23}_{-0.54}$ \\[0.15cm] 
$11.75\leq\mathcal{M}<12.00$ & $-$5.71 $^{+0.30}_{-2.52}$ &  & & $-$6.53 $^{+0.38}_{-\inf}$ &  &  \\ 
\hline
\end{tabular}
\label{tab:red_number_densities}
\end{table*}

\begin{figure*}
\begin{tabular}{cc}
\includegraphics[width=0.8\columnwidth]{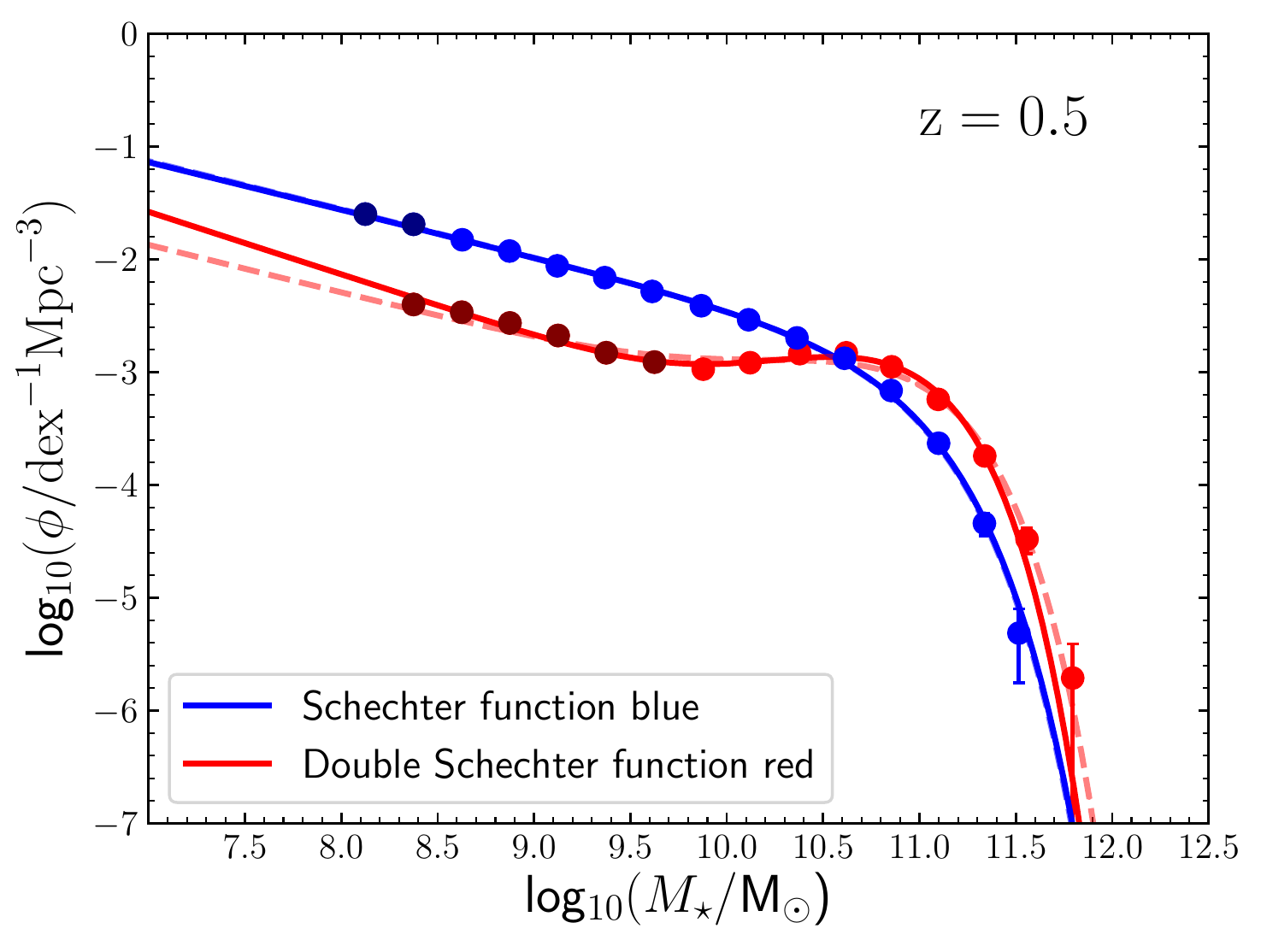} &
\includegraphics[width=0.8\columnwidth]{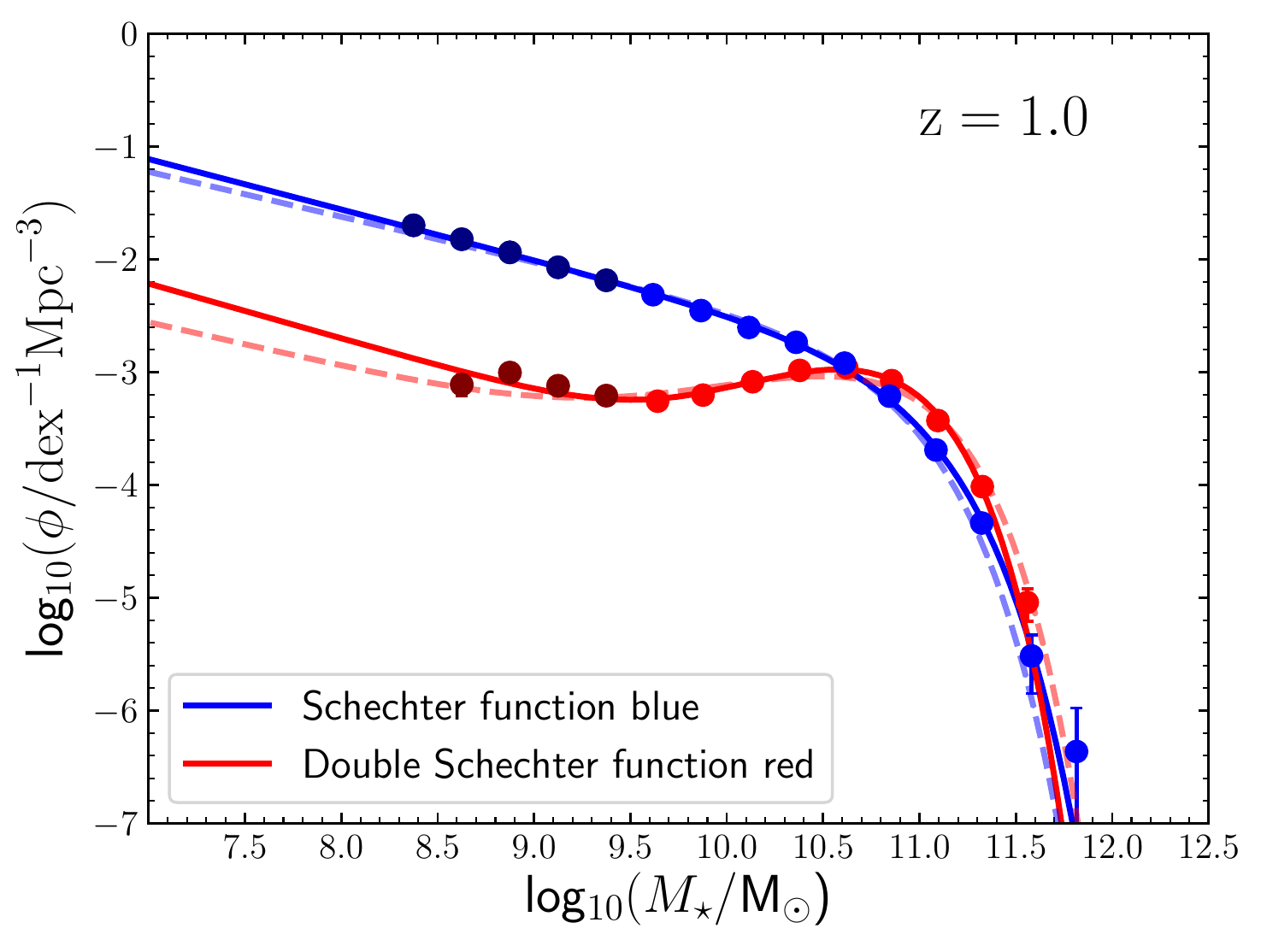} \\
\includegraphics[width=0.8\columnwidth]{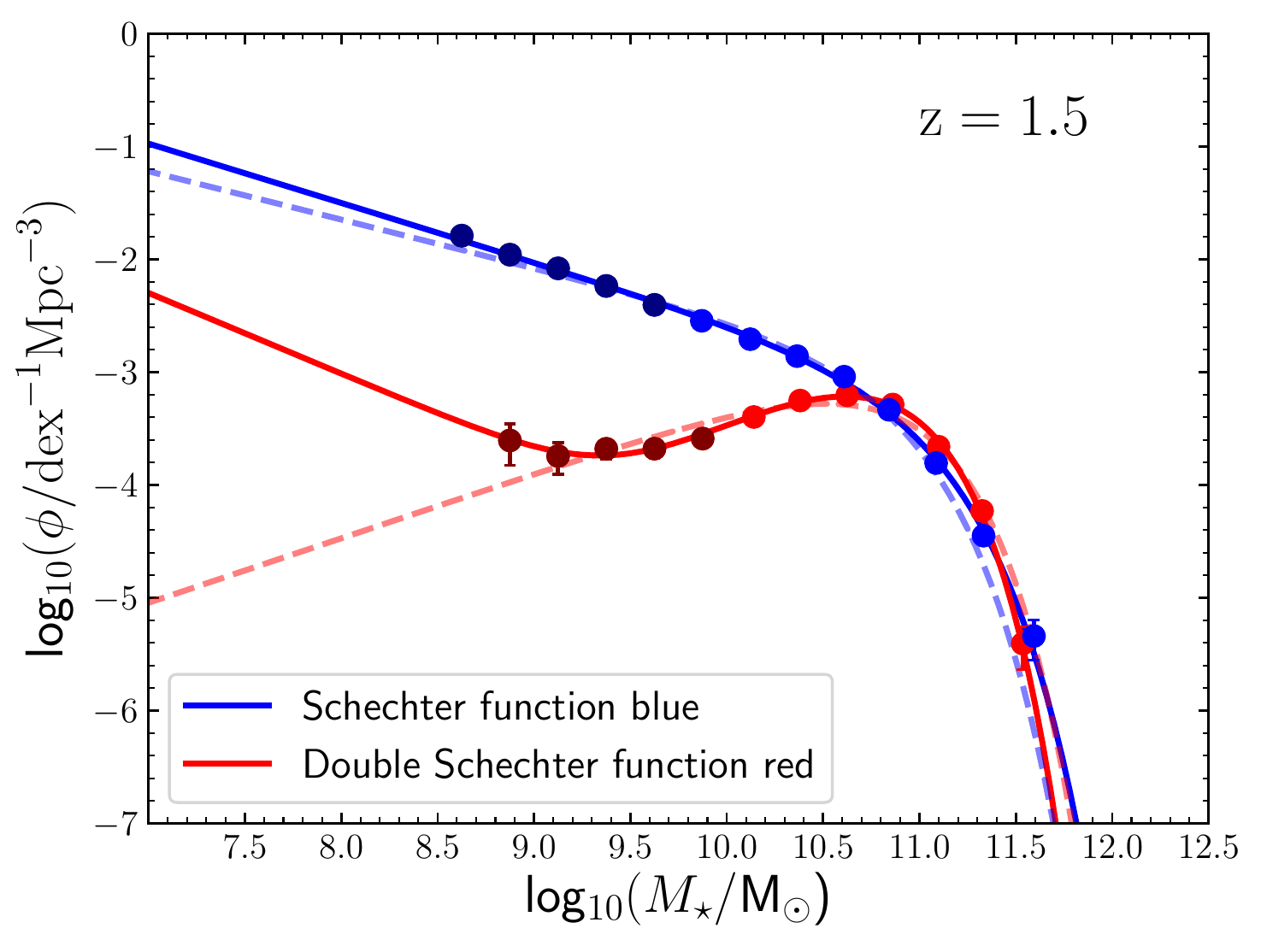} &
\includegraphics[width=0.8\columnwidth]{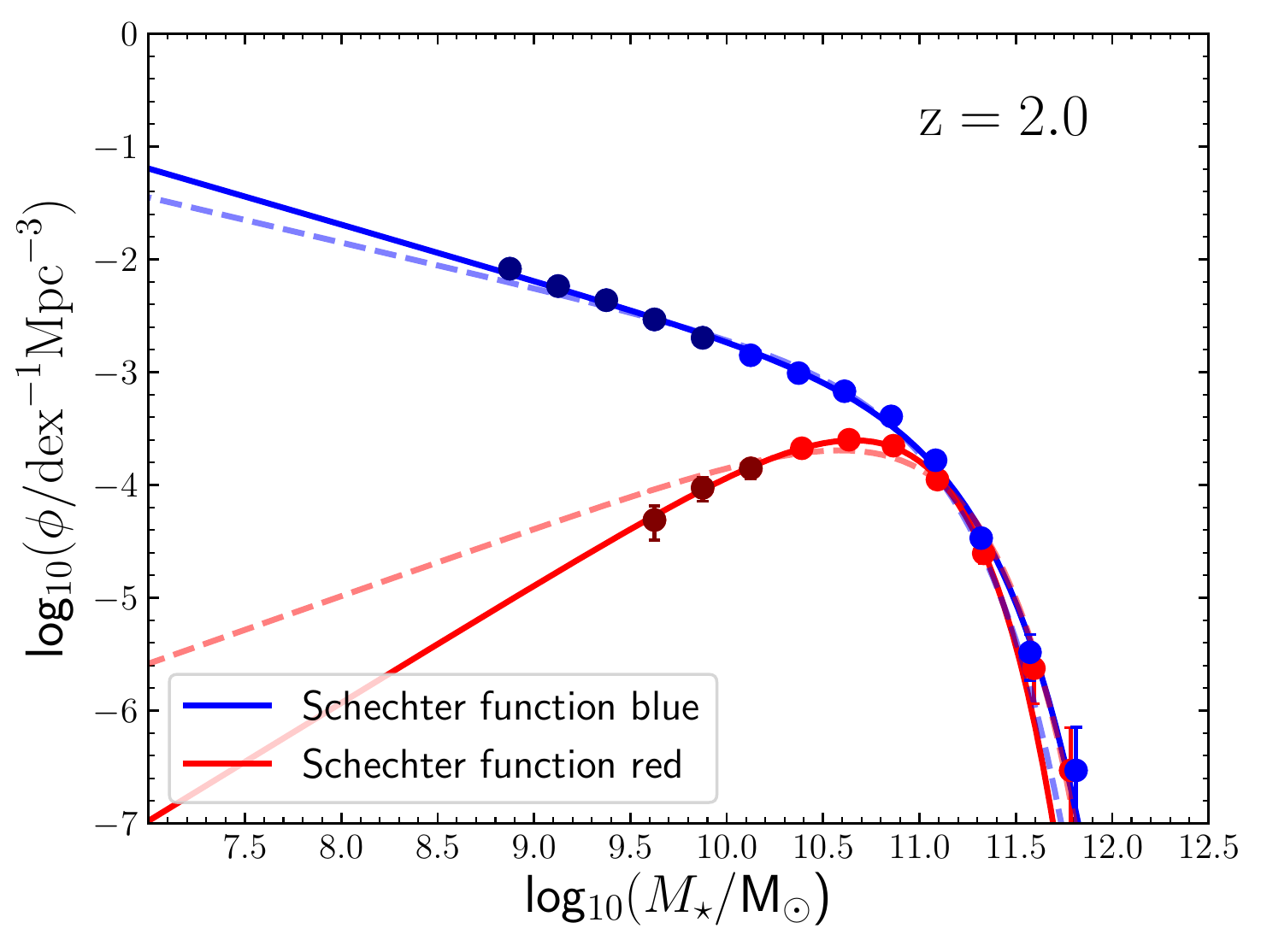} \\
\includegraphics[width=0.8\columnwidth]{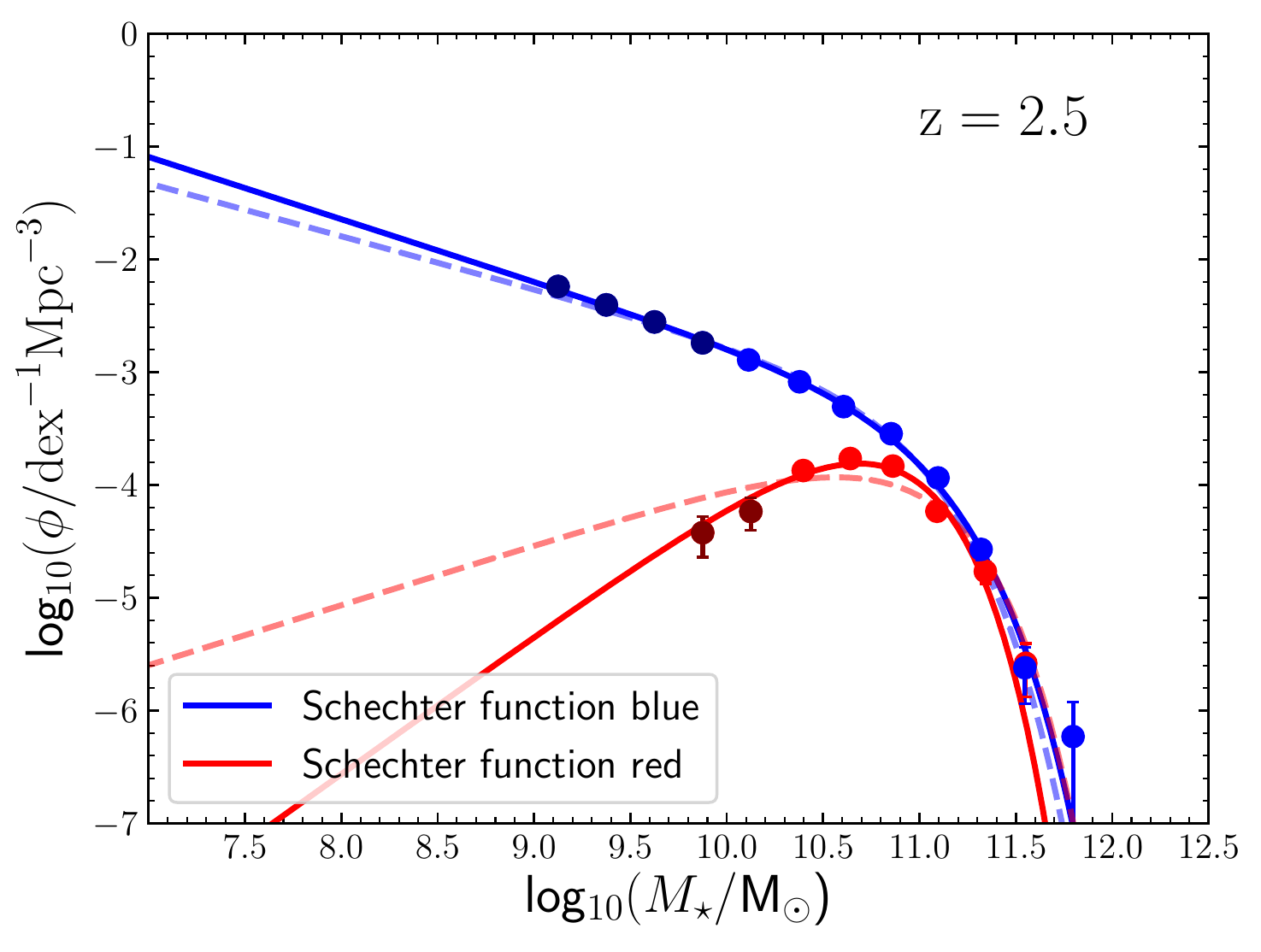} &
\includegraphics[width=0.8\columnwidth]{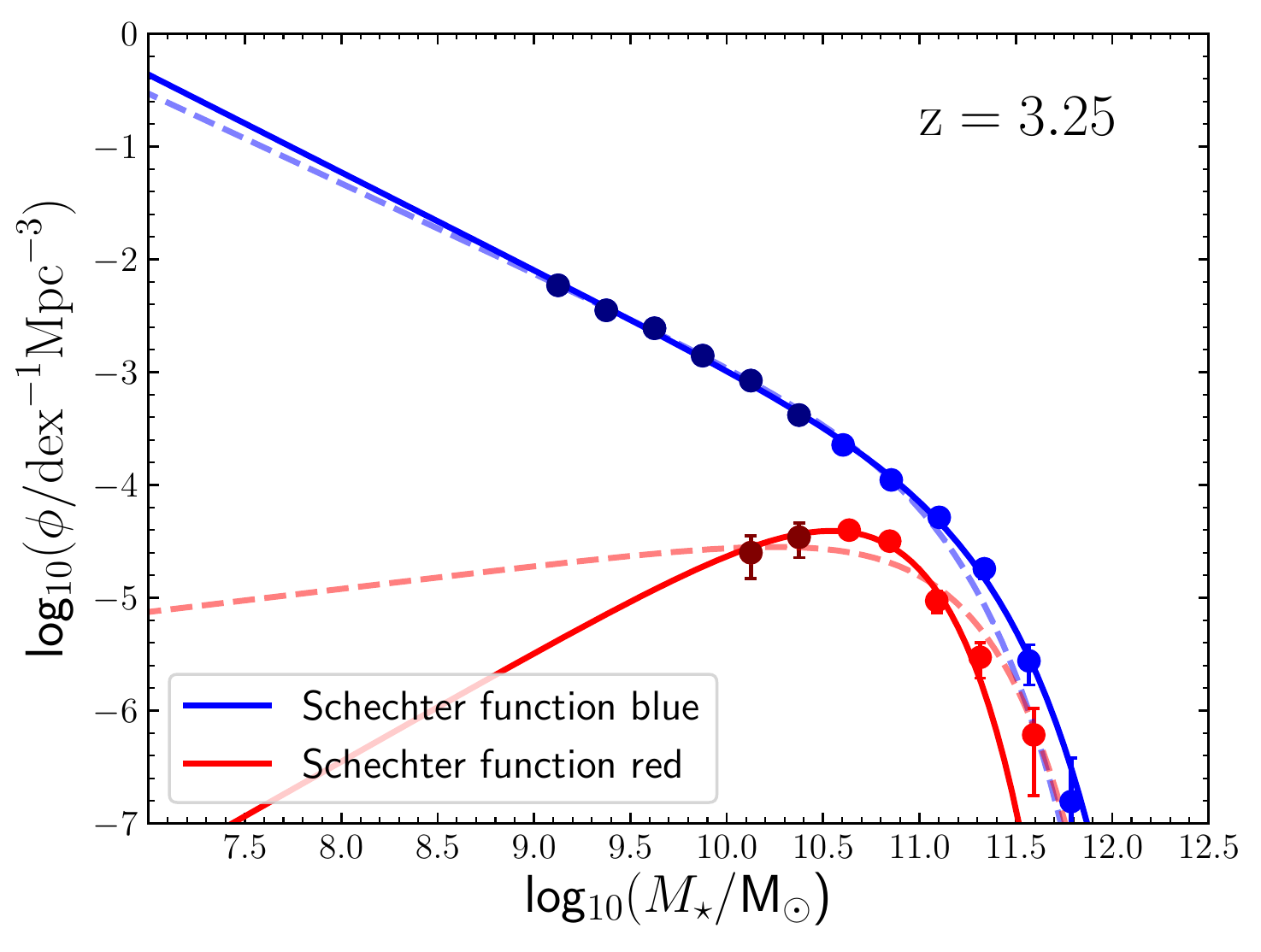}
\end{tabular}
\caption{The redshift evolution of the observed GSMF for star-forming (blue data points) and passive galaxies (red data points). A darker shade of blue/red is used for the datapoints that are determined using CANDELS}. The solid blue curves show the best-fitting single Schechter functions to the
star-forming GSMF. The red curves show the best-fitting double Schechter functions to the passive GSMF in the first three redshift bins and the best-fitting single Schechter functions to the passive GSMF in the final three redshift bins. In each redshift bin, the dashed blue and red curves show the best fits to the star-forming and passive GSMFs using the 5-parameter \protect\cite{Peng2010} model (see text for discussion).
\label{fig:UVJ_GSMFs}
\end{figure*}

\begin{figure*}
\begin{tabular}{cc}
\includegraphics[width=\columnwidth]{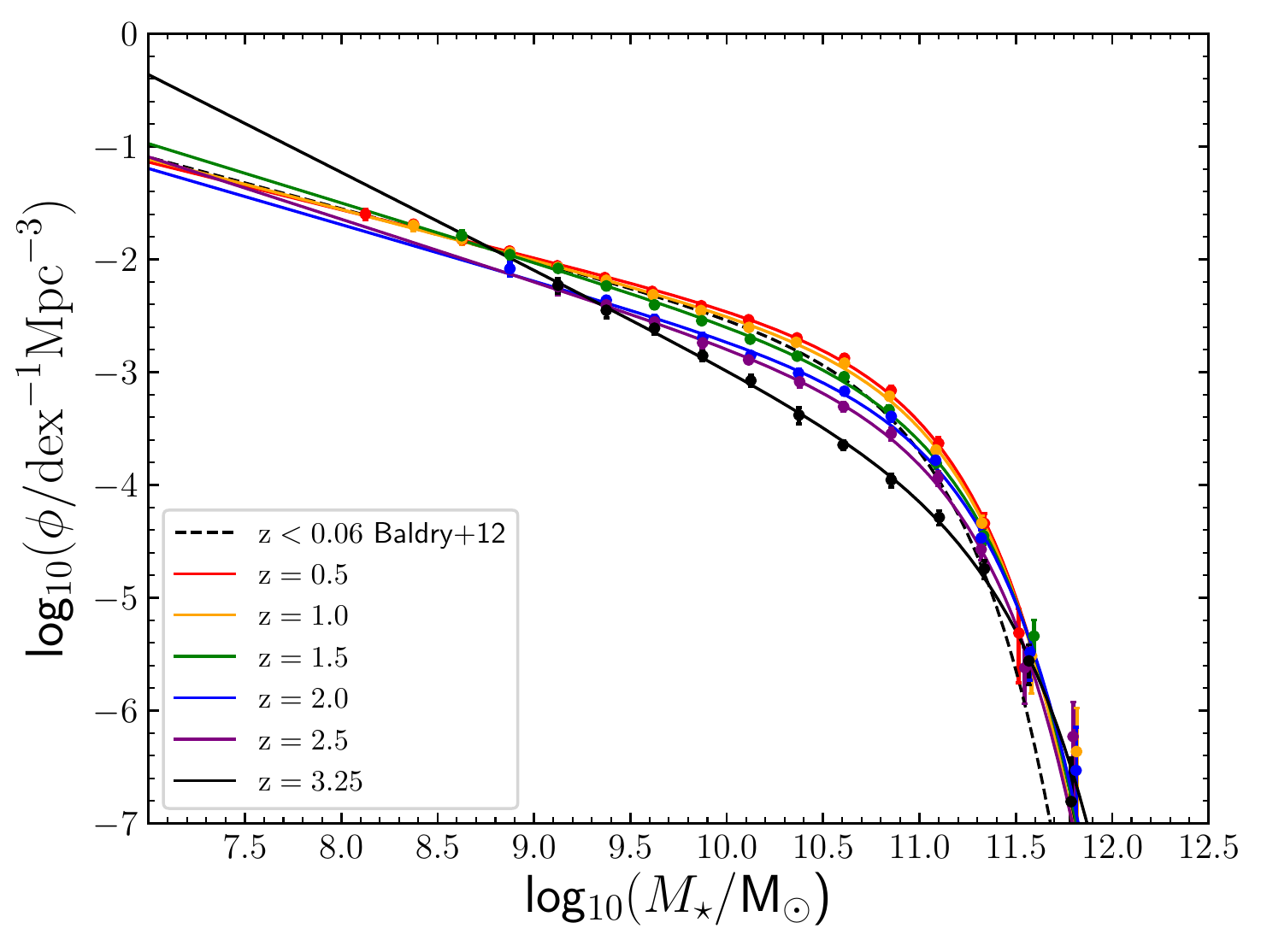} &
\includegraphics[width=\columnwidth]{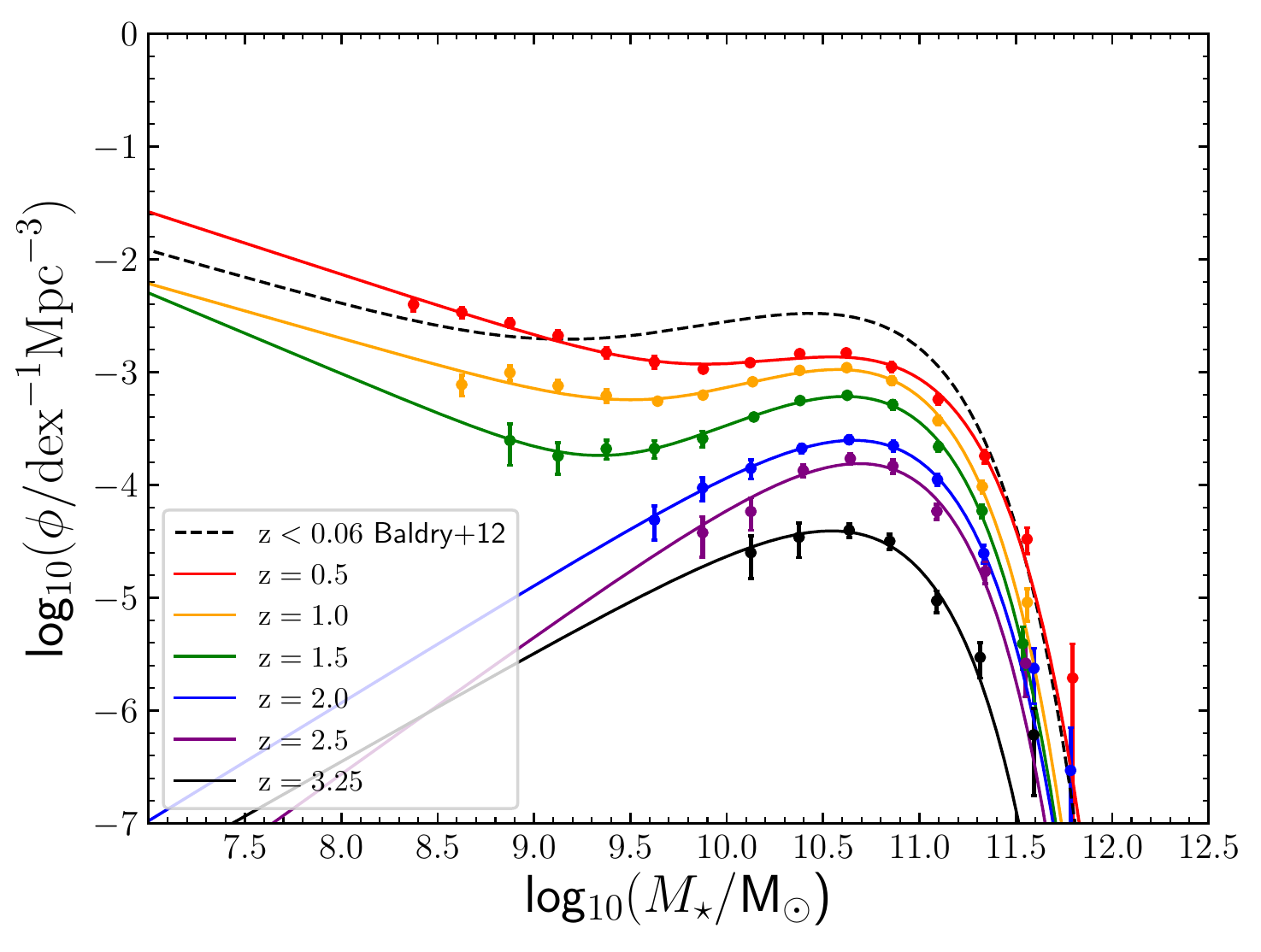}
\end{tabular}
\caption{The left-hand panel shows the observed GSMF data for star-forming galaxies, together with the best-fitting single Schechter functions. For comparison, we have also included our
single Schechter function fit to the star-forming galaxy data at $z<0.06$ from \protect\cite{Baldry2012}. The right-hand panel shows the equivalent information for the observed GSMF for passive galaxies.
In the $z=0.5, 1.0$ \& 1.5 redshift bins we plot the best-fitting double Schechter functions, whereas in the $z=2.0, 2.5$ \& 3.25 redshift bins we plot the best-fitting single Schechter functions. For comparison, we include our own double Schechter function fit to the \protect\cite{Baldry2012} passive GSMF data at $z<0.06$.}
\label{fig:observed_UVJ_GSMFs_allz}
\end{figure*}

\subsection{Schechter function fits}
We performed maximum likelihood fitting to the star-forming and passive galaxy number densities, using both single and double Schechter functional forms, as before. The best-fitting
Schechter function parameters are presented in Tables \ref{tab:blue_Schechters} \& \ref{tab:red_Schechters} for the star-forming and passive galaxies, respectively. In Fig. \ref{fig:UVJ_GSMFs} we plot the separate star-forming and passive GSMFs, along with the best-fitting single and double Schechter function fits.

\subsection{The star-forming GSMF}
We find that a single Schechter function provides a statistically acceptable description of both the observed and intrinsic star-forming galaxy GSMF, over the full $0.25 \leq z < 3.75$
redshift range covered by our data set. This is in contrast to some other studies (e.g. \citealt{Drory2009, Ilbert2013, Tomczak2014, Davidzon2017}), who concluded that
a double Schechter function is a superior fit to the star-forming GSMF, particularly at $z \leq 2$. Based on our data, we find that although double Schechter function fits do return lower
values of reduced $\chi^{2}$, the improvement in the quality of the fit is not generally sufficient to justify the inclusion of two additional degrees of freedom.

It can be seen from Table \ref{tab:blue_Schechters} and Fig. \ref{fig:UVJ_GSMFs} that the star-forming GSMF is remarkably stable.
Over the redshift range $0.0 \leq z < 1.25$ the intrinsic Schechter function parameters are effectively constant (within the errors). The same statement can be made about the observed Schechter function
parameters over this redshift range, with the $\simeq 2\sigma$ shift in $\mathcal{M}^{\star}$ between $z\simeq 0$ and $z=0.5$ being largely attributable to increased Eddington bias. The star-forming GSMF only evolves gradually over the redshift interval $1.25 \leq z < 2.75$, with both the observed and intrinsic Schechter function fits displaying a
$\simeq 0.2$ dex drop in $\phi^{\star}$, while the values of $\mathcal{M}^{\star}$ and $\alpha$ remain essentially unchanged.
It is only in the highest redshift bin at $2.75 \leq z < 3.75$ that we see a significant change in the shape of the star-forming GSMF, with both the observed and intrinsic Schechter functions
showing a $\simeq 0.2$ dex increase in $\mathcal{M}^{\star}$, a further $\simeq 0.7$ dex drop in $\phi^{\star}$ and a significant $\simeq 0.3$ steepening in the faint-end slope. The remarkably
gradual evolution of the star-forming GSMF is illustrated by the left-hand panel of Fig. \ref{fig:observed_UVJ_GSMFs_allz}, which shows an overlay of the data and best-fitting
Schechter function for all six redshift bins. 

\begin{table*}
\caption{The best-fitting observed (upper section) and intrinsic (lower section) Schechter function parameters for the star-forming GSMF, where $\mathcal{M}^{\star}~\equiv~\log_{10}(M^{\star}/\Msun)$ and the units of $\phi^{\star}$
are dex$^{-1}$ Mpc$^{-3}$. The final two columns list the $\chi^2$ and $\chi^2_{\nu}$ values of the fits, respectively. We have included the parameters from our own fits to the \protect\cite{Baldry2012}
data at $z<0.06$ for comparison. To derive the intrinsic Schechter function parameters for the \protect\cite{Baldry2012} data we assumed that $\sigma_{\mathcal{M}}=0.1$ dex \protect\citep{Wright2018}.}
\begin{tabular}{ | c | c | c | c | r | c |}
\hline
Redshift& $\mathcal{M}^{*}$ & $\log (\phi^{*})$  &
$\alpha$ & $\chi^{2}$ & $\chi^{2}_{\nu}$ \\ 
\hline
$z < 0.06$      & 10.74 $\pm$ 0.05 & $-$3.16 $^{+0.05}_{-0.05}$ & $-$1.46 $\pm$ 0.02 & 23.79 & 1.70 \\[0.15cm]
$0.25 \leq z < 0.75$ & 10.85 $\pm$ 0.02 & $-$3.12 $^{+0.03}_{-0.03}$ & $-$1.42 $\pm$ 0.02 & \phantom{0}6.34 & 0.53 \\[0.15cm]
$0.75 \leq z < 1.25$ & 10.86 $\pm$ 0.02 & $-$3.20 $^{+0.03}_{-0.03}$ & $-$1.45 $\pm$ 0.02 &  \phantom{0}8.25 & 0.69 \\[0.15cm]
$1.25 \leq z < 1.75$ & 10.91 $\pm$ 0.03 & $-$3.39 $^{+0.04}_{-0.04}$ & $-$1.53 $\pm$ 0.02 &  \phantom{0}8.61 & 0.86 \\[0.15cm]
$1.75 \leq z < 2.25$ & 10.93 $\pm$ 0.03 & $-$3.51 $^{+0.05}_{-0.05}$ & $-$1.50 $\pm$ 0.03 & 9.91 & 0.99 \\[0.15cm]
$2.25 \leq z < 2.75$ & 10.92 $\pm$ 0.04 & $-$3.62 $^{+0.06}_{-0.07}$ & $-$1.55 $\pm$ 0.04 &  \phantom{0}3.97 & 0.44 \\[0.15cm]
$2.75 \leq z < 3.75$ & 11.12 $\pm$ 0.05 & $-$4.29 $^{+0.08}_{-0.10}$ & $-$1.87 $\pm$ 0.04 &  \phantom{0}4.26 & 0.47 \\[0.15cm]
\hline
$z<0.06$      & 10.72 $\pm$ 0.05 & $-$3.15 $^{+0.05}_{-0.05}$ & $-$1.45 $\pm$ 0.02 & 26.03 & 1.86 \\[0.15cm]
$0.25 \leq z < 0.75$ & 10.77 $\pm$ 0.03 & $-$3.07 $^{+0.03}_{-0.04}$ & $-$1.41 $\pm$ 0.02 & 14.43 & 1.20 \\[0.15cm]
$0.75 \leq z < 1.25$ & 10.77 $\pm$ 0.02 & $-$3.13 $^{+0.03}_{-0.03}$ & $-$1.43 $\pm$ 0.02 & 15.20 & 1.27 \\[0.15cm]
$1.25 \leq z < 1.75$ & 10.83 $\pm$ 0.03 & $-$3.33 $^{+0.04}_{-0.05}$ & $-$1.51 $\pm$ 0.02 & 11.59 & 1.16 \\[0.15cm]
$1.75 \leq z < 2.25$ & 10.84 $\pm$ 0.03 & $-$3.43 $^{+0.05}_{-0.05}$ & $-$1.47 $\pm$ 0.03 & 15.84 & 1.58 \\[0.15cm]
$2.25 \leq z < 2.75$ & 10.82 $\pm$ 0.04 & $-$3.52 $^{+0.06}_{-0.07}$ & $-$1.52 $\pm$ 0.05 &  \phantom{0}7.22 & 0.80 \\[0.15cm]
$2.75 \leq z < 3.75$ & 11.02 $\pm$ 0.05 & $-$4.19 $^{+0.09}_{-0.11}$ & $-$1.85 $\pm$ 0.04 &  \phantom{0}6.28 & 0.70 \\[0.15cm]
\hline
\end{tabular}
\label{tab:blue_Schechters}
\end{table*}

\subsection{The passive GSMF}
In contrast to the star-forming GSMF, it can be seen from Table \ref{tab:red_Schechters} and Fig. \ref{fig:UVJ_GSMFs} that
the passive GSMF evolves dramatically over the redshift range studied here. As discussed in the introduction, it has long been established that a double Schechter function is required to match
the shape of the passive galaxy GSMF in the local Universe, due to a distinct upturn in the number densities of passive galaxies at low
stellar masses, that is usually interpreted as a clear signature of environmental quenching (e.g. \citealt{Peng2010}).

Our results indicate that the double Schechter functional form of the passive GSMF persists until at least $z\simeq 1.0$, and very likely until $z\simeq 1.5$. If confirmed, the upturn in the
number densities of low-mass passive galaxies seen in the $z=1.5$ redshift bin would argue that the impact of some form of environmental quenching, presumably galaxy-galaxy
mergers rather than satellite quenching, is becoming apparent at a look-back time of $\geq 9$ Gyr. This result is in agreement with the previous GSMF study of \cite{Tomczak2014}, who also concluded
that the passive GSMF required a double Schechter at $z\leq 1.5$. Moreover, although \cite{Mortlock2015} only fitted a single Schechter function to the passive GSMF at $z\geq 1$, there is an indication
of an upturn at low stellar masses in their $1.0<z<1.5$ redshift bin. In contrast, the recent study by \cite{Davidzon2017} only detects evidence of an upturn in the passive GSMF at $z\leq 0.8$, although
this is almost certainly explained by a lack of dynamic range in stellar mass.

At redshifts $z\geq 1.75$, our determination of the passive GSMF is well described by a single Schechter function. Due to the increase of our stellar-mass completeness limit with redshift,
based on the current data set, it is not possible to determine whether this change in shape is intrinsic, or simply due to insufficient dynamic range in stellar mass. Accurately determining
the shape of the passive GSMF is clearly a task which can be addressed with the unique near-IR sensitivity offered by the {\it James Webb Space Telescope} ({\it JWST}).

The dramatic evolution of the passive GSMF is illustrated by the right-hand panel of Fig. \ref{fig:observed_UVJ_GSMFs_allz}, which shows an overlay of the data and the best-fitting Schechter
function fits over the full redshift range. In addition to the change in shape, this figure also illustrates the dramatic ($\simeq 2$ dex) decrease in the number density of
$\mathcal{M}\simeq\mathcal{M}^{\star}$ passive galaxies from $z\simeq 0$ to $z\simeq 3$. The evolving contribution of passive galaxies to the integrated stellar-mass density is explored in Section 6.

\subsection{A comparison with the Peng et al. model}
In the introduction we discussed how the empirical model proposed by \cite{Peng2010} can provide useful insights into how different
quenching mechanisms control the shape of the GSMF. Given that the Peng et al. model can accurately reproduce the shape of local star-forming and passive GSMFs (e.g. \citealt{Baldry2012}), it is interesting
to explore how well the model continues to perform at higher redshifts.

To investigate this question,  we performed maximum likelihood fits to the star-forming and passive GSMF data in Tables \ref{tab:blue_number_densities} \& \ref{tab:red_number_densities} using the 5-parameter
Peng et al. model ($\mathcal{M}^{\star},\phi^{\star},\alpha,\phi_{1}^{\star},\phi_{2}^{\star}$). The first three parameters of the model describe the star-forming GSMF with a single Schechter function. The final two parameters of the model are the twin normalizations of the double Schechter function describing the passive GSMF, which is constrained to have the same $\mathcal{M}^{\star}$ as the star-forming GSMF, $\alpha_{2}=\alpha$ and $\alpha_{1}=\alpha+1.0$. The results of these constrained fits are plotted as the dashed blue and red curves in Fig. \ref{fig:UVJ_GSMFs}.

In the $z=0.5$ redshift bin the \cite{Peng2010} model continues to produce an excellent qualitative, and statistically acceptable, match to the star-forming and passive GSMFs. In the next two redshift bins, the Peng et al. model continues to produce a good qualitative match to the observed data, although the fits become progressively poorer in a statistical sense. In the final three redshift bins, the Peng et al. model struggles to match the shape of the passive GSMF, although it arguably still produces a respectable qualitative match to the observed data.

It is worth remembering that the comparison between the model and the observed data in the higher redshift bins is complicated by the increasing difficulty in cleanly separating the star-forming and passive galaxy populations. Moreover, while in this study we have adopted UVJ criteria
to separate the star-forming and passive populations, the Peng et al. model assumes the populations are separated based on
an evolving rest-frame $U-B$ colour cut. Once again, it is clear that the unique near-IR sensitivity provided by {\it JWST} will be
crucial for confirming or refuting the steep fall-off in the number density of
$\mathcal{M} \leq \mathcal{M}^{\star}$ passive galaxies currently indicated by our $z>2$ data.

\begin{table*}
\caption{The best-fitting observed (upper section) and intrinsic (lower section) double Schechter function parameters for the passive GSMF, where $\mathcal{M}^{\star}~\equiv~\log_{10}(M^{\star}/\Msun)$ and the units
of $\phi^{\star}$, $\phi^{\star}_{1}$, $\phi^{\star}_{2}$ are dex$^{-1}$ Mpc$^{-3}$. The final two columns list the $\chi^2$ and $\chi^2_{\nu}$ values of the fits, respectively. The intrinsic double Schechter
function parameters were derived assuming $\sigma_{\mathcal{M}}=0.15$ dex at all redshifts. We have included the parameters from our own fits to the \protect\cite{Baldry2012}
data at $z<0.06$ for comparison. To derive the intrinsic Schechter function parameters for the \protect\cite{Baldry2012} data we assumed that $\sigma_{\mathcal{M}}=0.1$ dex \protect\citep{Wright2018}.}
\label{tab:red_Schechters}
\begin{tabular}{ | c | c | c | c | c | c | r | c |}
\hline
Redshift & $\mathcal{M}^{*}$ & $\log (\phi_{1}^{*})$ & $\alpha_{1}$ & $\log (\phi_{2}^{*})$ & $\alpha_{2}$ & $\chi^{2}$ & $\chi^{2}_{\nu}$ \\ 
\hline
$z < 0.06$      & 10.70 $\pm$ 0.05 & $-$2.47 $^{+0.03}_{-0.03}$ & $-$0.41 $\pm$ 0.16 & $-$4.09 $^{+0.45}_{-\inf}$& $-$1.49 $\pm$ 0.36 & \phantom{0}7.58 & 0.69 \\[0.15cm]
$0.25 \leq z < 0.75$ & 10.74 $\pm$ 0.04 & $-$2.85 $^{+0.03}_{-0.03}$ & $-$0.21 $\pm$ 0.15 & $-$4.01 $^{+0.16}_{-0.24}$  & $-$1.55 $\pm$ 0.08 & 13.94 & 1.39 \\[0.15cm]
$0.75 \leq z < 1.25$ & 10.64 $\pm$ 0.03 & $-$2.92 $^{+0.02}_{-0.02}$ & $-$0.07 $\pm$ 0.13 & $-$4.35 $^{+0.22}_{-0.48}$  & $-$1.49 $\pm$ 0.16 & \phantom{0}12.20 & 1.52 \\[0.15cm]
$1.25 \leq z < 1.75$ & 10.62 $\pm$ 0.03 &$-$3.15 $^{+0.02}_{-0.02}$  & \phantom{$-$}0.00 $\pm$ 0.13 & $-$5.25 $^{+0.49}_{-\inf}$ & $-$1.72 $\pm$ 0.57 & \phantom{0}3.53 & 0.50 \\[0.15cm]
$1.75 \leq z < 2.25$ & 10.64 $\pm$ 0.03 & $-$3.53 $^{+0.02}_{-0.02}$ & \phantom{$-$}0.05 $\pm$ 0.11 & & & \phantom{0}2.50 & 0.36\\[0.15cm]
$2.25 \leq z < 2.75$ & 10.60 $\pm$ 0.05 & $-$3.75 $^{+0.03}_{-0.03}$ &  \phantom{$-$}0.22 $\pm$ 0.18 & & & \phantom{0}8.52 & 1.70\\[0.15cm]
$2.75 \leq z < 3.75$ & 10.56 $\pm$ 0.09 & $-$4.34 $^{+0.05}_{-0.05}$ & $-$0.03 $\pm$ 0.38 & & & \phantom{0}4.57 & 1.14\\[0.15cm]
\hline
$z < 0.06$      & 10.65 $\pm$ 0.06 & $-$2.44 $^{+0.03}_{-0.03}$ & $-$0.29 $\pm$ 0.20 & $-$3.80 $^{+0.39}_{-\inf}$ & $-$1.37 $\pm$ 0.30 & 7.68 & 0.70\\[0.15cm]
$0.25 \leq z < 0.75$ & 10.59 $\pm$ 0.04 & $-$2.79 $^{+0.03}_{-0.03}$ &    \phantom{$-$}0.19 $\pm$ 0.19 & $-$3.80 $^{+0.12}_{-0.17}$ & $-$1.49 $\pm$ 0.07 & 5.72 & 0.57 \\[0.15cm]
$0.75 \leq z < 1.25$ & 10.48 $\pm$ 0.04 & $-$2.91 $^{+0.03}_{-0.03}$ &    \phantom{$-$}0.41 $\pm$ 0.19 & $-$3.97 $^{+0.15}_{-0.23}$ & $-$1.32 $\pm$ 0.11 & 4.73 & 0.59 \\[0.15cm]
$1.25 \leq z < 1.75$ & 10.43 $\pm$ 0.04 & $-$3.19 $^{+0.07}_{-0.08}$ &    \phantom{$-$}0.69 $\pm$ 0.28 & $-$4.23 $^{+0.30}_{-1.58}$ & $-$1.13 $\pm$ 0.34 & 2.31 & 0.33 \\[0.15cm]
$1.75 \leq z < 2.25$ & 10.52 $\pm$ 0.04 & $-$3.51 $^{+0.02}_{-0.02}$ &    \phantom{$-$}0.32 $\pm$ 0.15 & & & 1.38 & 0.20\\[0.15cm]
$2.25 \leq z < 2.75$ & 10.45 $\pm$ 0.06 & $-$3.77 $^{+0.05}_{-0.05}$ &    \phantom{$-$}0.71 $\pm$ 0.28 & & & 3.37 & 0.67\\[0.15cm]
$2.75 \leq z < 3.75$ & 10.40 $\pm$ 0.11 & $-$4.33 $^{+0.09}_{-0.11}$ &    \phantom{$-$}0.49 $\pm$ 0.59 & & & 3.00 & 0.75\\[0.15cm]
\hline
\end{tabular}
\end{table*}

\section{The integrated stellar-mass density}

\begin{table*}
\caption{Integrated stellar-mass densities for the total, star-forming and passive GSMFs over the redshift interval $0.25 \leq z <3.75$. In all cases the GSMFs have been integrated between the limits of $\mathcal{M}=8$ and $\mathcal{M}=13$. For comparison, we have also included our calculation of the integrated stellar-mass densities for the data at $z<0.06$ from \protect\cite{Baldry2012}. The units of $\rho_{\star}$ are $\Msun$ Mpc$^{-3}$.}
\begin{tabular}{ccccccc}
\hline
  & total & total & star-forming  & star-forming & quiescent & quiescent\\
Redshift range & $\log(\rho_{\star}^{\rm{obs}})$ & $\log(\rho_{\star}^{\rm{int}})$ & $\log(\rho_{\star}^{\rm{obs}})$ & $\log(\rho_{\star}^{\rm int})$ & $\log(\rho_{\star}^{\rm obs})$ & $\log(\rho_{\star}^{\rm int})$ \\
\hline
$z < 0.06$        & 8.34 $\pm$ 0.02 & 8.34 $\pm$ 0.02 & 7.78 $\pm$ 0.02 & 7.76 $\pm$ 0.02 & 8.20 $\pm$ 0.03 & 8.19$\pm$ 0.03 \\[0.15cm]
$0.25 \leq z < 0.75$ & 8.20 $\pm$ 0.02 & 8.17 $\pm$ 0.02 & 7.90 $\pm$ 0.01 & 7.87 $\pm$ 0.01 & 7.91 $\pm$ 0.03 & 7.89$\pm$ 0.03 \\[0.15cm]
$0.75 \leq z < 1.25$ & 8.09 $\pm$ 0.02 & 8.07 $\pm$ 0.02 & 7.85 $\pm$ 0.01 & 7.82 $\pm$ 0.01 & 7.73 $\pm$ 0.02 & 7.71$\pm$ 0.02 \\[0.15cm]
$1.25 \leq z < 1.75$ & 7.94 $\pm$ 0.01 & 7.91 $\pm$ 0.01 & 7.77 $\pm$ 0.01 & 7.74 $\pm$ 0.01 & 7.48 $\pm$ 0.02 & 7.45$\pm$ 0.02 \\[0.15cm]
$1.75 \leq z < 2.25$ & 7.76 $\pm$ 0.02 & 7.74 $\pm$ 0.02 & 7.65 $\pm$ 0.01 & 7.62 $\pm$ 0.01 & 7.11 $\pm$ 0.03 & 7.08$\pm$ 0.03 \\[0.15cm]
$2.25 \leq z < 2.75$ & 7.68 $\pm$ 0.02 & 7.65 $\pm$ 0.02 & 7.57 $\pm$ 0.02 & 7.54 $\pm$ 0.02 & 6.90 $\pm$ 0.04 & 6.88$\pm$ 0.04 \\[0.15cm]
$2.75 \leq z < 3.75$ & 7.46 $\pm$ 0.03 & 7.43 $\pm$ 0.03 & 7.45 $\pm$ 0.02 & 7.41 $\pm$ 0.02 & 6.21 $\pm$ 0.06 & 6.19$\pm$ 0.06 \\
\hline\end{tabular}
\label{tab:SMD}
 \end{table*}

\begin{table*}
\caption{Integrated stellar-mass densities for the total, star-forming and passive GSMFs over the redshift interval $0.25 \leq z <3.75$, but this time including stellar remnants. In all cases the GSMFs have been integrated between the limits of $\mathcal{M}=8$ and $\mathcal{M}=13$. For comparison, we have also included our calculation of the integrated stellar-mass densities for the data at $z<0.06$ from \protect\cite{Baldry2012}. The units of $\rho_{\star}$ are $\Msun$ Mpc$^{-3}$.}
\begin{tabular}{ccccccc}
\hline
  & total & total & star-forming  & star-forming & quiescent & quiescent\\
Redshift range & $\log(\rho_{\star}^{\rm{obs}})$ & $\log(\rho_{\star}^{\rm{int}})$ & $\log(\rho_{\star}^{\rm{obs}})$ & $\log(\rho_{\star}^{\rm int})$ & $\log(\rho_{\star}^{\rm obs})$ & $\log(\rho_{\star}^{\rm int})$ \\
\hline
$z < 0.06$        & 8.47 $\pm$ 0.02 & 8.46 $\pm$ 0.02 & 7.89 $\pm$ 0.02 & 7.87 $\pm$ 0.02 & 8.34 $\pm$ 0.03 & 8.33$\pm$ 0.03 \\[0.15cm]
$0.25 \leq z < 0.75$ & 8.30 $\pm$ 0.02 & 8.28 $\pm$ 0.02 & 7.99 $\pm$ 0.01 & 7.96 $\pm$ 0.01 & 8.03 $\pm$ 0.03 & 8.00$\pm$ 0.03 \\[0.15cm]
$0.75 \leq z < 1.25$ & 8.17 $\pm$ 0.02 & 8.15 $\pm$ 0.02 & 7.92 $\pm$ 0.01 & 7.89 $\pm$ 0.01 & 7.82 $\pm$ 0.02 & 7.80$\pm$ 0.02 \\[0.15cm]
$1.25 \leq z < 1.75$ & 8.00 $\pm$ 0.01 & 7.98 $\pm$ 0.01 & 7.82 $\pm$ 0.01 & 7.79 $\pm$ 0.01 & 7.55 $\pm$ 0.02 & 7.52$\pm$ 0.02 \\[0.15cm]
$1.75 \leq z < 2.25$ & 7.81 $\pm$ 0.02 & 7.79 $\pm$ 0.02 & 7.70 $\pm$ 0.01 & 7.67 $\pm$ 0.01 & 7.18 $\pm$ 0.03 & 7.15$\pm$ 0.03 \\[0.15cm]
$2.25 \leq z < 2.75$ & 7.73 $\pm$ 0.02 & 7.70 $\pm$ 0.02 & 7.62 $\pm$ 0.02 & 7.59 $\pm$ 0.02 & 6.97 $\pm$ 0.04 & 6.95$\pm$ 0.04 \\[0.15cm]
$2.75 \leq z < 3.75$ & 7.50 $\pm$ 0.03 & 7.47 $\pm$ 0.03 & 7.49 $\pm$ 0.02 & 7.45 $\pm$ 0.02 & 6.28 $\pm$ 0.06 & 6.26$\pm$ 0.06 \\
\hline\end{tabular}
\label{tab:SMD}
 \end{table*}
\begin{figure*}
\begin{tabular}{cc}
\includegraphics[width=\columnwidth]{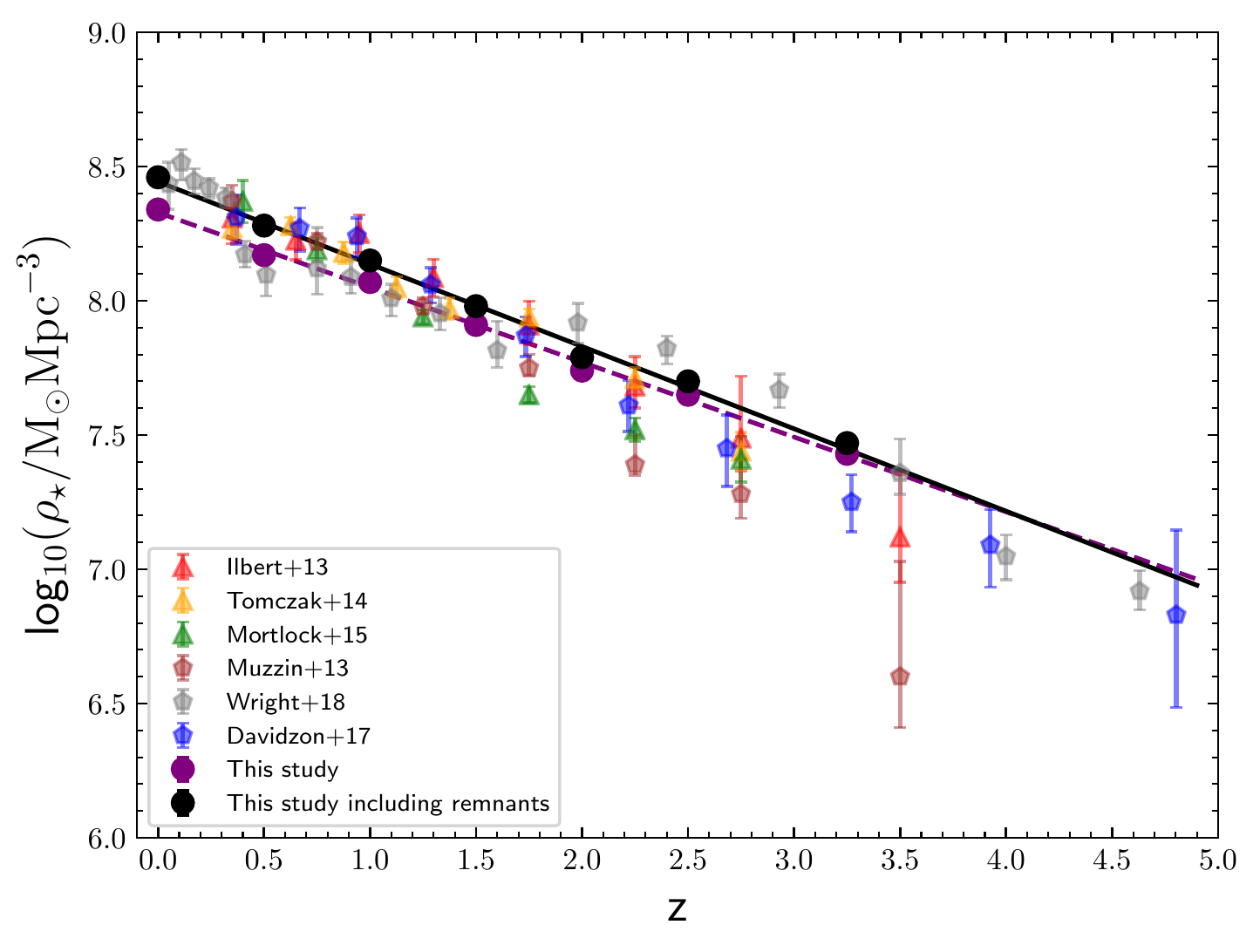}
\includegraphics[width=\columnwidth]{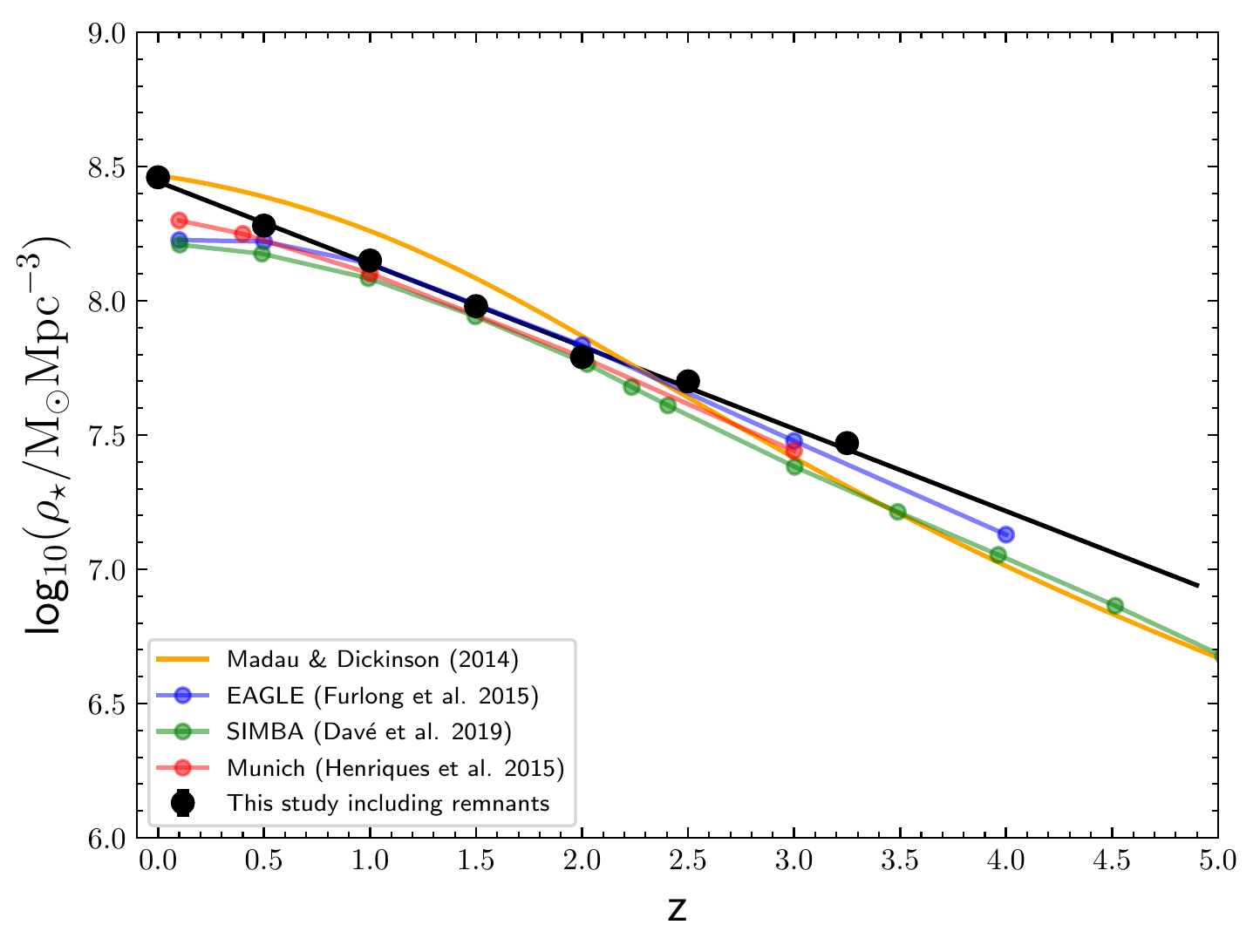}
\end{tabular}
\caption{The left-hand panel shows the redshift evolution of the integrated stellar-mass density (purple data points) based on integrating our derivation of the total intrinsic GSMF between the limits of $\mathcal{M}=8$ and $\mathcal{M}=13$. The dashed purple line is a log-linear fit to the purple data points and has the functional form: $\log_{10}(\rho_{\star}/\Msun\rm{Mpc}^{-3})=-0.28(\pm 0.01)z+8.33(\pm 0.01)$. For comparison, we also plot the results of \protect\cite{Ilbert2013,Muzzin2013, Tomczak2014, Mortlock2015, Davidzon2017, Wright2018}. Where necessary, the literature results
have been converted
to a Chabrier IMF and recalculated to match our adopted integration limits. The purple data point at $z=0$ is based on the local GSMF derived by \protect\cite{Baldry2012}. We also show the evolution of $\log_{10}(\rho_{\star})$ when including stellar remnants, which follows the relation $\log_{10}(\rho_{\star}/\Msun\rm{Mpc}^{-3})=-0.31(\pm 0.01)z+8.44(\pm 0.01)$ (black points, solid line). The right-hand panel shows a comparison
between the integrated stellar-mass density (including stellar remnants) derived here and the predictions of the SIMBA \protect\citep{Dave2019}, Munich \protect\citep{Henriques2015} and
EAGLE \protect\citep{Furlong2015} theoretical models. For comparison, we also plot the predicted stellar-mass density from \protect\cite{Madau2014}, based on integrating their fitting function to the evolving
cosmic star-formation rate density. The \protect\cite{Madau2014} curve has been converted to a Chabrier IMF.}
\label{fig:SMD_total_GSMF}
\end{figure*}

\begin{figure*}
\begin{tabular}{cc}
\includegraphics[width=0.99\columnwidth]{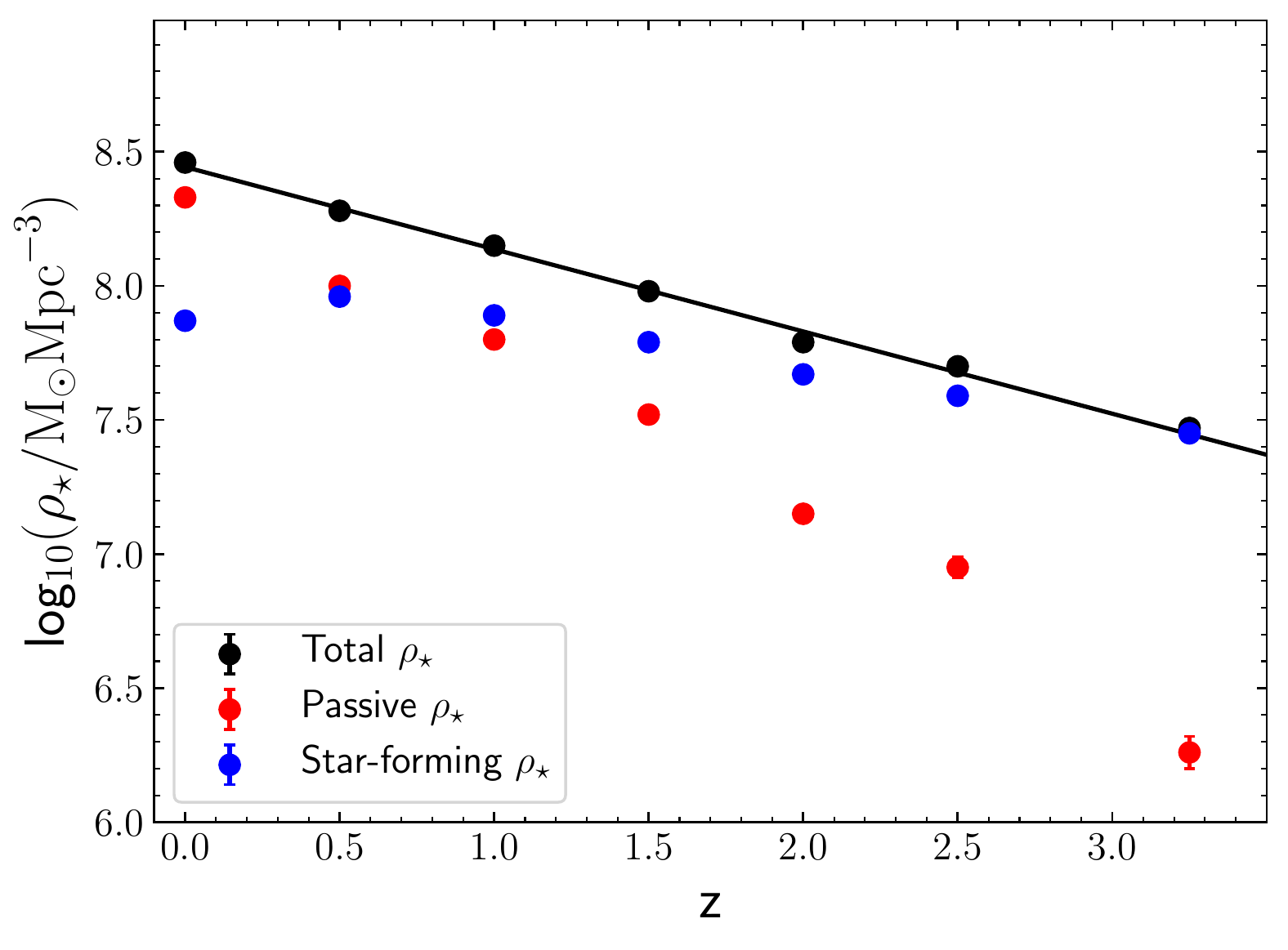} &
\includegraphics[width=0.99\columnwidth]{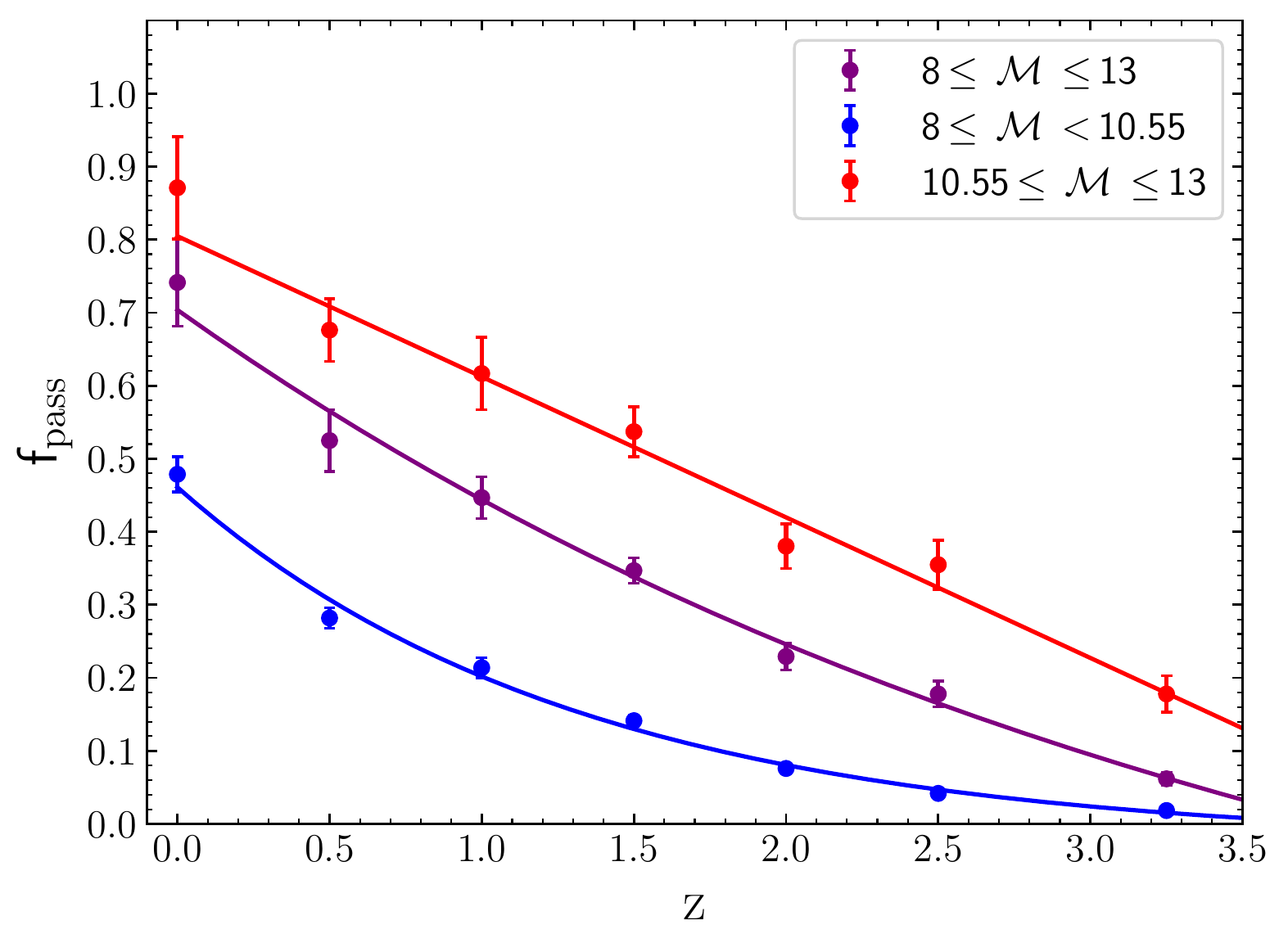} \\
\includegraphics[width=0.99\columnwidth]{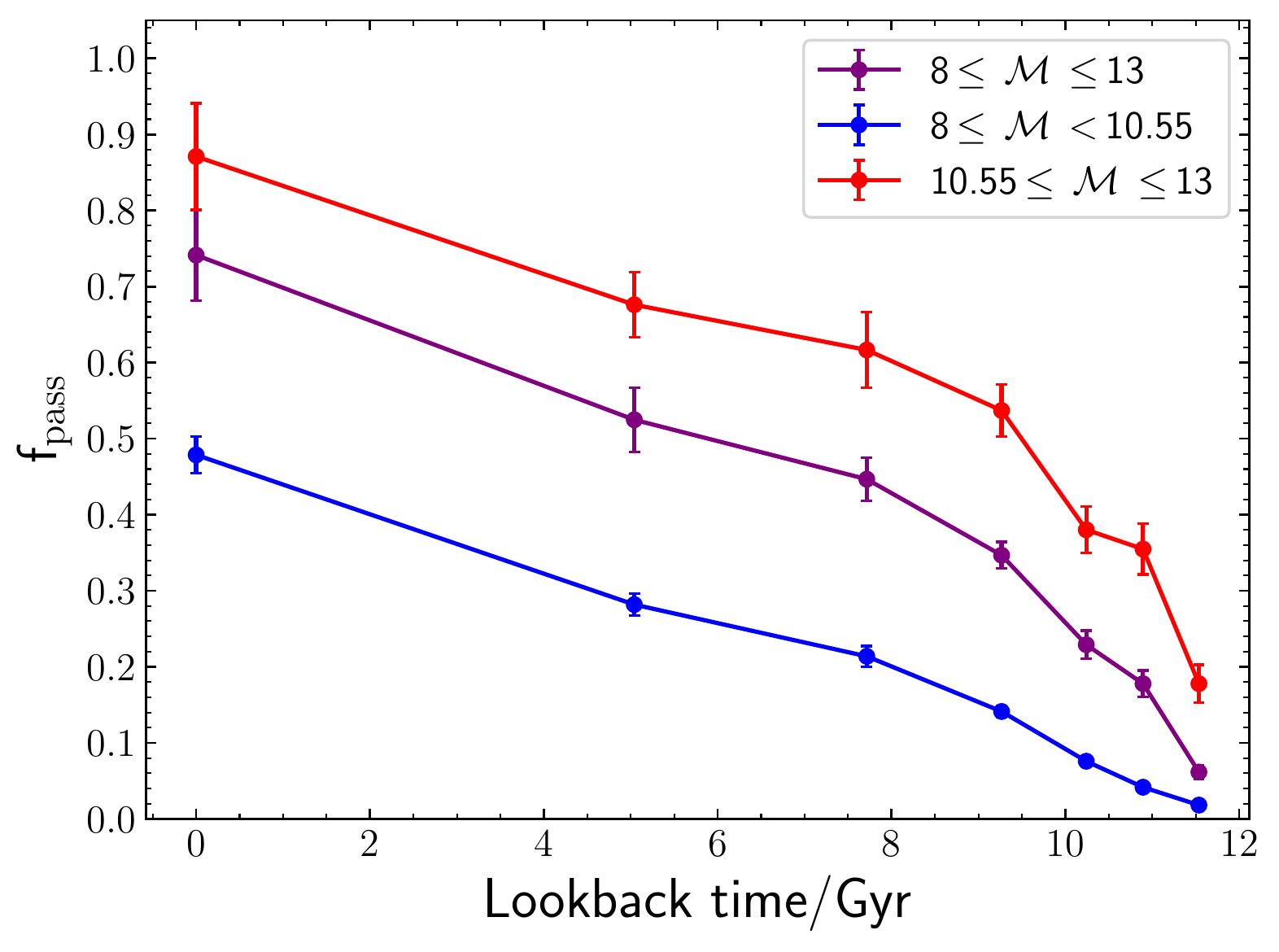} &
\includegraphics[width=0.99\columnwidth]{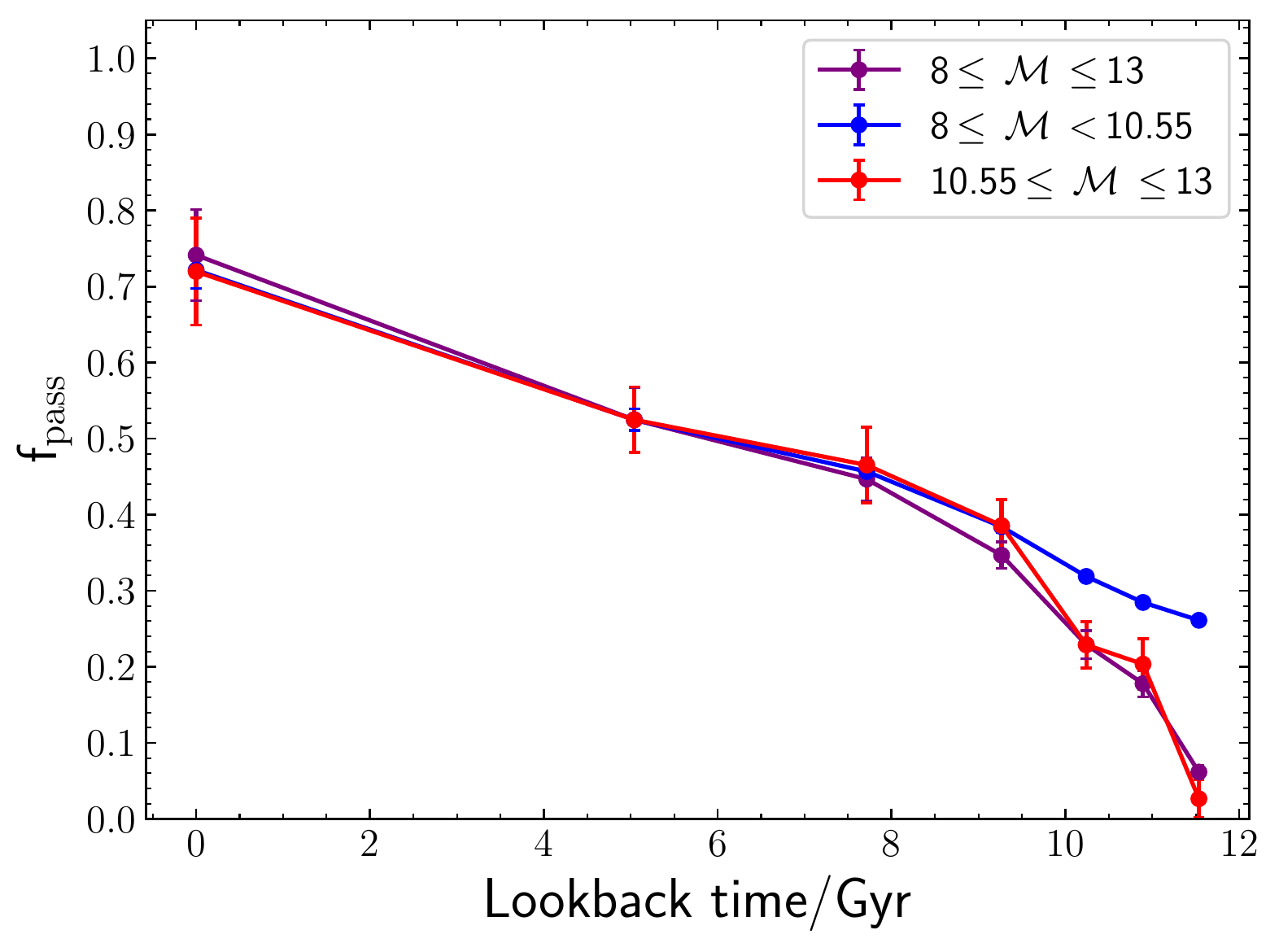}
\end{tabular}
\caption{The top left-hand panel shows the redshift evolution of the integrated stellar-mass density for the total (black), star-forming (blue) and passive (red) galaxy populations. In all cases the relevant GSMFs have been integrated between the limits of $\mathcal{M}=8$ and $\mathcal{M}=13$ and include stellar remnants. The top right-hand panel shows the redshift evolution of the fraction of the total stellar-mass density comprised of passive galaxies, in three different stellar-mass ranges. The purple data points correspond to the passive fraction over the full $8 \leq \mathcal{M} \leq 13$ stellar-mass range, whereas the red data points are the passive fraction in the stellar-mass range $10.55 \leq \mathcal{M}\leq 13$ and the blue data points are the passive fraction within the stellar-mass range $8 \leq \mathcal{M}< 10.55$ (see text for discussion). 
The sample has been split at $\mathcal{M}=10.55$ because for the double Schechter function fits to the total GSMF, $\mathcal{M}^{\star}=10.55\pm0.1$ over the full redshift range studied here (see Tables 5 \& 6). 
The bottom-left panel shows the same information as the top-right panel, but plotted against lookback time rather than redshift. The bottom-right panel again shows passive fraction versus lookback time, but here the red and blue data points have been shifted by the required amount to match the passive fraction of the purple data points at $z=0.5$ (i.e. lookback time $\simeq 5$ Gyr).}
\label{fig:SMD_UVJ}
\end{figure*}

In order to investigate the redshift evolution of the assembled stellar-mass density, we integrate our intrinsic double Schechter function fits \footnote{We will focus here on
the intrinsic stellar-mass densities, but note that these are only $\simeq 0.02$ dex lower than the observed values.} presented in Table \ref{tab:intrinsic_Schechter_parameters} between the stellar-mass limits $\mathcal{M}=8$ and $\mathcal{M}=13$. To calculate the
local stellar-mass density we integrated the double Schechter fit provided by \cite{Baldry2012} between the same limits. We show a comparison between our stellar-mass density results (purple data points) and
those from comparable previous literature studies in the left-hand panel of Fig.~\ref{fig:SMD_total_GSMF}. Where necessary, we have converted the literature results to a Chabrier IMF and the same integration limits.

It is important to note that stellar-mass densities based on integrating the Schechter function parameters presented in Tables 4, 5, 9 \& 10 are living main sequence stellar masses densities, and do not include stellar remnants. To illustrate the difference, the black data points in the left-hand panel of Fig.~\ref{fig:SMD_total_GSMF} show our stellar-mass density results including the contribution of stellar remnants. 

It can be seen from Fig.~\ref{fig:SMD_total_GSMF} that our stellar-mass density results are consistent with the majority of previous studies but, thanks to the large volume and dynamic
range in stellar mass provided by the current data set, carry significantly smaller uncertainties. The dashed purple line shown in the left-hand panel of Fig.~\ref{fig:SMD_total_GSMF} is a log-linear fit to our living main sequence stellar-mass densities, and has the form:
\begin{equation}
\log_{10}(\rho_{\star}/\Msun\rm{Mpc}^{-3})=-0.28(\pm 0.01)z+8.33(\pm 0.01).
\end{equation}
We note that this relationship is very similar to the previous determination of \cite{Tomczak2014}, but with a somewhat shallower slope. The solid black line in the left-hand panel of
Fig.~\ref{fig:SMD_total_GSMF} is a log-linear fit to our stellar-mass densities including stellar remnants, and has the form:
\begin{equation}
\log_{10}(\rho_{\star}/\Msun\rm{Mpc}^{-3})=-0.31(\pm 0.01)z+8.44(\pm 0.01).
\end{equation}

In the right-hand panel of Fig.~\ref{fig:SMD_total_GSMF} we show a comparison between our stellar-mass density results (including remnants) and the predictions of the
SIMBA \citep{Dave2019}, Munich \citep{Henriques2015} and EAGLE \citep{Furlong2015} theoretical models. On the same plot, we also show the
predicted stellar-mass densities from \cite{Madau2014}, based on integrating their fitting function to the evolving
cosmic star-formation rate density and converting to a Chabrier IMF.

Overall, it can be seen that there is good agreement between our observational results and the predictions from the
latest hydrodynamical and semi-analytic galaxy evolution models. Over the majority of the redshift
range explored by this study, the observational and theoretical results are in agreement to within $\simeq 0.15$ dex although, as discussed in Section 4.5, some difference do exist with regard to the precise shape of the evolving GSMF.

Given that the stellar-mass densities predicted by integrating the \cite{Madau2014} fit to the cosmic star-formation rate density include the contribution from stellar remnants, it
is clearly of interest to compare them to our direct results. 
It can be seen from the right-hand panel of Fig.~\ref{fig:SMD_total_GSMF} that the \cite{Madau2014} curve begins to overshoot our observed data at $z\simeq 2.0$, reaching a maximum
discrepancy of $\simeq 0.1$ dex at $z\simeq 1.0$, before closing again to fall into excellent agreement at $z=0$. This effect is well known, and the potential
reasons for the discrepancy are discussed at length in  \cite{Madau2014}. However, it is noteworthy that based on our new observational results, using a Chabrier IMF
and including the contribution of stellar remnants, the discrepancy is much smaller than has often been reported in the literature.

Given the excellent agreement between our observational data, the \cite{Madau2014} curve and all three theoretical models at $z=2.0-2.5$, combined with the continued agreement
between our results and the theoretical models at $z\leq 2.0$, it is tempting to speculate that the SFR estimates used to study the evolution of the cosmic star-formation rate density are
systematically over-estimated in the redshift interval $0.5 \leq z \leq 2.5$. Within this context, we note that an evolving off-set between observed and intrinsic SFRs of this form
is predicted by the {\sc universemachine} model of \cite{Behroozi2019}. Alternatively, it is clearly possible that our stellar masses could be systematically under-estimated, perhaps
due to a failure to correctly account for the contribution of older stellar populations (e.g. \citealt{Leja2020, Carnall2019}). However, if this is the case, any systematic increase
in the stellar masses must be limited to a relatively modest $\simeq 0.1$ dex.

\subsection{The evolving stellar-mass density of passive galaxies}
In the top-left panel of Fig.~\ref{fig:SMD_UVJ} we compare the redshift evolution of the integrated stellar-mass density with the evolution of the separate star-forming and passive galaxy contributions. As with the total stellar-mass densities, the star-forming and passive contributions have been calculated by integrating the best-fitting intrinsic Schechter function fits between the stellar-mass limits $\mathcal{M}=8$ and $\mathcal{M}=13$. This panel illustrates the dramatic rise in the stellar-mass density of passive galaxies, from providing an essentially negligible contribution at $z\geq 3$, to reaching parity with the star-forming galaxy population between $z=1.0$ and $z=0.5$, to dominating the stellar-mass density in the local Universe. We note that the data shown in the top-left panel of Fig.~\ref{fig:SMD_UVJ} is in good agreement with the results derived by \cite{Tomczak2014}.

In the top-right panel of Fig.~\ref{fig:SMD_UVJ} we explore this issue further by plotting the redshift evolution of the passive fraction, $f_{\rm pass}~=~\rho_{\star}^{\rm pass}/\rho_{\star}^{\rm tot}$, in three different stellar-mass ranges. The purple data points show the evolution of $f_{\rm pass}$ calculated within the full $8 \leq \mathcal{M} \leq13$ stellar-mass range. The purple curve is a fit of the form: $f_{\rm pass}=a\exp{(-bz)}+c$, with best-fitting parameters $a=1.095\pm 0.183, b=0.271\pm 0.093$ and $c=-0.392\pm 0.215$. Based on the fitted curve, passive galaxies dominate the total stellar-mass density budget at $z\leq 0.75$, but contribute $\lesssim 10\%$ by $z\simeq 3$.

As discussed in Section 4.4, the characteristic stellar mass of the double Schechter function fit to the intrinsic GSMF is remarkably stable at $\mathcal{M}^{\star}=10.55\pm{0.1}$, over the full redshift range studied here. 
To capture the different behaviour of the passive fraction either side of $\mathcal{M}^{\star}$, the blue data points in the top-right panel of Fig.~\ref{fig:SMD_UVJ} show $f_{\rm pass}$ integrating over the stellar-mass range $8 \leq \mathcal{M}<10.55$ (i.e. $\mathcal{M}<\mathcal{M}^{\star}$) and the red data points show $f_{\rm pass}$ integrating over the stellar-mass range $10.55 \leq \mathcal{M} \leq 13$ (i.e. $\mathcal{M} \geq \mathcal{M}^{\star}$). Comparison of the red and blue data points shows a clear downsizing signature, with a significantly higher $f_{\rm pass}$ amongst the $\mathcal{M} \geq \mathcal{M}^{\star}$ galaxies at all redshifts. The fit to the red data points is a linear relation of the form: $f_{\rm pass}=az + b$, with best-fitting parameters of $a=~-~0.193\pm 0.014$ and $b=0.805\pm 0.031$. This relation indicates that the $\mathcal{M} \geq \mathcal{M}^{\star}$ galaxies already have $f_{\rm pass}\simeq 0.15$ by $z=3.5$ and reach a passive fraction of $f_{\rm pass}=0.5$ by $z\simeq 1.6$. The curve fitted to the blue data points has the form: $f_{\rm pass}=a\exp{(-bz)}+c$, with best-fitting parameters $a=0.487\pm 0.017, b=0.759\pm 0.064$ and $c=-0.026\pm 0.011$. In contrast to the high-mass galaxies, this indicates that the $\mathcal{M}<\mathcal{M}^{\star}$ galaxies have a passive fraction of essentially zero at $z=3.5$ and only reach approximate parity between the stellar-mass contributions of the star-forming and passive populations at $z\simeq 0$.

In the bottom-left panel of Fig.~\ref{fig:SMD_UVJ} we re-plot the $f_{\rm pass}$ data as a function of lookback time ($t_{\rm lb}$). Plotting the data in this fashion serves to highlight the rapid build-up of the passive fraction at high stellar-masses, where the dominant quenching mechanism is thought to be mass quenching (c.f. \citealt{Peng2010}). Moreover, plotting the data versus lookback time also highlights the fact that the increase in $f_{\rm pass}$ at lookback times of $t_{\rm lb}\lesssim 8$ Gyr appears to follow a very similar slope in all three stellar-mass ranges. This is confirmed by the bottom-right panel of Fig.~\ref{fig:SMD_UVJ}, in which the red and blue data points have been shifted vertically in order to match the passive fraction calculated over the full stellar-mass range at a lookback time of $t_{\rm lb}\simeq 5$ Gyr (i.e.~$z=0.5$).

The results shown in the bottom-right panel of Fig.~\ref{fig:SMD_UVJ} indicate that
the rate of increase in $f_{\rm pass}$ appears to be largely independent of stellar mass at lookback times of $t_{\rm lb}\lesssim 8$ Gyr. When combined with the observed stability of the star-forming GSMF over this epoch (see Section 5.3), this suggests that the quenching rates (i.e. $\frac{d\rho_{\star}}{dt}$) at the low and high-mass end of the GSMF must be broadly comparable.

\section{Conclusions}
In this paper we have presented a new derivation of the GSMF over the redshift interval $0.25 \leq z \leq 3.75$,
based on a near-IR selected galaxy sample covering a raw survey area of 3 deg$^{2}$ and spanning $\geq 4$ dex in stellar mass.
The powerful combination of a large dynamic range in stellar mass and a large, non-contiguous, cosmological volume 
has allowed us to robustly constrain both the high and low-mass end of the GSMF. Moreover, by carefully accounting for Eddington bias, we
have been able to derive best-fitting Schechter function parameters for both the observed and intrinsic GSMF. By splitting our galaxy sample into its
constituent parts, we have investigated the differential evolution of the star-forming and passive GSMFs and explored their evolving contribution
to the integrated stellar-mass density. Where appropriate, we have compared our new results with previous observational constraints
and the predictions of both phenomenological models and galaxy evolution simulations. Our main conclusions can be summarized as follows:

\begin{enumerate}

\item{We find that a double Schechter function is a better fit to both the observed and intrinsic GSMF over the full redshift
range explored in this study, although by $z\simeq 3.25$ the single and double Schechter function fits are indistinguishable.}

\item{The redshift evolution of the GSMF is remarkably smooth in general, and we find no evidence for significant evolution in $\mathcal{M}^{\star}$.
Over the full redshift range explored, the best-fitting values of $\mathcal{M}^{\star}$ are consistent with $\mathcal{M}^{\star}=10.75\pm{0.1}$ and
$\mathcal{M}^{\star}=10.55\pm{0.1}$ for the observed and intrinsic GSMFs, respectively.}

\item{Motivated by the smooth evolution of the GSMF, we derive a simple evolving parameterization that can provide an accurate estimate of either the observed or intrinsic GSMF at any desired redshift within the range $0 \leq z \leq 4$.}

\item{Our new determination of the GSMF is in generally good agreement with the predictions of the EAGLE, SIMBA and Munich galaxy evolution models although, in detail, differences still exist. In particular, all SIMBA and Munich models have a tendency to over-predict the number densities of
high-mass ($\mathcal{M}\geq 11.5$) galaxies over the full redshift range.}

\item{Splitting our galaxy sample into its constituent star-forming and passive galaxy components, we find that the star-forming GSMF is adequately
described by a single Schechter function at all redshifts. Moreover, we find that the star-forming GSMF has not evolved significantly since $z\simeq 2.5$.}

\item{In contrast, we find that the passive GSMF has evolved significantly over the redshift range explored by this study, both in normalization and
functional form. We find that the passive GSMF is best described by a double Schechter function at $z\leq 1.5$, but can be described by a single
Schechter function at higher redshifts. Based on our current data set, it is not possible to determine if this change in functional form is intrinsic, or
the result of insufficient dynamic range in stellar mass at $z\geq 1.5$.}

\item{We find that the \cite{Peng2010} phenomenological model does a qualitatively good job of reproducing the functional form of our star-forming
and passive GSMFs at $z\leq 1.5$, but appears to perform less well at higher redshifts. That said, more dynamic range in stellar mass will
be required to robustly confirm that the components of the high-redshift GSMF deviate significantly from the predictions of this continuity-based model.}

\item{Based on our new determinations of the evolving GSMF, we find that the redshift evolution of the integrated stellar-mass density (including stellar remnants) is well described by a log-linear relation of the form: $\log_{10}(\rho_{\star}/\Msun\rm{Mpc}^{-3})=-0.31  (\pm 0.01)z+8.44 (\pm 0.01)$ out to $z\simeq 4$. This functional form is in agreement with, although much better constrained than, previous literature results, and in excellent agreement with the predictions of recent
theoretical galaxy evolution models.}

\item{We find that the passive galaxy contribution to the integrated stellar-mass budget ($f_{\rm pass}=\rho_{\star}^{\rm pass}/\rho_{\star}^{\rm tot}$)
evolves by an order of magnitude over the redshift range explored in this study. Within the stellar-mass range $8\leq \mathcal{M} \leq 13$, we find that passive galaxies dominate the total integrated stellar-mass budget at $z\leq 0.75$, but only contribute $\lesssim 10\%$ at $z\simeq 3$.}

\item{By exploring the evolution of $f_{\rm pass}$ within low stellar-mass ($\mathcal{M}<\mathcal{M}^{\star}$) and high stellar-mass ($\mathcal{M}>\mathcal{M}^{\star}$) sub-samples, we find that the rate of increase in the passive fraction appears to be largely independent of stellar mass at lookback times of $t_{\rm lb}\lesssim 8$~Gyr. This suggests that at this epoch the quenching rates (i.e. $\frac{d\rho_{\star}}{dt}$) at the low and high-mass end of the GSMF are broadly comparable.}

\end{enumerate}

\section*{acknowledgements}
We would like to thank the anonymous referee for their comments and suggestions which improved the paper. F. Cullen acknowledges support from the UK Science and Technology Facilities Council.
The authors would like to thank Romeel Dav\'{e} for providing the SIMBA data and useful discussions.
This work is based in part on data products from observations made with ESO Telescopes at the La Silla Paranal Observatory under ESO programme ID 179.A-2005 and
ID 179.A-2006 and on data products produced by CALET and the Cambridge Astronomy Survey Unit on behalf of the UltraVISTA and VIDEO consortia. This work is based in part on
data obtained as part of the UKIRT Infrared Deep Sky Survey. This work is based in part on observations made with the NASA/ESA {\it Hubble Space Telescope}, which is
operated by the Association of Universities for Research in Astronomy, Inc; under NASA contract NAS5-26555.
This work is also based in part on observations made with the {\it Spitzer Space Telescope}, which is operated by the Jet Propulsion Laboratory, 
California Institute of Technology under NASA contract 1407.
This work is based in part on observations obtained with MegaPrime/MegaCam, a joint project of CFHT and CEA/IRFU, at the Canada-France-Hawaii Telescope (CFHT) which is operated by the National Research Council (NRC) of Canada, the Institut National des Science de l'Univers of the Centre National de la Recherche Scientifique (CNRS) of France, and the University of Hawaii. This work is based in part on data products produced at Terapix available at the Canadian Astronomy Data Centre as part of the Canada-France-Hawaii Telescope Legacy Survey, a collaborative project of NRC and CNRS.

\bibliographystyle{mnras}
\bibliography{GSMF_McLeod_2021.bib}

\section*{data availability}
The data underlying this article will be shared on reasonable request to the corresponding author.

\begin{appendix}
\section{The evolving fit to the GSMF}
\begin{figure*}
\includegraphics[width=16.0cm]{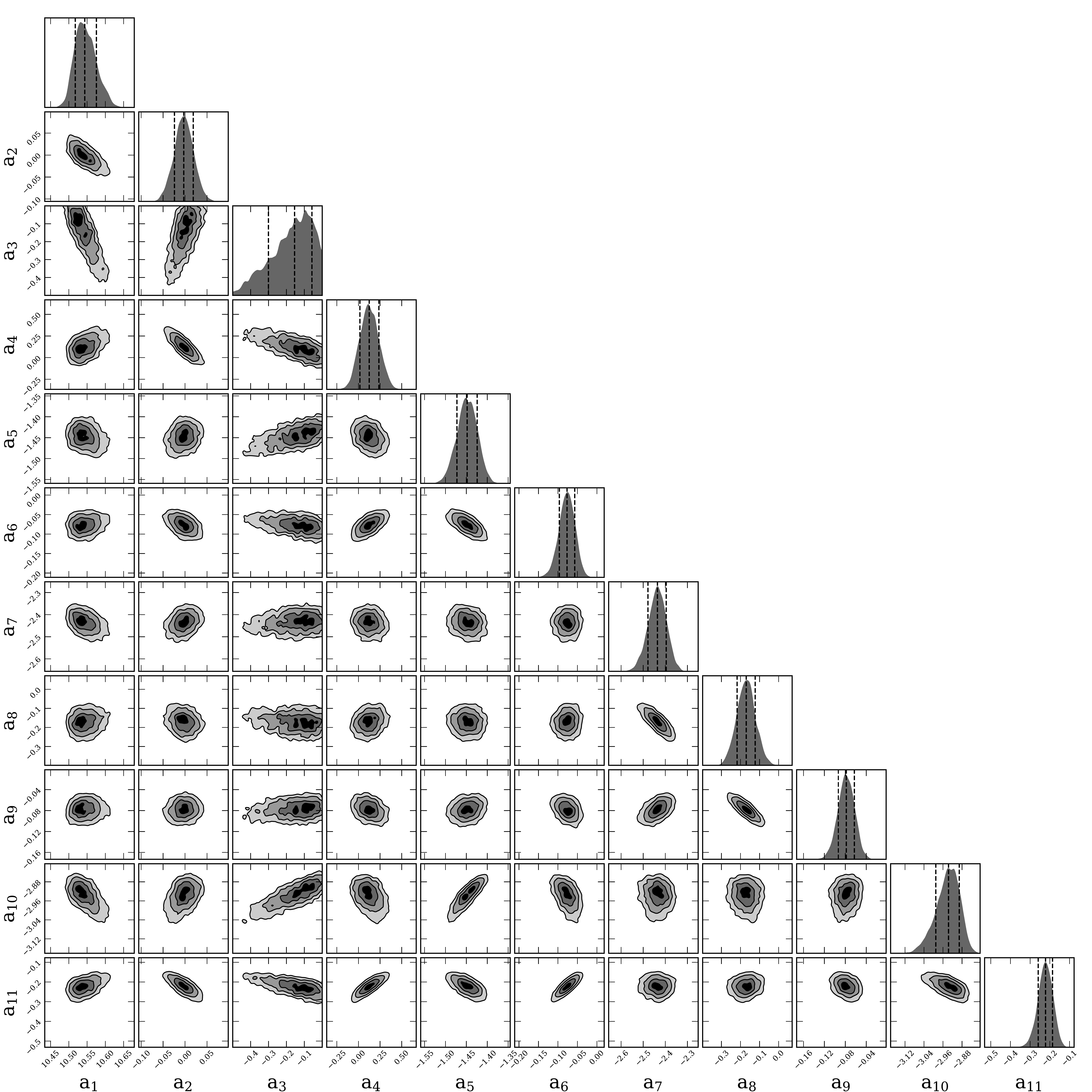}
\caption{A corner plot showing the 1-D and 2-D marginalized posteriors for our 11-parameter
evolving model for the intrinsic GSMF. The 16\%, 50\% and 84\% percentiles (i.e. the 1-sigma constraints) are shown by
the vertical dashed lines.}
\label{fig:corner}
\end{figure*}

In Section 4.6 we presented an 11-parameter model designed to describe the evolving form of either
the observed or intrinsic GSMF. Based on this model, it is possible to produce accurate estimates of both the observed and
intrinsic GSMFs at any redshift within the range $0 \leq z \leq 4$. For completeness, in Fig.~\ref{fig:corner} we
present a corner plot showing the 1-D and 2-D marginalized posteriors for the 11-parameter fit describing the evolution of the intrinsic GSMF.

\end{appendix}

\bsp	
\label{lastpage}

\end{document}